\documentclass{aastex}
\usepackage{emulateapj5}

\shorttitle{MIR Observations of Star Forming Regions with Water Masers}
\shortauthors{De Buizer et al.}

\begin{document}

\title{Observations of Massive Star Forming Regions with Water Masers: Mid-Infrared Imaging}

\author{J. M. De Buizer\altaffilmark{1}}
\affil{Gemini Observatory, Casilla 603, La Serena, Chile}
\affil{and Cerro Tololo Inter-American Observatory, Casilla 603, La Serena, Chile\altaffilmark{2}}
\email{jdebuizer@gemini.edu}
\author{J. T. Radomski, C. M. Telesco, and R. K. Pi\~{n}a\altaffilmark{1,3}}
\affil{Department of Astronomy, University of Florida, Gainesville, FL 32611, USA}

\altaffiltext{1}{Visiting Astronomer at the Infrared Telescope Facility, which is operated by the University of Hawaii under Cooperative Agreement no. NCC 5-538 with the National Aeronautics and Space Administration, Office of Space Science, Planetary Astronomy Program.}
\altaffiltext{2}{CTIO is operated by AURA, Inc., under contract to the National Science Foundation.}
\altaffiltext{3}{Present address: Photon Research Associates, Inc., 5720 Oberlin Drive, San Diego, CA 92121, USA}

\begin{abstract}
We present here a mid-infrared imaging survey of 26 sites of water maser emission. Observations were obtained at the InfraRed Telescope Facility 3-m telescope with the University of Florida mid-infrared imager/spectrometer OSCIR, and the JPL mid-infrared camera MIRLIN. The main purpose of the survey was to explore the relationship between water masers and the massive star formation process. It is generally believed that water masers predominantly trace outflows and embedded massive stellar objects, but may also exist in circumstellar disks around young stars. We investigate each of these possibilities in light of our mid-infrared imaging. We find that mid-infrared emission seems to be more closely associated with water and OH maser emission than cm radio continuum emission from UC \ion{H}{2} regions. We also find from the sample of sources in our survey that, like groups of methanol masers, both water and OH masers have a proclivity for grouping into linear or elongated distributions. We conclude that the vast majority of linearly distributed masers are not tracing circumstellar disks, but outflows and shocks instead.
\end{abstract}


\section{Introduction}

The discovery of the first astronomical maser was the hydroxyl (OH) maser by Weaver et al. (1965), followed shortly thereafter by the first detection of an astronomical water maser by Cheung et al. (1969). In the years following these major discoveries, astronomers found that these maser species were closely associated with phenomena related to the formation of massive stars, in particular they appeared to be directly associated with regions of cm radio continuum and far infrared emission. The extremely reddened far infrared colors of these regions as seen with IRAS seemed to confirm the idea that the masers are located in the hot, dusty environments of star formation containing copious amounts of radio emission. Masers, therefore, became known as signposts of massive star formation. However, advances in higher resolution radio imaging and accurate astrometry in recent decades have led to observations that show that not all masers are directly coincident with cm radio continuum emission. Genzel \& Downes (1977) conducted the first water maser study toward known \ion{H}{2} regions and discovered that while OH masers tend to be in or projected on \ion{H}{2} regions, water masers were typically offset. Other observations followed (e.g. Forster \& Caswell 1989; Hofner \& Churchwell 1996), weakening the direct physical link between water masers and \ion{H}{2} regions. If masers are close to, but not directly coincident with young massive stars, what phenomena or processes do these masers trace?

Masers in general are excited to emit from both radiative and collisional processes. In the star forming environment there are several possible processes and locations that have been suggested where masers can exist. It has been suggested (e.g., Elitzur 1992) that the cool, dense layer of gas between the ionization and shock fronts in the expanding \ion{H}{2} regions around young massive stars may provide a habitable zone for masers (OH masers, in particular). The bulk motion and relatively high density of molecular material caught up in a well-collimated bipolar outflow or jet from a young star may, in principle, be a good location for maser emission (e.g., Torrelles et al. 1997). Even if masers are not taking part in the outflow from a young stellar object, the shock created by an outflow as it impinges on the ambient medium or on knots of material in the immediate vicinity of an outflowing star, also seem to be good locations for masers (water masers, in particular). The idea that masers are excited by embedded protostellar objects was originally suggested by Mezger and Robinson (1968) for OH masers. Specifically, these masers would exist in the accreting envelopes of massive protostars and be excited by the energy from accretion shocks at stages of early evolution \textit{before} the onset of an \ion{H}{2} or ``ultra-compact \ion{H}{2}'' (UC \ion{H}{2})  region (Forster \& Caswell 2000). Water and OH masers have also been suggested to exist in the warm and dense environment of circumstellar disks (e.g., Torrelles et al. 2002, and references therein).

It is difficult to link any of these phenomena of massive star formation to the maser emission when it is not known where the locations of all the associated young stellar objects are in a particular region with respect to the masers. Clearly, traditional methods of searching for massive stars by imaging in the cm radio continuum will not reveal pre-ionizing protostars in embedded envelopes (so-called ``high mass protostellar objects'' or HMPOs). Furthermore the jets, outflows, and winds from lower mass non-ionizing stars have been found to be associated with water maser emission (e.g., Wilking \& Classen 1987). Since massive stars are believed to form in clusters, one would expect there to be stars of a range of masses and phases of early stellar evolution present. Consequently, these regions need to be imaged at wavelengths other than in the cm continuum to find all the stellar sources within the regions of maser emission. However, massive star forming regions are generally located deep within giant molecular clouds, and are thus obscured at visible wavelengths. Dust absorbs and scatters visible light in these regions giving rise to significant extinction. However, infrared radiation is much less affected by extinction than visible radiation, and thus infrared imaging can probe through the cool obscuring dust enshrouding the massive stellar environment.

Infrared observations of these regions must be performed at moderate to high spatial ($\lesssim$1$\arcsec$) resolutions. Given that the average distances to massive star forming regions are several kiloparsecs away, and that massive stars tend to form in a highly clustered way, resolution is an issue when trying to determine which young stellar object is most closely associated with the maser emission. However, large ground-based telescopes with mid-infrared detectors yield both the required resolution and the ability to penetrate the significant obscuration in these regions. Mid-infrared radiation traces the warm dust close to young stellar sources, allowing one to observe the spatial relationship between masers and young stars. Combined with comparable spatial resolution cm radio continuum maps of these regions, mid-infrared images will allow identification of the vast majority of stellar and protostellar sources in each region of maser emission with none of the effects of confusion from foreground or background objects.

We present here a mid-infrared imaging survey of 26 water maser sites taken mostly from a list of sources from Forster \& Caswell (1989) and Hofner\& Churchwell (1996). Most of these sources are imaged here at $\sim$1$\arcsec$ resolution in the mid-infrared for the first time. The general goal of this work is to try to determine the relationship between water masers and massive star formation processes. A small subset of this survey was already published by De Buizer et al. (2003), which was concentrated on specific mid-infrared sources believed to be associated with water maser emission from HMPO candidates. This paper is a broader and larger-scale mid-infrared survey designed to study the full array of phenomena water (and to a small extent OH masers) may trace in these massive star forming regions. This mid-infrared survey of water maser sites is meant to be complementary to our results from our mid-infrared (De Buizer, Pi\~{n}a, \& Telesco 2000) and near-infrared (De Buizer 2003) surveys of massive stars associated with methanol maser emission.

In \S2 we will discuss the observations, and in \S3 the data reduction. We will discuss what is already known about each of the individual regions of the survey, and what new things we have learned in light of our mid-infrared observations in \S4. In \S5 we will discuss what we have learned from this survey as a whole, and we will compare and contrast this to the results of De Buizer, Pi\~{n}a, \& Telesco (2000) for massive star forming regions associated with methanol masers. We will end with our conclusions in \S6.

\section{Observations}

Exploratory observations were performed using the University of Florida mid-infrared camera and spectrometer, OSCIR, in 1997 September at the 3-meter NASA Infrared Telescope Facility (IRTF). This instrument employs a Rockwell 128 $\times$ 128 Si:As BIB (blocked impurity band) detector array, which is optimized for wavelength coverage between 8 and 25 \micron. The field of view of the array is 29$\arcsec$ $\times$ 29$\arcsec$, for a scale of 0.223$\arcsec$$\cdot$pixel$^{-1}$. Observations were centered on the H$_{2}$O maser reference feature coordinates given in Table 1, with 30 second on-source exposure times taken through a broad-band \emph{N} filter ($\lambda _{0}$ = 10.46 $\micron$, $\Delta \lambda $ = 5.1 $\micron$) and the \emph{IHW18} (International Halley Watch, $\lambda _{0}$ = 18.06 $\micron$, $\Delta \lambda $ = 1.7 $\micron$) filter. Unfortunately, cirrus clouds terminated the survey before all sites were observed through both filters, however most sites were observed through the \emph{N} filter. The standard star for all observations was $\gamma$ Aql, for which the flux density was taken to be 74.9 Jy in the \emph{N} filter, and 25.7 Jy in the \emph{IHW18}, both of which were derived from the templates of Cohen et al. (1999). Many sites contained sources barely detected in the 30 second exposures, so we decided to expand the observations and use longer exposure times.

The full survey was performed in 2002 June, again at the IRTF, but this time using the Jet Propulsion Laboratory mid-infrared camera, MIRLIN. This instrument employs a Boeing HF-16 128 $\times$ 128 Si:As BIB (blocked impurity band) detector array. The pixel-scale is 0.475$\arcsec$$\cdot$pixel$^{-1}$, for a field of view of 61\arcsec\ $\times$ 61\arcsec. Observations were taken through the \emph{N4} ($\lambda_0$ = 11.70 $\micron$, $\Delta\lambda$ = 1.11 $\micron$) and \emph{Q3} ($\lambda_0$ = 20.81 $\micron$, $\Delta\lambda$ = 1.65 $\micron$) filters with exposures times of 184 and 192 sec, respectively. Twenty-six water maser sites were imaged, including all those observed previously with OSCIR. All observations were taken at airmasses $<$1.5 under clear skies with low relative humidity ($<$25\%). The standard stars used throughout the observations were $\gamma$ Aql, for which the flux densities were taken to be 61.0 Jy in the \emph{N4} filter and 19.7 in the \emph{Q3}, and $\beta$ Gem, for which the flux densities were taken to be 281.0 Jy and 102.0 Jy in the \emph{N4} and \emph{Q3} filters, respectively. These values were also derived from the templates by Cohen et al. (1999).

Before imaging each maser site, the telescope was slewed between two or three stars with accurate coordinates obtained from the Hipparcos Main Catalogue. These reference stars all lied within 15\arcmin\ of the target position. Slewing between these references stars showed that the telescope slewed very accurately; each star appeared centered in the visual camera to within a few 1/10ths of an arcsecond of the centering crosshairs. The visual camera and MIRLIN were aligned so that when a source is centered in the crosshairs of the visual camera, it is also centered on the MIRLIN array in the 11.7 $\micron$ filter. From the closest reference star, the telescope was slewed to the H$_{2}$O maser coordinates and images were obtained. Therefore, the water maser reference position on the mid-infrared images is always the center (pixel x = 64, y = 64) of the array. This same technique was used for both these observations and the earlier OSCIR observations. The combined OSCIR and MIRLIN observations confirm that the absolute pointing of the telescope is good to $<$1.0$\arcsec$, and we quote this as the accuracy in the astrometry between the mid-infrared images and the positions of the water maser spots.

Point-spread function (PSF) stars were imaged through each filter near the positions of most of the targets. Error in the PSF size was taken to be the standard deviation of the size of the PSF stars imaged throughout the night. A target object was considered to be resolved if the measured full width at half maximum (FWHM) was greater than three standard deviations from its closest PSF FWHM. The average PSF FWHM, and hence the resolution of the observations is 1$\farcs$2 at \emph{N} and 1$\farcs$6 at 18.1 $\mu$m using OSCIR, and 1$\farcs$3 at 11.7 $\mu$m and 1$\farcs$7 at 20.8 $\mu$m when MIRLIN was used. Table 3 has labels showing all sources that are believed to be resolved or unresolved. Most of the images shown in Figures 1-20 have a modest amount of smoothing made by convolving the image with a gaussian of a certain FWHM (typically 0$\farcs$5 to 1$\farcs$0), and this information is given in the figure captions.

\section{Results and Data Reduction}

Of the 26 sites we observed in the mid-infrared, we had seven sites that yielded no mid-infrared source within 5$\arcsec$ of the maser positions (Table 1). Table 2 lists the sites where detections were made, and the corresponding observed flux densities. The sites where no detections were found are listed with a 3-$\sigma $ upper limit for a point source flux density. Sources marked with a `w' or an `h' in Tables 2 and 3 are those closest to or those thought to be associated with the water  and hydroxyl masers, respectively. Not all sites were observed through all four filters, and for those sites that were, not all sources were detected at all four wavelengths. For each site where there were mid-infrared sources detected, representative images at a shorter and longer wavelength are presented in Figures 1-20. Many sites contained multiple sources which are labeled ``1'', ``2'', ``3'', etc.\footnote{These labels are the IAU recommended names which are in the form Glll.ll$\pm $b.bb:DRT04 \#. For instance, we find that G12.68-0.18 has two mid-infrared sources, whose names in the full form are G12.68-0.18:DRT04 1 and GG12.68-0.18:DRT04 2. Some sources already have names, as shown in the tables and discussed in Section 4.} in the figures so they can be addressed individually. Individual water masers are plotted in the figures as crosses, and OH masers as triangles. In most cases these maser positions are from Forster \& Caswell (1989) or Hofner \& Churchwell (1996), unless otherwise noted. For some fields the locations of known near-infrared sources from Testi et al. (1994) and Testi et al. (1998) are plotted as boxes in the figures. For most sources radio continuum or molecular line maps from the literature are also shown overlaid on one of the mid-infrared images. Details about the near-infrared and radio observations are discussed in more detail in \S4 and \S5.

\subsection{Observed Flux Densities}

The calibration factor (ratio of accepted flux in janskys to analog-to-digital converter units (ADUs) per second per pixel) derived from the standard star observations varied throughout the course of each night mostly as a result of changes in atmospheric conditions. There was an overall trend as a function of airmass only on the nights with MIRLIN. Therefore, air mass corrections were made to the MIRLIN 11.7 and 20.8 $\mu$m observations only. We estimate the absolute photometric accuracy for the OSCIR night from the standard deviation of the mean observed standard star flux density throughout the night. This was found to be 2.1\% in the \emph{N} filter and 11.2\% in the 18.1 $\mu$m filter. The absolute photometric accuracy of the MIRLIN nights was estimated from the standard deviation of the standard star flux densities from the least-squares airmass fit. These were found to be 8.9\% at 11.7 $\mu$m and 13.1\% at 20.8 $\mu$m.

In addition to the flux calibration error, there is also the statistical error from the aperture photometry due to the standard deviation of the background array noise. For the MIRLIN 11.7 and 20.8 $\mu$m data, the detector was extremely noisy and therefore the statistical error for the MIRLIN images is quite large in comparison to the OSCIR data at \emph{N} and 18.1 $\mu$m. However, the image-to-image variations of the standard deviation of the background array noise through a particular filter were very small, so the average of this value can be used to characterize the typical noise of the detector at each wavelength observed. From this we can state that the typical 3-$\sigma$ upper limit on a point source detection is 0.03, 0.12, 0.37, and 0.75 Jy, for the \emph{N}, 11.7 $\mu$m, 18.1 $\mu$m, and 20.8 $\mu$m filters, respectively. These are the quoted values for non-detections in Table 2.

The errors in the measured flux densities in Table 2 are the flux calibration error and background array noise added in quadrature, and represent the 1-$\sigma$ total error of the quoted flux density. There are several sources in common with this work and in our previous paper, De Buizer et al. (2003). In that paper, we supplied flux density estimates of the sources, but only gave an estimate of the flux calibration error. This flux calibration error was simply taken to be the largest deviation of the standard star flux from a set of observations temporally coincident to the scientific target observations. While the flux calibration error is, in general, the dominant source of error in mid-infrared observations, when the source flux in faint, it is instead dominated by the statistical errors associated with the background array noise. Therefore, for the sources in common to both works, you will see the same flux densities quoted in Table 2 of this paper, however, the quoted errors are different. The new error presented here gives one a better feel for the statistical significance of a faint detection. Furthermore, due to a better understanding and characterization of the detector noise in MIRLIN, the 3-$\sigma$ upper limits on a point source detection through the 11.7 and 20.8 $\mu$m filters presented here much larger and should be considered a revision to those presented in De Buizer et al. (2003).

\subsection{Derived Dust Temperatures and Optical Depths}

Dust color temperatures and emission optical depth values in Table 3 were derived from the mid-infrared flux densities and were obtained by numerically integrating the product of the Planck function, emissivity function (given by $1-e^{\tau _{\lambda }}$, where $\tau_{\lambda }$ is given by the Mathis 1990 extinction law), filter transmission, solid angle subtended by the source, and model atmospheric transmission through the filter bandpasses. For resolved sources, the source sizes were taken to be the $N$-band FWHMs subtracted in quadrature from the median standard star FWHM. For unresolved sources, calculations were made using lower limit (blackbody limiting) and upper limit (resolution limiting) sizes. The resolution limiting size was calculated to be 0$\farcs$63 from the 3-$\sigma$ variation of the standard star FWHMs throughout the night at 11.7 $\mu$m. For extremely low S/N sources, we can not be sure what the sizes of the sources are. We therefore performed our calculations in the limits where the sources are optically thick (blackbody limit) and optically thin. Both the unresolved and low S/N objects are noted in Table 3.

Visual extinctions associated with the mid-infrared emitting dust were found for the sources in the survey and are listed in Table 3. These were calculated by using our derived emission optical depth values at 11.7 $\micron $ and the Mathis (1990) extinction law, which yields the relation $A_{V}=34.97\cdot \tau _{11.7\mu m}$. We find that more than half of the sources in our survey have $A_{V}\gtrsim 2.5$ in the emitting regions. Thus, $>$90\% of the visual radiation from the star is absorbed by the surrounding dust and converted into mid-infrared radiation, assuming 4$\pi $ steradian coverage.

\subsection{Source Luminosities and Spectral Types}

Mid-infrared luminosities in Table 3 were computed by integrating the Planck function from 1 to 600 microns at the derived dust color temperature and emission optical depth for each source, again using the above emissivity function and assuming emission into 4$\pi $ steradians. If we assume that all of the shorter wavelength flux has been absorbed by the dust and reradiated as mid-infrared emission, then our derived mid-infrared luminosities can be considered reasonable lower limit estimates to the bolometric luminosities for these sources. We used those estimates of the bolometric luminosities to estimate zero-age main sequence spectral types for the sources using the tables of Doyon (1990), which are based on stellar atmospheric models by Kurucz (1979). However, because our luminosity measurements are lower limits, the true spectral types of the sources are likely earlier than their calculated spectral types (Table 4). The three main problems with this method of deriving estimates to the bolometric luminosity are (1) if the dust is anisotropically distributed around the source, the derived luminosity would depend on this dust distribution because some of the stellar flux will escape unprocessed through the unobstructed regions; (2) heavy obscuration could lead to non-negligible reprocessing by dust of the mid-infrared photons into far-infrared and sub-mm photons; and (3) dust is in competition with gas for the short wavelength photons, which ionize the gas and produce UC \ion{H}{2} regions. All of these processes would lead to underestimates of the bolometric luminosities from mid-infrared fluxes; however, it is hard to quantify exactly how each contribute. For these reasons we believe that the derived bolometric luminosities represent good lower limits to the true bolometric luminosities. We also caution that some of the detected mid-infrared sources may not be centrally heated. Therefore, the derived ZAMS spectral types in reality will not apply, and
the luminosities given in Table 3 are a better indication of the infrared luminosities of the sources, rather than the bolometric luminosities of the central stellar sources.

For the sources in this survey that have measured radio continuum fluxes from the literature, one can derive radio spectral types to compare with the spectral types derived from the mid-infrared observations. The Lyman continuum photon rates can be derived from the standard equation for free-free emission:

\begin{eqnarray}
S_{\nu} [\textrm{mJy}] = 3.09\times10^{-57}\cdot\left(\frac{N_{Lyc}^{H}(1-f)}{\textrm{sec}^{-1}}\right) \left(\frac{\textrm{sec}^{-1}\cdot \textrm{cm}^3}{\alpha_{2}}\right)\cdot \nonumber\\
\left(\frac{\nu}{\textrm{GHz}}\right)^{-0.1}\left(\frac{T_{e}}{10^{4} \textrm{K}}\right)^{-0.35}
\left(\frac{a}{0.994}\right)\left(\frac{D}{\textrm{kpc}}\right)^{-2} \nonumber
\end{eqnarray}

where $S_{\nu}$ is the flux density in mJy at radio wavelength $\nu$, $T_e$ is the electron temperature which is taken to be 10,000 K from observations of typical \ion{H}{2} regions (Dyson \& Williams 1980), and $D$ is the distance to the source in kpc. The other parameters are $a$, which is a slowly varying function of frequency and electron temperature that has values very close to unity, and $\alpha_2$ is the recombination coefficient ignoring recombinations to the ground level (Case B recombination), which has a value of $2.6\times10^{-13}$ cm$^3$$\cdot$sec$^{-1}$. This equation is solved for $N_{Lyc}^{H}$, the Lyman continuum photon rate under the assumption that the fraction of ionizing photons absorbed by dust, $f$, is zero. From there the tables of Doyon (1990) were used to find the spectral type corresponding to that Lyman photon rate, and we present these spectral types in Table 4\footnote{The Lyman continuum photon rates were miscalculated from the radio fluxes in De Buizer, Pi{\~ n}a, \& Telesco (2000). The net effect being that the actual spectral type estimates are 0.5-1.5 spectral types earlier than those listed in Table 3 of that work. The correct radio-derived spectral types for all the sources in De Buizer, Pi{\~ n}a, \& Telesco (2000) can be found in Phillips et al. (1998).}. These spectral types are a much more accurate estimate of the true stellar spectral types than the mid-infrared derived spectral types because cm radio emission is not as effected by dust extinction.

\subsection{Adopted Distances}

Most of the distances given in this study are kinematic distances. These distances are derived from some measurement of the radial velocity of the region in question. Radio recombination lines, atomic transitions like HI, molecular line transitions like formaldehyde, and even masers themselves can yield a radial velocity estimate for a region in space. When this radial velocity information is combined with a model for the rotation of our Galaxy, distances to sources may be determined. The main errors associated with this distance determination method are: 1) The distance will be dependent upon the Galactic rotation curve used. Most models are simple power laws, and do not reflect accurately the true rotation of our Galaxy. From one rotation law to another, one may expect a difference in the distance estimates to be as high as 1 kpc, in the extreme; 2) The values used for the Galactocentric distance and orbital velocity of the Sun will affect the results. The present IAU accepted values of $\Theta_o$ = 220 km/sec and $R_o$ = 8.5 kpc are used in this work,  however, the older values of $\Theta_o$ = 250 km/sec and $R_o$ = 10 kpc are quite prevalently used in the distance determinations in the literature; and 3) It is not known how accurately the radial velocities derived from atomic and molecular transitions mimic the holistic velocity of the region. For instance, Forster \& Caswell (1989) used the radial velocities from OH masers to calculate the distances to the associated regions. However, if the OH masers are tracing some other dynamic process, the radial velocity measured will most certainly not be appropriate for determining the distance to the region. Other uncertainties include small fluctuations due to turbulence, larger variations due to peculiar velocities and the fact that rotation models do not account for velocity variations due to galactic latitude and non-circular orbital motions.

Another problem arises when the source or region in question lies within the solar circle. When this is the case, the distance to the source cannot be simply determined from its radial velocity. If simple circular orbits are assumed around the Galactic center, a line of sight will cross an orbit at two points with the same velocity but different distances from the Sun. This leads to the kinematic distance ambiguity for sources within the solar circle, as they may lie at either the near or far distance given by a radial velocity. The only exception is when the source lies at a point in its orbit where it is tangent to the line of sight. This is where the radial velocity for a source it at its maximum, and there is no distance ambiguity.

There are some methods for determining the actual distance to a source. For instance, for nearby stellar sources, one can determine a star's spectral type and UBV flux. In this way, accurate spectrophotometric distances can be obtained. However, if one only has a near and far kinematic distance, the distance ambiguity may be resolved in four ways. First, if a \ion{H}{2} region can be seen optically, it is believed to be evidence for it being at the near distance. However, absence of optical emission does not necessarily imply the far distance because the regions such as the ones in this survey suffer heavy optical obscuration. Second, massive star-forming regions are mostly located in or near the galactic plane. If a region has a galactic latitude greater than ~0.5°, it is most likely at the near distance, otherwise it would be located too far out of the plane of the Galaxy. Third, absorption components of radio spectral lines at smaller velocities than that of the radial velocity determined for the region or source, means the near distance is most likely. For instance, this method is employed by Kuchar \& Bania (1990, 1994) using HI absorption towards Galactic plane \ion{H}{2} regions. They first make the reasonable assumption that the line of sight to an \ion{H}{2} region in the plane of the Galaxy will cross several HI clouds. The HI in front of the \ion{H}{2} region will absorb the thermal continuum from the \ion{H}{2} region. The distance ambiguity can be resolved by measuring the maximum velocity of the HI absorption. The HI gas at higher radial velocity than the \ion{H}{2} region will be behind the \ion{H}{2} region and will not contribute to the absorption spectra. Therefore the absorption spectrum will only show absorption up to the velocity of the \ion{H}{2} region (as determined from recombination lines or masers). Likewise, absorption components with velocities greater than the velocity at the tangent point is evidence for it being located at the far kinematic distance. Forth and finally, one can make an argument based upon maser luminosities, as outlined in Caswell et al. (1995).  The usual assumption is employed that the maser emission beamed in our direction is representative of the intensity in other directions, and can be considered quasi-isotropic. The peak maser luminosity is defined as $F \cdot D^2$, where $F$ is the peak maser flux density in Jy and $D$ is the distance in kpc. Caswell et al. (1995) argues that the maser source in their survey with the highest flux density is G9.62+0.20 at 5090 Jy. At the well-determined near distance (from absorption measurements) of 0.7 kpc, its luminosity is 2500 Jy kpc$^2$. The highest luminosity sources in the survey of Caswell et al. (1995) are around 80,000 Jy kpc$^2$. Some sources in our survey can be excluded from the far distance because their maser luminosities would be much larger than 80,000 Jy kpc$^2$.

For the sources where the information was available, the HI absorption observations of Kuchar \& Bania (1990, 1994) or the formaldehyde absorption measurements of Downes et al. (1980) and Watson et al. (2003) were employed to determine whether to use the near or far kinematic distance. For sources where this information is unavailable, one of the other above methods was used. In \S4 we discuss our choice of distance for some problematic sources, and Table 3 tabulates our adopted distances for the sources with references. For those sources where the distance was calculated from the radial velocity using the older values of $\Theta_o$ and $R_o$, we correct these distances with the IAU accepted values using the galactic rotation curve model of Wouterloot \& Brand (1989) given by $\Theta_\star$ = $\Theta_o$($R_\star$/$R_o$)$^{0.0382}$.

\section{Individual Fields}

Out of 26 maser sites observed in the mid-infrared, there were six fields with no detections (G12.21 -0.10, G32.74-0.07, G33.13-0.09, G35.03+0.35, G45.47+0.13, and G75.78+0.34). Of the 20 sites containing mid-infrared emission, 14 contain double or multiple sources associated with the maser group. Six are single sources, but of them, five are extended. Figures 1 - 20 shows our mid-infrared maps of these regions while flux densities for each source are given in Table 2. The following sections discuss our results as they pertain to the individual fields.

\subsection{G00.38+0.04 (IRAS 17432-2835)}

This site contains water, OH, and methanol masers, but no UC \ion{H}{2} region. Forster \& Caswell (1989) failed to detect 1.36 cm continuum here with an upper limit of 70 mJy, and Forster \& Caswell (2000) did not detect a UC \ion{H}{2} region here either with a upper limit of 0.7 mJy at 3.5 cm. The site also lacks the molecular core signatures of CS and NH$_3$ (Anglada et al. 1996). Mid-infrared source DRT04 1 (Figure 1) is seen at both 11.7 $\micron$ and 20.8 $\micron$ and is located $>$20$\arcsec$ from the masers. At 20.8 $\micron$ a possible mid-infrared source is located $\sim$10$\arcsec$ from the maser clump, but is at a S/N ratio of $<$3 and may simply be due to noise. We therefore cannot draw any conclusions as to what is exciting the masers on this field.

\subsection{G00.55-0.85 (IRAS 17470-2853)}

This site is also known as RCW 142 and contains water, OH, and methanol masers, as well as radio continuum (Walsh et al. 1998; Forster \& Caswell 2000). Plume, Jaffe, \& Evans (1992) detected CS and CO toward this site. Anglada et al. (1996) confirm the detection of CS and find NH$_3$ as well. Forster (1990) found that the water masers here are linearly distributed, and advances a disk or ring hypothesis to explain their distribution. Caswell (1998) suggests that since the various masers are spread over a large area, this may be a case where the masers are tracing an extended source rather than the masers tracing a cluster of individual sites.

This site was observed at \emph{N} and 18.1 $\micron$ (Figure 2) as well as 11.7 $\micron$ and 20.8 $\micron$ with a total of 4 sources detected. The OH masers hug the contours of DRT04 3 and possibly trace the shock region of the ionization front in its UC \ion{H}{2} region. DRT04 3 is slightly extended with a hint of double peak which could possibly be due to a heavily embedded double source.  The closest mid-infrared source to the water masers is DRT04 1 which is easily seen in the \emph{N}-band image and at 11.7 $\micron$, but not detected at 18.1 $\micron$. It is present, but barely resolved from DRT04 2 at 20.8 $\micron$. This line of water masers appears to emanate radially from DRT04 1 and thus may be tracing an outflow from the source. The near infrared source given by Testi et al. (1994) is probably associated with DRT04 2. The final source on the field, DRT04 4, is very amorphous.

There is some confusion as to what the distance to this site is. Walsh et al. (1997) adopt a distance of 9.1 kpc to this site from the radial velocities of the methanol masers, whereas Forster \& Caswell (1999) believe it is either 2.0 or 18.0 kpc, depending on near or far kinematic distance, from the OH maser velocities. This disagreement in distance is somewhat confusing since the OH and methanol masers at this location are not only coincident spatially, but overlap in velocity as well. Both molecules have radial velocities that span from +8 to +20 km/sec. Kinematic distances were independently derived here using these velocities and the galactic rotation curve of Wouterloot \& Brand (1989), as discussed in \S3.4. It was found that radial velocities in this range should yield a tangent distance close to 9 kpc, so in this paper the distance of Walsh et al. (1997) of 9.1 kpc is adopted here.

\subsection{G9.62-0.19 (IRAS 18032-2032)}

G9.62+0.19 and its surrounding environment have been well studied at a variety of wavelengths (see Testi et al. 2000 and references therein). A detailed discussion and analysis of mid-infrared emission from this source is given in De Buizer et al. (2003). In summary, this complex region contains a wealth of high-mass sources of different evolutionary states, from the hot molecular core (HMC) phase to well-developed \ion{H}{2} regions. A total of nine mid-infrared sources are detected (Table 2). Figure 3 shows a mid-infrared map of the region at 11.7 $\micron$.

A study of the centimeter continuum emission from this region by Garay et al. (1993) yielded the designation of radio sources labeled A to E (Figure 3b). Sources A and B are large, extended ($\sim$30$\arcsec$) regions of centimeter radio continuum emission, and D is a bright, compact radio continuum source just to the east of B. Source D is the southernmost component to a string of radio continuum sources that run to the northwest, ending with source C, approximately 20$\arcsec$ from D. The HMC, which lies in this string and is nearest to D, was first observed in thermal ammonia line emission by Cesaroni et al. (1994) and has been given the designation F. Masers of several species (H$_2$O, OH, CH$_3$OH, and NH$_3$) lie along this string of radio sources. The HMC is coincident with several water masers and is most likely responsible for their excitation. Though mid-infrared emission is associated with many radio sources on this field, there is no detectable mid-infrared emission from the location of the HMC.

\subsection{G10.62-0.38 (IRAS 18075-1956)}

This is the site of a UC \ion{H}{2} region that is part of the W31 complex. It also contains water and OH masers. The UC \ion{H}{2} region has been imaged at 1.36 cm (Forster \& Caswell 1989), 2 cm (Hofner \& Churchwell 1996; Wood \& Churchwell 1989), 3.5 and 4.5 cm (Walsh et al. 1998), and 6 cm (Wood \& Churchwell 1989). Several molecular species have also been detected here, such as CS, CO, HC$_3$N, and CH$_3$CN (Olmi \& Cesaroni 1999; Wyrowski, Schilke, \& Walmsley 1999; Hauschildt et al. 1993; Plume, Jaffe, \& Evans 1992; Churchwell, Walmsley, \& Wood 1992), indicating that this may be the site of a HMC.

Low resolution ($\sim$18$\arcsec$) MSX satellite observations of the 10$\arcmin$$\times$10$\arcmin$ region show multiple sources at 21 $\micron$ (Crowther \& Conti 2003). High resolution ($<$2$\arcsec$) mid-infrared observations of the central $\sim$30$\arcsec$ at \emph{N} and 18.1 $\micron$ are shown in Figure 4. Five sources are observed and the water maser reference feature is located between two sources, DRT04 2 and DRT04 4. The OH masers seem to be associated with DRT04 4, and are positioned in a region where one would expect an ionization front to be located. Testi et al. (1994) detected two near-infrared sources in this field, one of which seems to be associated with source DRT04 4. The second near-infrared source does not seem to be associated with any mid-infrared source. Sources DRT04 1 is a lower S/N point-like source, whereas DRT04 3 is low S/N and amorphous. Source DRT04 5 does not appear in Figure 4, but lies $\sim$27$\arcsec$ NE of DRT04 2.

The linear distribution of the water maser spots could be interpreted as participating in some sort of outflow. It appears that there may be an outflow as traced by the $\sim$18000 AU water maser chain extending to the southeast from DRT04 2. The UC \ion{H}{2} region seen in the 2 cm maps from Hofner \& Chuchwell (1996) has no coincident mid-infrared emission. It may be that the this region is optically thick to mid-infrared emission (but not radio continuum emission), and we are seeing mid-infrared sources only where there are more optically thin holes.

\subsection{G11.94-0.62 (IRAS 18110-1854)}

This site has a cometary-shaped UC \ion{H}{2} region present, with isolated water masers located out in front of the cometary arc (Hofner \& Churchwell 1996). There are two clusters of water masers near the UC \ion{H}{2} region, and both are offset to the west. The westernmost water maser group is located $\sim$10$\arcsec$ away from the UC \ion{H}{2} region and represents a possible embedded high mass protostellar object detected in the mid-infrared. This is discussed in more detail in De Buizer et al. (2003). In this paper we address the mid-infrared region as whole.

Figure 5 shows our 11.7 $\micron$ and 20.8 $\micron$ maps of the central $\sim$30$\arcsec$ of G11.94-0.62. Seven mid-infrared sources are identified, though the extended mid-infrared emission in this region is widespread and spectacular. The other maser group not associated with the high mass protostellar object candidate DRT03 1 lies on a ridge of emission seen at 20.8 $\micron$. Interestingly, the cometary shape seen in the centimeter radio continuum images of Hofner \& Churchwell (1996) is not evident in the mid-infrared (Figure 5c). This is also the case for the 2.7 mm images of Watt \& Mundy (1999). Low resolution ($\sim$18$\arcsec$) MSX satellite observations of the 10$\arcmin$$\times$10$\arcmin$ region shows perhaps more extended emission surrounding the region at 21 $\micron$ (Crowther \& Conti 2003).

\subsection{G12.21-0.10 (IRAS 18097-1825A)}

This is the site of another cometary UC \ion{H}{2} region with several water maser groups out in front of the cometary arc. Because this region has been observed to contain several tracers of hot and dense chemically enriched gas, such as high-excitation NH$_3$ (Cesaroni et al. 1992; Anglada et al. 1996), CH$_3$CN (Millar \& Hatchell 1997), and CS (Plume, Jaffe, \& Evans 1992), the location of the isolated water masers was considered a prime location for a HMPO/HMC candidate and was discussed in De Buizer et al. (2003).

Observations of G12.21-0.10 using MIRLIN at 11.7 $\micron$ or 20.8 $\micron$ did not reveal the presence of mid-infrared emission from a HMPO at the maser location. Even more surprising is that the UC \ion{H}{2} region, detected at centimeter radio wavelengths (Hofner \& Churchwell 1996) and in the submillimeter (Hatchell et al. 2000), was not detected at either mid-infrared wavelength using MIRLIN. MSX satellite observations of the 10$\arcmin$$\times$10$\arcmin$ region however show low-level extended emission surrounding the region at 21 $\micron$
(Crowther \& Conti 2003). The most likely reason why we do not see this emission in the MIRLIN data, may simply be because for extended emission the coarser sampling of MSX would yield a higher flux per pixel than for MIRLIN. Also, the region of extended emission is large enough that we might have been chopping onto extended emission in our reference beam, thereby canceling out the extended emission in our target beam.

\subsection{G12.68-0.18}

This is the only site in our survey with no associated IRAS source. It contains OH, water, and methanol masers, however it does not contain a UC \ion{H}{2} region. This region does contain a rather large and diffuse radio continuum source covering $\sim$20 square arcseconds (Codella, Testi, \& Cesaroni 1997; Forster \& Caswell 2000) known as W33B. The OH and water masers do not appear to be contained within this large radio source. Molecular tracers are found toward this site, such as CS, CO, and NH$_3$ (Plume, Jaffe, \& Evans 1992; Anglada et al. 1996; Codella, Testi, \& Cesaroni 1997). Mid-infrared observations at \emph{N} and 18.1 $\micron$ revealed two sources (Figure 6). DRT04 1 is located very close ($\sim$2$\arcsec$) to the water maser reference position. DRT04 2 appears as a point-like condensation surrounded by diffuse emission at \emph{N} and at 11.7 $\micron$, however, at 18.1 $\micron$ this central emission condensation disappears and only the very faint diffuse emission remains. At 20.8 $\micron$ the diffuse emission is brighter and larger, now extending over an area with a $\sim$4$\arcsec$ radius. Goedhardt, van der Walt, \& Gaylord (2002) present near-infrared and MSX observations of this region that show that this field is on the outer edge of a large (radius$\sim$30$\arcsec$) mid-infrared emitting region. We can see a hint of gradation through our \emph{N}-band field consistent with this. The near-infrared source given by Testi et al. (1994) does not seem to be associated with any of the mid-infrared sources seen at this site and this near-infrared source is not readily apparent in the observations of Goedhardt, van der Walt, \& Gaylord (2002).

The OH and water masers lie between the two mid-infrared sources and it is not clear how they may be excited. The maser positions displayed in Figure 6 are from Forster \& Caswell (1989) and consistent with those shown in Forster \& Caswell (2000). Goedhardt, van der Walt, \& Gaylord (2002) state that they too used the Foster \& Caswell (1989) OH maser positions, however they appear to plotted in the wrong positions in their paper.

There is considerable discrepancy between the values for the distance to this site. Codella et al. (1997) give a distance of 11.5 kpc based upon the radial velocity of NH$_3$ lines. Braz \& Epchtein (1983) give a distance of 4.5 kpc, based upon the velocity of the brightest water maser line. Forster \& Caswell (1989) adopt a distance of 6.4 kpc based on the OH maser velocities. When the kinematic distances were recalculated based on the NH$_3$ velocity given by Codella et al. (1997), it was found that the near and far distances were 4.9 and 11.6 kpc, respectively. Codella et al. (1997) may have chosen the far distance based upon the fact that this was the only site in their survey in which they did not detect a NH$_3$ molecular core. Since the value of 4.9 kpc is close to the value of Braz \& Epchtein (1983) and to that of Forster \& Caswell (1989) (when corrected for the now accepted values for the solar distance and velocity), this value is adopted here.

\subsection{G16.59-0.05 (IRAS 18182-1433)}

This source is the site of water, OH, and methanol masers, and contains a very weak radio continuum source. Forster \& Caswell (2000) detect an extremely weak (0.3 mJy at 3.5 cm) UC \ion{H}{2} region that is just above there 3-$\sigma$ detection level, and is coincident with the OH and water maser positions.  This is also the site of molecular emission from NH$_3$ and CS (Codella, Testi, \& Cesaroni 1997; Bronfman, Nyman, \& May 1996). Two mid-infrared sources are detected at \emph{N} and 18.1 $\micron$ (Figure 7).

A faint mid-infrared source, DRT04 1, was detected at the location of the water masers, coincident with the weak UC \ion{H}{2} region seen by Forster \& Caswell (2000). A larger amorphous mid-infrared DRT04 2 was also detected, and is located to the southeast. Testi et al. (1994) also claim to have detected a faint near-infrared source at the location of the masers as well. This may be an embedded molecular core which is just beginning to show signs of weak radio emission from the central stellar source. Though weak, DRT04 1 is detected at a $\sim$3-$\sigma$ level at \emph{N}, 18.1 $\micron$ and 20.8 $\micron$. In addition, its coincidence ($\sim$1$\arcsec$) with both water and OH masers as well as the near-infrared source detected by Testi et al (1994) and the weak UC \ion{H}{2} region detected by Forster \& Caswell (2000) lend further credibility to the detection.

The distance to this site as given by Codella et al. (1997) is 4.7 kpc. This is close to the distance given by Forster \& Caswell (1989) when corrected for a solar distance of 8.5 kpc from the galactic center. The distance of 4.7 kpc will therefore be adopted for the purposes of this paper.

\subsection{G19.61-0.23 (IRAS 18248-1158)}

This is an extremely interesting and complex site. It contains OH, water, and methanol masers, and a grouping of UC \ion{H}{2} regions and extended radio continuum.  The radio continuum emission from this region comes from five main sources, all of which are discussed in detail in a paper by Garay et al. (1998). There are molecular tracers here in the form of CS, NH$_3$, and CO (Larionov et al. 1999; Plume, Jaffe, \& Evans 1992; Garay et al. 1998). Extended mid-infrared emission associated with the region is also detected by the MSX satellite (Crowther \& Conti 2003)

Figure 8 shows 11.7 $\micron$, 20.8 $\micron$, and 18.1 $\micron$ emission associated with six mid-infrared sources. The brightest mid-infrared source is DRT03 3, whose radio continuum is labeled A by Garay et al. (1998) and is also the brightest radio source. DRT03 3 is bright and cometary shaped, with a tail of emission sweeping to the west. It seems to be diamond shaped, and therefore might contain several stellar objects embedded it the same material. The near-infrared source of Testi et al. (1994) seems to be coincident with DRT03 3, but not very close to the mid-infrared peak. Perhaps it is coincident with a source in a different part of the extended object seen in the mid-infrared.

The water and OH masers seen here by Forster \& Caswell (1989) lie close to this source, and most of the masers are located in the in the outer traces of DRT03 3. The water masers seem to be lined in a string pointing radially away from DRT03 3. Garay et al. (1998) find extended ammonia emission here that has peaks located at both $\sim$5$\arcsec$ to the northwest, and $\sim$8$\arcsec$ to the west of DRT03 3 (Figure 8c). The water masers in the string may be associated with the northeast ammonia peak. No clear mid-infrared source was detected at this location. There is a solitary water maser coincident with the western ammonia peak, as well, and may be marking the location of an embedded molecular core. Again, there was no detection of any mid-infrared emission from this location. Another isolated clump of three water masers is located 2$\arcsec$ to the west of the string of water masers. The northeast ammonia peak and emission are elongated towards this location. There might be another embedded molecular clump here as well, that is unresolved from the clump responsible for the northeastern peak in the VLA observations. Though a faint 3-$\sigma$ clump of emission can bee seen at this location at 18.1 $\micron$ in Figure 8(c), more sensitive follow-up observations presented in De Buizer et al. (2003) at this wavelength showed no significant detectable mid-infrared emission from this location.

In addition to DRT03 3, five other mid-infrared sources populate our field. Sources DRT03 4, 5, and 6, are bridged together with DRT03 3 by extended low-level emission. Offset from these bridged sources are sources DRT03 1 and 2.

\subsection{G28.86+0.07 (IRAS 18411-0338)}

This site contains OH, water, and methanol masers, but does not have any detectable UC \ion{H}{2} region. Codella, Testi, \& Cesaroni (1997) find no 1.3 cm continuum emission here with an upper limit of 0.34 mJy. However, this site does contain molecular emission from NH$_3$, CS, and CO (Codella, Testi, \& Cesaroni 1997; Anglada et al. 1996; Plume, Jaffe, \& Evans 1992). Two mid-infrared sources are detected at \emph{N} and 18.1 $\micron$ (Figure 9). The brightest source, DRT04 1, is within $\sim$1$\arcsec$ of the water maser emission and $\sim$2$\arcsec$ of the OH maser emission. The OH masers lie at the outer traces of the mid-infrared source. To the south, DRT04 2 is an amorphous clump of emission connected to DRT04 1. The near-infrared source from Testi et al. (1994) could be associated with either source DRT04 1 or 2, but lies 3-4$\arcsec$ from both objects. It is difficult to say what may be exciting the masers in this field.

\subsection{G32.74-0.07 (IRAS 18487-0015)}

Water and OH masers were found at this location by Forster \& Caswell (1989), yet they did not detect a UC \ion{H}{2} region here at 22 GHz with a 3-$\sigma$ upper limit of 80 mJy. However, radio continuum emission was later found at this location at 5 GHz coming from a weak 7 mJy source by Becker et al. (1994). In addition, methanol masers have also been detected here (Menten 1991).

There have been few observations of this region searching for thermal emission from molecules, one of the few being the study by Anglada et al. (1996) where NH$_3$ was detected here. None of our observations of this region at 11.7 $\micron$, 18.1 $\micron$ or at \emph{N} yielded the detection of any significant mid-infrared emission.

\subsection{G33.13-0.09 (IRAS 18496+0004)}

This site has all the tracers of a star forming region yet has no mid-infrared emission detectable in our \emph{N}, 11.7 $\micron$, or 18.1 $\micron$ observations. Along with the water and OH masers of Forster \& Caswell (1989), this site also contains methanol maser emission originally found by Menten (1991). Molecular emission is also here in the form of CS (Larinov et al. 1999; Bronfman, Nyman, \& May 1996) and CO (Shepherd \& Churchwell 1996), however NH$_3$ was not detected here in the survey of Anglada et al. (1996). The $^{12}$CO observations of Shepherd \& Churchwell (1996) show broadened line wings that are interpreted as being due to high velocity gas possibly from outflow.

This field also contains a UC \ion{H}{2} region, observed by Kurtz, Churchwell, \& Wood (1994) at 3.6 and 2.0 cm and found to be cometary-shaped. Follow-up radio observations by Kurtz et al. (1999) show that this UC \ion{H}{2} region has little extended radio continuum emission.

\subsection{G34.26+0.15 (IRAS 18507+0110)}

Because this site harbors a known HMC and a spectacular cometary-shaped UC \ion{H}{2} region, it has been the target of numerous studies at a multitude of wavelengths, including the mid-infrared. The mid-infrared-bright sources in this region have been studied in detail at several mid-infrared wavelengths by Campbell et al. (2000) and De Buizer et al. (2003). Because of the wealth of data on this location, we will concentrate on the relationship between the mid-infrared emission and masers.

Observations by Hofner and Churchwell (1996) of this site at 2 cm shows a very bright cometary UC \ion{H}{2} region coincident with the mid-infrared source DRT04 1 (Figure 10). The 2 cm contours of this object are extended and sweep away from the location of source DRT03 5 and the water masers. A compact 2 cm source also lies at the location of DRT03 5, and another compact 2 cm source lies farther north as well, but has no mid-infrared counterpart. The mid-infrared emission from DRT04 1 actually breaks up into 4 separate sources in the high angular resolution mid-infrared observations of De Buizer et al. (2003). For several of these sources, Campbell et al. (2000) has performed in-depth radiative transfer analysis of the sources to derive and study their physical properties in a more rigorous way than that presented here. Interestingly, this mid-infrared emission, as seen in the higher resolution images, does not appear to be as dispersed as the radio continuum, and there is no evidence of a cometary shape in the dust distribution.

Looking at the arrangement of the individual water maser spots for this site in Figure 5, one sees a band of masers running north to south located in between DRT04 1 and DRT03 5. This band of masers extends over $\sim$18,000 AU and the the OH masers are distributed in a fashion that hugs the outer traces of the mid-infrared source DRT04 1.

The concentrated cluster of water masers just east of the mid-infrared peak of DRT04 1 is near the location of a knot of ammonia emission as seen in the observations of Keto et al. (1992). It was claimed by Keto et al. (1992) that perhaps these water masers are tracing a HMC at this location. As described in De Buizer et al. (2003), and unlike the claim of Keto et al. (1992), no mid-infrared emission can be detected at the location of this HMC. However, other HMC tracers have been found, such as CH$_3$CN (Pankonin et al. 2001), HNCO and C$^{18}$O (Zinchenko, Henkel, \&Mao 2000), confirming that existence of a HMC. However, it appears that there is more than one non-mid-infrared emitting HMC present, as evidenced by the far-infrared observations of the region by Campbell et al. (2004).

It is plausible that some of the masers present at this site may be excited to emit by the embedded high mass protostellar object in the hot molecular cores. However, since a majority of the OH and water masers lie in a line or arc along the eastern edge of the cometary UC\ion{H}{2} region, it has been suggested by Gasiprong, Cohen, \& Hutawarakorn (2002) that these masers may be tracing a bow shock. They suggest that OH is formed by dissociation of H$_2$O in this shock, and that this would account for the fact that the OH masers are found systematically closer to the cometary UC\ion{H}{2} region than the H$_2$O masers.

\subsection{G35.03+0.35 (IRAS 18515+0157)}

This appears to be another site with all the typical tracers of a star forming region, but again we detect no mid-infrared emission in the field. Kurtz, Churchwell, \& Wood (1994) observed  an unresolved UC \ion{H}{2} region at this location at 2 and 3.6 cm. The site contains maser emission from water and OH (Forster \& Caswell 1989), as well as methanol (Caswell et al. 1995). Several molecular tracers have also been found towards this region, including CS (Bronfman, Nyman, \& May 1996; Anglada et al. 1996; Larinov et al. 1999), NH$_3$ (Jijina, Myers, \& Adams 1999), and H$_2$CO which was used by Watson et al. (2003) to unambiguously determine that this region lies at the far kinematic distance of 10.0 kpc. Furthermore, the $^{12}$CO observations of Shepherd \& Churchwell (1996) show broadened line wings that are interpreted as being due to high velocity gas possibly from outflow.

\subsection{G35.20-0.74 (IRAS 18556+0136)}

This site has garnered a lot of interest because observations indicate that this region may contain a massive (proto) star with a jet driven outflow, something rarely observed towards massive young stellar objects. Like most of the sites in our survey, this site contains water and OH masers (Forster \& Caswell 1989), methanol masers (Caswell et al. 1995), and many tracers of molecular material like CH$_3$CN (Kalenskii et al. 2000), HCO+ and HCN (Gibb et al. 2003).

The cm radio continuum emission at this site was resolved by Heaton \& Little (1988) into three compact sources arranged north-south (Figure 11c). A hypothesis was put forth in that paper that the central radio source, which is where the masers are concentrated, is an unresolved UC \ion{H}{2} region and that the northern and southern radio sources are the the ends of a bipolar radio jet coming from this central UC \ion{H}{2} region. Sub-arcsecond VLA observations of this field by Gibb et al. (2003) at 3.5 and 6.0 cm show that these three concentrations of radio continuum emission break up into 11 individual sources all lying along this north-south position angle, with the central source itself breaking up into two sources separated by $\sim$0$\farcs$8. The northern of the two central sources (their source 7) has a spectral index typical of a UC \ion{H}{2} region, and is claimed to be the most likely driving source of the radio jet.

Gibb et al. (2003) also discuss that the large scale CO outflow seen by Dent et al. (1985) is not at the same position angle (132$^{\circ}$) as this radio jet (0$^{\circ}$). Their observations also reveal that there are likely several sources in the field that have outflows that contribute to the large-scale CO distribution.

This region was also observed recently at near-infrared \emph{K} (2.2 $\micron$) and \emph{L$^{\prime}$} (3.8 $\micron$) wavelengths by Fuller, Zijlstra, \& Williams (2001). At both wavelengths they detect a lobe of emission extending north of the central UC \ion{H}{2} region location. At both wavelengths it looks very reminiscent of a outflow lobe like those seen in near-infrared H$_2$ emission from outflowing low-mass young stellar objects. They claim that this \emph{K} and \emph{L$^{\prime}$} structure is near-infrared emission from a jet. They also claim that the OH masers as seen by Hutawarakorn \& Cohen (1999) trace a circumstellar disk because they lie near the base of the jet and are linearly distributed at an angle perpendicular to the jet.

Our mid-infrared observations of this region reveal an extended source that is highly elongated in the north-south direction with the peak offset to the south (Figure 11). The source has a cometary shape with the masers concentrated around the head. Though lacking the convex ``terminal bow shock'' structure seen in the \emph{L$^{\prime}$} images of Fuller, Zijlstra, \& Williams (2001), the mid-infrared images also look very reminiscent of a outflow lobe like those seen in near-infrared H$_2$ emission. The high polarization of the \emph{K} emission seen from this source (Fuller, Zijlstra, \& Williams 2001), shows that this light is predominantly scattered light, most likely scattering off the walls of a cavity carved out by the jet through the nascent molecular envelope surrounding the UC \ion{H}{2} region. Fuller, Zijlstra, \& Williams (2001) claim that the \emph{L$^{\prime}$} emission is also predominantly scattered light. Since the brightness of scattered light will have a $\lambda$$^{-4}$ wavelength dependence (assuming Rayleigh scattering), they use this relationship to find that in comparison to the \emph{L$^{\prime}$} emission, the \emph{K} emission is underluminous if solely created by scattering. They therefore estimate that there must be an extinction of A$_V$ of $\sim$25 magnitudes to account for the deficiency of \emph{K} band emission. However, for wavelengths longer than $\sim$3 $\micron$, one would expect thermal emission, not scattered light, to dominate the overall emission. An alternative reason why the flux is higher than expected by scattering alone at \emph{L$^{\prime}$} (compared to \emph{K})is that thermal emission, caused by the heating by the central (proto)star of the dust on the outflow cavity wall, is contributing reradiated emission to the total flux seen at this wavelength.

Evidence that this is the case comes from our mid-infrared observations. The expected color in magnitudes of the scattered light, as approximated by Rayleigh scattering is $K - N = -6.8$, and $K - 18.1 \micron = -9.2$. Therefore, extrapolating from the \emph{K} emission, the expected flux density for the mid-infrared emission from scattered light at \emph{N} for G35.20-0.74 would be 69 mJy, and 2.4 mJy at 18.1 $\micron$. However, this is only 2\% the flux density we measure at \emph{N} (3 Jy) and 0.003\% of what we measure at 18.1 $\micron$ (77 Jy). Therefore, we conclude that is source is dominated at wavelengths $>$3 $\micron$ by reradiated thermal emission from the outflow cavity rather than scattered light. In either case, the near- and mid-infrared emission appears to indeed be tracing the outflow from the UC \ion{H}{2} region. It also appears that a majority of the masers are involved in this outflow. Our observations do not have sufficient angular resolution to see thermal dust emission from the disk hypothesized to be present at the location of the southern OH masers by Fuller, Zijlstra, \& Williams (2001). Mid-infrared emission is found at this location, however it is unresolved from the extended emission.

\subsection{G35.20-1.74 (IRAS 18592+0108)}

This site contains water, OH, and methanol masers, and contains a large cometary UC \ion{H}{2} region. The UC \ion{H}{2} region is known as W48A and lies about 20$\arcsec$ away from the location of the water and OH masers. This region has been well-studied. There is CO in the area with a peak to the west of W48A  (Zeilik \& Lada 1978; Vallee \& MacLeod 1990) and the whole region is within extended ($\sim$1$\arcmin$) amorphous submillimeter emission. W48A is a very interesting and complex source, however there are no masers associated with it, and it seems unlikely that there is any relationship between it and the masers located 20$\arcsec$ away.

The mid-infrared images in this survey reveal a bright point source at the location of the water masers, as well as the UC \ion{H}{2} region to the south-east (Figure 12). A paper by Persi et al. (1997), presents observations of G35.20-1.74 at 11.2 $\micron$ and at near-infrared wavelengths. They claim to see six mid-infrared sources in the area, and detect the same source seen in this survey coincident with the masers at both 11.2 $\micron$ and \emph{K}. The field of our mid-infrared observations encompass four of the six mid-infrared sources. The source Persi et al. (1997) labeled MIR3 corresponds to the UC \ion{H}{2} region W48A, and can be see quite predominantly in the mid-infrared images (Figure 12). The source labeled MIR1 corresponds to the source in this survey that is seen coincident with the maser features. \emph{K} images of this source by Persi et al. (1997) show the source to have an elongation or tail pointing to the southwest. This elongation is not seen in either of our \emph{N}, 11.7 $\micron$, or 20.8 $\micron$ images, nor in the 11.2 $\micron$ image of Persi et al. (1997). Though the group of water masers given by Forster \& Caswell (1989) are found to be be offset slightly to the northeast of the MIR1, the water maser of Hofner \& Churchwell (1996) is coincident with it. The OH masers appear to be pointing radially away from MIR1 towards the northeast, in the opposite sense as the near infrared tail of emission. Perhaps in this case the OH and water masers are tracing an outflow in the northeastern lobe, and the near infrared emission is reflected light coming from the inner wall of the cavity produced by the southwestern outflow lobe. Unfortunately, the only available CO maps are those of Zeilik \& Lada (1978) which have a 3$\arcmin$ resolution and it is difficult to tell if there is any semblance of outflow on such a crude scale.

\subsection{G35.58-0.03 (IRAS 18538+0216)}

This site contains radio continuum emission, water and OH masers, but no methanol masers (Caswell et al. 1995). Kurtz, Churchwell, \& Wood (1994) find the radio continuum here to be coming from two extremely close UC \ion{H}{2} regions, which appear just resolved from each other with a $\sim$2$\arcsec$ separation in their 3.6 and 2 cm maps (Figure 13c). Kurtz et al. (1999) have 21 and 3.6 cm maps showing the UC \ion{H}{2} regions to be lying in large scale extended continuum emission. They consider the two UC \ion{H}{2} regions separate sources: G35.578-0.030 to the west, and G35.578-0.031 to the east.

Plume, Jaffe, \& Evans (1992) detect CO and CS toward this site. However, Anglada et al. (1996) did not detect NH$_3$ emission here. The 350 $\micron$ dust emission contour maps of Mueller et al. (2002) show a compact source centered on the masers.

There is good agreement in position between the western UC \ion{H}{2} region of Kurtz, Churhwell, \& Wood (1994) and the water and OH masers located here. In our survey we find that the mid-infrared source is situated close to the masers. This source is relatively bright and elongated to the northwest (Figure 13). Shirley et al. (2003) mapped the region in CS and find the integrated CS emission to be elongated at the same position angle as the mid-infrared emission and centered on the maser location. Perhaps the elongation is due to an outflow of molecular material. By comparing the mid-infrared image with the 3.6 cm radio image of Kurtz et al. (1999), it can be seen that there is a little extension in the western UC \ion{H}{2} region at approximately the same position angle as the mid-infrared source elongation. It can therefore be concluded with some confidence that the mid-infrared source is in fact the same source as the western UC \ion{H}{2} region G35.578-0.030.  There are, however, no signs of mid-infrared emission from the eastern UC \ion{H}{2} region, G35.578-0.031.

\subsection{G40.62-0.14 (IRAS 19035+0641)}

This site contains all three maser types and a UC \ion{H}{2} region. The OH and water masers here are coincident and well mixed spatially. The UC \ion{H}{2} region detected here at 6 cm by Hughes \& MacLeod (1993), is unresolved with their 3$\arcsec$ resolution. It is also a weak radio source, having an integrated flux density of only 3.8 mJy. The molecular species that have been detected toward this site are CS and CO, but not NH$_3$ (Larionov et al. 1999; Beuther et al. 2002; Anglada et al. 1996; Plume, Jaffe, \& Evans 1992). Beuther et al. (2002) find a 1.2 mm continuum source here peaked on the maser location and extended to the north.

A fairly bright and slightly elongated source was detected in the mid-infrared at this site. Even though most of the masers are within the contours of the mid-infrared source, the water maser reference feature is offset from the mid-infrared source peak. The mid-infrared source peak is 1$\farcs$7 west and 1$\farcs$0 south of the water maser reference feature. The UC \ion{H}{2} region peak of Hughes \& MacLeod (1993) is also off by 1$\farcs$9 west and 1$\farcs$1 south of the water maser feature coordinates. This offset is therefore most likely real and that the UC \ion{H}{2} region is the radio continuum component of the source seen in the mid-infrared. Osterloh et al. (1997) found a \emph{K$^{\prime}$} source situated 3$\arcsec$ to the northeast of the mid-infrared source. It is elongated in the northeastern direction as well. They state that it may appear elongated because the near infrared light is being reflected off of an outflow cavity from the star within the UC \ion{H}{2} region which is too heavily embedded to have a near-infrared component of its own. If this is true, the OH and water masers, which are distributed somewhat linearly and at the correct position angle, may be tracing the outflow into this cavity.

\subsection{G43.80-0.13 (IRAS 19095+0930)}

This site contains water, OH, and methanol masers, as well as radio continuum emission. Kurtz, Churchwell, \& Wood (1994) state that the radio source here is unresolved, however there are in fact two sources that can be seen in their maps at 3.6 and 2.0 cm. There are also molecular lines of CO and CS detected here (Osterloh et al. 1997; Larionov et al. 1999, Plume, Jaffe, \& Evans 1992), but no NH$_3$ was found (Anglada et al. 1996).

Two distinct sources were detected in the mid-infrared. The eastern source DRT04 2 is very low S/N and has no associated cm radio continuum or masers. The western mid-infrared source, DRT04 1, is kidney bean shaped and is coincident with the OH and water masers (Figure 15). By overlaying the radio continuum maps of Kurtz, Churchwell, \& Wood (1994) it was found that the double peaked UC \ion{H}{2} region matches the kidney bean shape seen in the mid-infrared (Figure 15c). Given the absolute astrometry of the radio continuum map of Kurtz, Churchwell, \& Wood (1994), it was found that the water maser positions given by Forster \& Caswell (1989) are offset from the UC \ion{H}{2} region by 2$\farcs$5. This is nearly the same offset seen in the mid-infrared image given the pointing accuracy alone between the masers and the mid-infrared source. However, Forster \& Caswell (1989) claim to see a UC \ion{H}{2} region coincident with the water maser reference feature. It can be concluded that the UC \ion{H}{2} region seen by Forster \& Caswell (1989) and the UC \ion{H}{2} region seen by Kurtz, Churchwell, \& Wood (1994) are most likely the same source. This would indicate that the absolute astrometry given by Forster \& Caswell (1989) for this particular set of masers is in error. Similar absolute astrometrical errors have been pointed out by Forster \& Caswell (1999) for other sources in the Forster \& Caswell (1989) survey. Figure 15 is shown with this correction to the maser positions.

The mid-infrared source DRT04 1 is elongated east-west. Interestingly, the OH masers lie predominantly in an east-west fashion, whereas the water masers lie in an elongated distribution pointing to the northwest. Shirley et al. (2003) mapped the region in CS and find the integrated emission to be elongated at the same position angle as the mid-infrared emission (E-W) and and slightly offset from the water maser location. Lekht (2000) investigated the radial velocities and properties of the water maser emission in G43.80-0.13 and concludes that these maser are most likely stimulated to emit by a shock wave arising from interaction of a molecular outflow or jet with the surrounding molecular material. We postulate that the diffuse, fan-shaped mid-infrared source DRT04 2 may be tracing the inner heated walls of an outflow cavity whose direction is situated E-W. This would imply that DRT04 1 is the source of the outflow, and given that the masers (particularly the OH masers) are situated parallel to this axis, they too may be excited by this outflow.

\subsection{G45.07+0.13 (IRAS 19110+1045)}

This is a well-studied site which contains all three masers types and a UC \ion{H}{2} region. The water and OH masers are intermixed and exist in two groups separated by about 1$\farcs$7. Hofner \& Churchwell (1996) present a 2 cm map of this site, which shows a solitary unresolved UC \ion{H}{2} region. The southern group of water masers lie across the continuum of this UC \ion{H}{2} region in a linear fashion at a position angle of approximately -45$^{\circ}$ (Figure 16c). A multi-wavelength study (CO, CS, millimeter, submillimeter) of G45.07+0.13 was performed by Hunter, Phillips, \& Menten (1997). They find this site to be the center of a molecular outflow. Their CS maps show the outflow is centered close to the UC \ion{H}{2} region, and that the bipolar outflow axis is roughly the same as that of the water masers. Hofner \& Churchwell (1996) suggest that the water masers here trace the outflow.

Hunter, Phillips, \& Menten (1997) believe that G45.07+0.13 is a single star early in its evolutionary phases. To the contrary, we find that there are three mid-infrared sources within 8$\arcsec$ of the water maser reference feature (Figure 16). Mid-infrared source DRT03 2 is coincident with the southern group of masers and the UC \ion{H}{2} region. DRT03 3 is the northernmost mid-infrared source and is coincident with the northern group of masers. De Buizer et al. (2003) describes in detail the hypothesis that DRT03 3 is an embedded high mass protostellar object whose position is traced by the masers at this location. So it would appear for this region the masers are tracing the locations of two young stellar objects, one an embedded protostar and the other a more evolved young massive star in a phase of outflow. A third mid-infrared source is present on the field (DRT03 1) but does not appear to be associated with either masers or radio continuum. All three of these sources have also been observed in the mid-infrared by Kraemer et al. (2003).

\subsection{G45.47+0.05 (IRAS 19120+1103)}

This source contains water and OH masers, as well as methanol masers and a UC \ion{H}{2} region. Molecular tracers such as CS, HCO, SiO, and NH$_3$ have been detected towards this site (Olmi \& Cesaroni 1999; Cesaroni et al. 1992). There is also a submillimeter and millimeter source coincident with the UC \ion{H}{2} region here. The UC \ion{H}{2} region, as seen by Wood \& Churchwell (1989), has a kidney bean shape and is coincident with the water and OH masers.

Cesaroni et al. (1992) found emission and red-shifted absorption towards this site in NH$_3$. They claimed this is due to a cloud surrounding the UC \ion{H}{2} region collapsing onto the UC \ion{H}{2} region itself. Further observations were performed at higher resolution by Wilner et al. (1996) in HCO and SiO. They claim that the cloud core as seen in HCO emission appears clumpy and fragmented. They say that it is likely that this site is in an early stage of forming an OB star cluster. They claim that their HCO images provide no evidence for a spherical collapse localized to the UC \ion{H}{2} region as suggested by the observations of Cesaroni et al. (1992). However, the most recent observations of NH$_3$ by Hofner et al. (1999) suggests that this source is indeed undergoing infall of the remnant molecular core onto a UC \ion{H}{2} region where a massive star has recently formed.

The mid-infrared data may not be able to help solve this controversy, however it is interesting to note that the NH$_3$ in this area, as seen by Hofner et al. (1999), is elongated at a position angle of  about -40$^{\circ}$. The mid-infrared observations of this site show an elongated source lying at a position angle of -40$^{\circ}$ (Figure 17). If the material is indeed falling into the UC \ion{H}{2} region, which is thought to exist in the center of the NH$_3$ emission, then mid-infrared dust emission we see may be an accretion disk around the central stellar source.

The relative astrometry between the UC \ion{H}{2} region seen by Wood \& Churchwell (1989), the masers seen by Forster \& Caswell (1989), and the mid-infrared source is uncertain. Whereas the UC \ion{H}{2} region seems to be located coincident with the OH masers and only slightly offset from the water maser reference feature ($\sim$1$\arcsec$), the astrometry places the mid-infrared source peak $\sim$2.5$\arcsec$ northwest from the UC \ion{H}{2} region peak. It seems unlikely that the mid-infrared astrometry could be that far off, but an error in astrometry can not be ruled out. However, given the astrometry there is a hint of radio emission from the location of the peak of the mid-infrared source as seen in Figure 17c, and at 20.8 $\micron$ the source extends all the way to the maser location. This mid-infrared source can be seen in the low resolution MSX images of Crowther \& Conti (2003) as a point source and lies within an arcminute of the expansive mid-infrared source of G45.46+0.06.

There is a fair amount of variation on the distance to this site in the literature, from 6.0 kpc (Hatchell et al. 1998) to 9.7 kpc (Downes et al. 1980). However, the recent distance estimate of 8.3 kpc made by Kuchar \& Bania (1994) is adopted here, which is based upon the method of using HI absorption measurements.

\subsection{G45.47+0.13 (IRAS 19117+1107)}

This region is also known as K47 and is located approximately 5$\arcmin$ from G45.47+0.05 and G45.46+0.06. The site contains water and OH masers (Forster \& Caswell 1989), as well as methanol masers (Menten 1991). CS was found in the region by Larinov et al. (1999), NH$_3$ and H$_2$O thermal emission was also detected by Churchwell, Walmsley, \& Cesaroni (1990), however later NH$_3$ observations of Anglada et al. (1996) failed to detect the molecule. Wood \& Churchwell (1989) detect a large elongated (25$\arcsec$$\times$5$\arcsec$) region of radio continuum with multiple peaks, probably missing most of the extended emission due to the lack of short baselines in their interferometric observations. Likewise, a large region of mid-infrared emission (radius$\sim$40$\arcsec$) as seen by MSX was detected in the northwest part of the field of G45.47+0.05 in the paper by Crowther \& Conti (2003), though this particular source was not labeled on their figure. Interestingly, we do not detect any sources at \emph{N}, 11.7 $\micron$, or 20.8 $\micron$ on this field. Given the extended nature of the emission in the high angular resolution radio continuum images, perhaps the mid-infrared emission seen by MSX is also wide-spread through out region, thus leading to non-detections in the observations of this survey.

\subsection{G48.61+0.02 (IRAS 19181+1349)}

This site contains water and OH masers but no methanol masers. This site also contains a complex arrangement of radio continuum that was imaged at 3.6 and 2 cm by Kurtz, Churchwell, \& Wood (1994), who identified three ultracompact components in a field of extended continuum. G48.606+0.023 lies to the southeast, G48.606+0.024 lies in the middle, and the diffuse and extended G48.609+0.027 lies to the north (Figure 18c). This area was also found to have CO and CS, but no NH$_3$ (Anglada et al. 1996; Plume, Jaffe, \& Evans 1992; Shirley et al. 2003).

There are three low S/N mid-infrared sources located at this site, and a brighter but extended source located to the north (Figure 18). The water and OH masers seem to be located between sources DRT04 3 and DRT04 4, which are separated by about 5$\arcsec$. A relatively good match is found for the extended mid-infrared source DRT04 1 in the northwest and the UC \ion{H}{2} region designated G48.609+0.027, as well as a good match between G48.606+0.024 and DRT04 2 (Figure 18c). What relationship the mid-infrared sources have with the masers, and what causes the excitation of the masers in this field can not be determined. Verma et al. (2003) observed and studied this region at thirteen wavelengths between 3.3 $\micron$ and 210 $\micron$ using balloon-borne telescope, ISO, and IRAS observations. MSX observations of this region are also presented in Crowther \& Conti (2003). However the angular resolution of all of these observations was too coarse to help in the interpretation of the region.

The far kinematical distance is adopted to this site, following the precedent started by Solomon et al. (1987) based upon his results from studying HI absorption towards this location. The general agreement for the far distance is 11.8 kpc, as given by Forster \& Caswell (1989).

\subsection{G49.49-0.39 (IRAS 19213+1424)}

This site lies in an extensive (diameter$\sim$12$\arcmin$) star forming region named W51. The closest IRAS point source is 19213+1424, however this is coincident with W51-North, which is about 1$\arcmin$ away from where the water maser location. Martin (1972) observed this region in centimeter continuum emission and named the eight components here as W51 a through h. The strongest radio continuum emission is from W51e and d. Scott (1978) found the two small UC \ion{H}{2} regions, in addition to the larger \ion{H}{2} regions e and d. Because of their proximity to W51e, they were named e1 and e2. More recently, Gaume, Johnston, \& Wilson (1993) discovered two more compact continuum components near the large e \ion{H}{2} region at 3.6 cm, which were named e3 and e4. Most recently Zhang and Ho (1997) discovered a source named e8 at 1.3 cm that lies between e4 and e1. The water and OH masers observed in this survey are spread over a 10$\arcsec$ region, within which these small \ion{H}{2} regions e1, e2, e3, e4, and e8 are located.

Our observations in the mid-infrared for this region show a single unresolved point source lying $\sim$15$\arcsec$ south from a large region of extended emission (Figure 19). This large mid-infrared region is the southern portion of W51e. Given the astrometry of this survey, it was found that none of the masers in this area correspond to the point source DRT04 1 seen in the mid-infrared. However, by overlaying the radio continuum maps of several authors, it was found that the mid-infrared source lies only 2$\arcsec$ west of e1 (Figure 19c). It is uncertain if the source seen in the mid-infrared is e1 or another embedded source.  All of the masers in this area lie close to, or are coincident with, the UC \ion{H}{2} regions e2, e4, and e8 . Furthermore, there are two ammonia clumps in this area (Ho et al. 1983), one coincident with e2, and the other with e4 and e8 (Figure 19c). These molecular clumps can also be seen at 2 mm, and in CS, and CH$_3$CN. It is therefore likely that the maser are excited by stars embedded in these molecular clumps and with the sources already seen because of their radio continuum emission.

\subsection{G75.78+0.34 (IRAS 20198-3716)}

This field is also known as ON2-N and Cygnus 2 N. It houses a UC \ion{H}{2} region as seen at 6 cm by Hofner \& Churchwell (1996), however the peak of the radio continuum emission is offset $\sim$2$\arcsec$ from the maser location. The UC \ion{H}{2} region is cometary shaped and the masers are located in front of the cometary arc. This site has been probed for many molecular species like CS and CO (Anglada et al. 1996; Shepherd, Churchwell, \& Wilner 1997; Shirley et al. 2003), but also contains tracers of dense and hot molecular material like NH$_3$, HCO+, and CH$_3$CN (Hatchell et al. 1998; Anglada et al. 1996; Shepherd, Churchwell, \& Wilner 1997; Pankonin et al. 2001). The location of the water masers appears to be marking a the site of a hot molecular core because of all of these molecular tracers, and more importantly, the detection of a 7 mm continuum source at the maser location by Carral et al. (1997).

As described in De Buizer et al. (2003), this site was imaged a mid-infrared wavelengths to search for hot dust emission from the HMC, however the field yielded no detection of a mid-infrared source, including from the location of the UC \ion{H}{2} region. Given the presence of a HMC, UC \ion{H}{2} region, and four outflows in this region (Shepherd, Churchwell, \& Wilner 1997), it certainly appears to be an active star forming region so should in fact contain copious amounts of dust emission. The maps of the molecular species in the region (e.g., CO, HCO+) by Shepherd, Churchwell, \& Wilner (1997) show that the dense molecular material present here is widely distributed. Therefore, the reason that we did not detect mid-infrared emission is perhaps that the whole region is masked by an optically think envelope. The MSX observations in Crowther \& Conti (2003) show the the masers are located at the northeast edge of a large (radius$\sim$1$\arcmin$) mid-infrared source. Again it may be that the mid-infrared emission from this region is spread out so that we do not detect it in the higher angular resolution images presented here.

\subsection{Cepheus A HW2 (IRAS 22543+6145)}

Cepheus A HW2 (source 2 from the original cm study of Hughes and Wouterloot 1984) is now a well-studied case in which the linearly distributed water masers have been claimed to be tracing a circumstellar disk. Observations by Rodriguez et al. (1994) identified a radio jet from this potentially massive protostar, and observations by Torrelles et al. (1996, 1998) also discovered that the water masers in this region were distributed more or less perpendicular to the radio jet (Figure 20b inset). They suggested that the water masers were situated in a circumstellar disk with a radius of $\sim$300 AU. Later observations at milliarcsecond resolution with the VLBA (Torrelles et al. 2001) showed that these water masers were not distributed in a simple elongated pattern, but instead they were resolved into individual filaments made up of many masers distributed in ``microstructures'' that are linearly distributed. Torrelles et al. (2001) revise their claim that the masers exist in a disk, and instead describe how some of these maser microstructures are created by shock fronts from the wide-angled stellar wind from the stellar source at the HW2 center, while other microstructures are claimed to be associated with unidentified nearby protostars.

In summary, the masers of HW2 are no longer believed to be tracing a single circumstellar disk. This simple scenario has been replaced with a much more complex one to describe the complex distribution seen at high angular resolution. However, it is interesting that the mid-infrared emission (Figure 20) is elongated parallel to and perpendicular to the radio jet seen by Rodriguez et al. (1994). In figure 20, one can see that the mid-infrared peak is located $\sim$3$\arcsec$ from the water maser location. Looking to the VLA cm maps of the entire Cepheus A East region (Garay et al. 1996), we see that HW2 is at the center of a string of cm continuum sources running from the northeast to the southwest. Given the astrometry of the mid-infrared observations, the peak of the mid-infrared source lies almost exactly between source HW2 and HW4, with the extended mid-infrared emission creating a ``bridge'' between the two cm continuum sources. Therefore the mid-infrared emission as seen with our observations from the IRTF may be one or several unresolved mid-infrared sources lying along this string of cm sources, but having no cm emission of its (their) own.

\section{Discussion}

\subsection{Relationships between Mid-Infrared Sources and Masers}

This mid-infrared survey contains 26 sites centered on their water maser reference features. We detected on our fields mid-infrared sources in 20 cases. This is a detection rate of 77\% towards water maser emission. This is comparable to the results of Churchwell et al. (1990) who searched for water maser emission from a large number of UC \ion{H}{2} regions. They used the Effelsburg 100 m radio telescope, which has a HPBW of 40$\arcsec$, and detected water maser emission in 67\% of the cases.

Fortunately, the astrometry of the mid-infrared observations in this survey is accurate enough to allow even further analysis into the spatial relationship of the mid-infrared sources with respect to the water (and OH) maser distributions. Though we targeted 26 water maser sites, several of these sites have maser distributions that are confined to several centers of emission. We define a ``center'' of maser emission as a group of one or more masers that are clustered together and are separated from other maser centers by more than one arcsecond. For instance, G19.61-0.23 has three centers of water maser emission but only one center of OH maser emission (Figure 8), whereas G34.26+0.15 has 4 centers of water maser emission and 4 centers of OH maser emission (Figure 10). We use the separations from mid-infrared sources to maser centers in our analysis rather than the separations to each individual maser because individual masers have proper motions, they vary in intensity and appear and disappear with time. However the centers of maser emission appear to be longer-lived entities and more importantly mark the areas conducive to maser emission. There are a total of 44 water maser centers and 31 OH maser centers in this survey and we measured the projected angular distances between the mid-infrared peak and the centroid of the maser centers.

Radio observations of Hofner \& Churchwell (1996) found that the median separation from UC \ion{H}{2} region peaks to water maser centers is 2.9$\arcsec$ or 0.091 pc (18770 AU). We found that the median angular separation between the water maser centers and the mid-infrared source peaks is 2.1$\arcsec$, and the median physical separation is 0.042 pc (8730 AU). A similar result was found for the coincidence of OH masers with the mid-infrared sources (1.8$\arcsec$, 7800 AU). It can be concluded that for the sources in this survey, water (and OH) maser emission is more closely associated with mid-infrared emission than cm radio continuum emission. This close association between thermal dust emission and water maser emission may be expected in light of new models by Babkovskaia \& Poutanen (2004) of the water maser pumping mechanism. Their models not only require the presence of warm dust, but and they find that the strength of the water masers depends, in-part, on the dust-to-water mass ratio and the difference between the dust and gas temperatures.

Figure 21 presents histograms showing the number of water (Figure 21a and b) and OH (Figure 21c and d) masers given the separation from the mid-infrared source center. These results can be compared with those from Hofner \& Churchwell (1996), for the distribution of water masers with UC \ion{H}{2} region centers (Figure 21e and f). Though the histograms showing angular separations do not appear to be very much different from one another, the histograms showing the physical separations seem to confirm the idea that water masers are indeed found more often at locations physically closer to mid-infrared rather than cm radio continuum emission. Included in the above median separation calculations, but not shown in Figure 21, are the maser centers with angular separations $>$25$\arcsec$ or projected linear distances $>$100000 AU. These sources are likely to be associated with different centers of star formation\footnote{Only two water maser separations are left off of Figures 21 a and b, one OH maser separation is left off Figures 21 c and d, and 6 are left off Figures 21 e and f (Hofner \& Churchwell 1996).}.

\subsubsection{The Possible Association of Water Masers with Embedded Sources}

Since there are a significant number of water masers that are not coincident with radio continuum sources, there have been two main hypotheses suggested to resolve this puzzle. The first hypothesis is that the water masers are excited by young and embedded massive stellar sources that are energetic enough to stimulate the masers, but too young or embedded to show signs of ionized emission. These young stellar objects would be invisible at cm radio continuum wavelengths, but their outer envelopes would be visible in molecular line emission. These envelopes should be extremely bright in their dust emission (especially in the far-infrared to sub-mm). Only in the recent decade have these types of objects gained attention as observations of such sources have become more common.

The idea that massive stars begin their lives in such a phase gained momentum through the work of Cesaroni et al. (1994), even though their work involved molecular ammonia observations toward only four UC\ion{H}{2} regions. They were looking for the relationship between the dense molecular material and the UC \ion{H}{2} regions in each massive star forming region. What they found was that in two of the cases, the molecular material as traced by ammonia emission was not coincident with the UC \ion{H}{2} region, but was offset and coincident with clusters of water masers. Cesaroni et al. (1994) claimed that these ``hot molecular cores'' (HMCs) that they were observing in these regions are most likely the precursors to massive stars and their UC\ion{H}{2} regions. Therefore HMCs are considered to be a subclass of HMPOs (high mass protostellar objects), which is the general name given to all the formative phases of massive stellar birth up to the moment the star begins forming a UC \ion{H}{2} region. The HMC phase can be considered the last phase in of an HMPO, just before onset of a UC \ion{H}{2} region.

A subset of the sources from this survey was published already in the article by De Buizer et al. (2003). This work identified 7 sources from Hofner \& Churchwell (1996) that fulfilled the minimal requirements for being HMPOs: 1) the fields could be identified as massive star forming regions by the presence of one or more UC \ion{H}{2} regions; 2) water masers were present in locations where there was no cm radio continuum emission, i.e. the masers were offset from the UC \ion{H}{2} regions; and 3) the regions were known to contain molecular tracers present in known molecular clumps and HMCs. The sources in this survey that were identified in De Buizer et al. (2003) as HMC/HMPO candidates were G9.62+0.19, G11.94-0.62, G12.21-0.10, G19.61-0.23, G34.26+0.15, G45.07+0.13, and G75.78+0.34. Water maser and UC \ion{H}{2} region positions for all of these sources are given in Hofner \& Churchwell (1996). For G12.21-0.10 and G75.78+0.34, we did not detect any mid-infrared emission in the field, either associated with the water maser position or the UC \ion{H}{2} region. For G9.62+0.19, G19.61-0.23, and G34.26+0.15 no mid-infrared source was found at the known/supposed location of the HMC. Only in the cases of G11.94-0.62 and G45.07+0.13 were mid-infrared sources found coincident with the position of the water masers thought to be marking the location of an HMPO. One of these sources, G45.07+0.13, has the equivalent luminosity of a B0 star and thus appears to be a legitimate HMPO.

There are other sources that might be potential HMPO candidates in this survey that were not pointed out in De Buizer et al. (2003). There are several other sources in the survey that are directly coincident with water masers (with the errors of astrometry) and have no radio continuum emission of their own. In the case of G16.59-0.05, there is a weak and compact source (DRT04 1) detected at the location of the water and OH masers. This location might contain an extremely weak radio continuum source ($\sim$0.2 mJy at 3 cm) as seen by Forster \& Caswell (2000), however this source is only a 2-$\sigma$ detection. Molecular tracers such as NH$_{3}$ and CS have also been observed towards this site. Like G11.94-0.62, a point that does not favor the embedded source hypothesis is that there is near-infrared emission observed here at \emph{J}, \emph{H}, and \emph{K} (Testi et al. 1994). We do not expect emission at \emph{J} to be present if the source is so embedded. Furthermore, the mid-infrared lower limit estimate on the bolometric luminosity of this source is the equivalent to a B8-B9 star (also like G11.94-0.62).

Source G28.86+0.07:DRT04 1 is a mid-infrared source that is directly coincident with a small group of water masers. This site contains molecular tracers such as NH$_{3}$, CS and CO, but has no detectable radio continuum. Testi et al. (1994) find a near-infrared source in this region, but it is located almost 5$\arcsec$ from either the water masers or the peak of source 1, and therefore is probably not related to the mid-infrared emission. The mid-infrared lower limit estimate on the bolometric luminosity for this source is that equivalent to a B2-B3 star, so it may indeed be a legitimate HMPO. Sources G35.20-1.74:DRT04 1 and G40.62-0.14:DRT04 1 are at the limits of being spatially coincident with the water masers near them. With estimated lower limit spectral types of B3 and B6, respectively they might be massive enough to be considered HMPOs as well.

One could argue that all of these sources mentioned are HMPOs. First, this survey has relatively good astrometry and these sources are directly coincident with clusters of water masers which are generally thought to be tracers of massive star formation. Second, all of the sources are bright in the mid-infrared, implying that they are indeed embedded in dust. Third, none of the sources have cm radio continuum emission themselves (or very little cm radio continuum emission). And finally, there are observations providing evidence of molecular emission coming from each of these regions, and HMPOs are believed to be molecularly rich. This evidence combined points to the conclusion that these are indeed young stellar sources embedded in molecular cores.

However, unlike G45.07+0.13, the rest of these sources may have luminosities consistent with young intermediate mass, non-ionizing stars. This may indicate that many of these sources with maser emission and no cm radio continuum emission may simply be intermediate mass stars.  It has been argued (i.e., Forster \& Caswell 2000) that masers cannot be associated with all lower-mass stars or there would be many more masers sites than what we observe. However, observations towards intermediate mass Herbig Ae/Be stars have found water masers are associated with these stars at a rate of $\sim$10\% (Palla \& Prusti 1993). Given the fact that the population of intermediate mass stars is much less than the population of all stars later than A type, and the fact that water masers appear to be associated with a smaller percentage of intermediate mass stars than high mass stars (water masers are found toward UC \ion{H}{2} regions in $\sim$33\% by Forster \& Caswell 2000), it may well be that the majority of masers offset from UC \ion{H}{2} regions could be accounted for by excitation from a small subset of the population of young intermediate mass stars. This idea is further supported by the fact that massive stars are believed to form in clusters with stars of a wide range of masses.

\subsubsection{The Possible Association of Water Masers with Outflows}

The second hypothesis for the excitation of water masers away from radio continuum sources is the hypothesis that they are somehow associated with outflow from young stellar objects. There are two ways in which one can get maser emission from an outflow. The first is when masers trace the high-collimated outflow or jet near the stellar source, and the second is when the outflow impinges upon the ambient medium or on a pre-existing knot of molecular material more distant from the stellar source.

The water masers that trace outflow may be associated with a jet. A jet is well-collimated flow, with an opening angle no more than a few degrees, and can exist concurrently with the wider angled bipolar molecular outflow. Gwinn, Moran, \& Reid (1992) speculate that the masers may be gas condensations or `bullets' that move out from a common center. A later paper by Mac Low et al. (1994) propose that the masers come from the shell of swept-up gas that is driven out by a jet. These cocoons are very elongated and expand quickly along the ends in the direction of the jet, but slow along the transverse sides. Specifically, the water masers are formed in the outer shock of the expanding jet cocoon. One source in the water maser selected survey has been viewed as an example of this phenomena in the literature. G45.07+0.13 has a known CO outflow (Hunter, Phillips, and Menten 1997) and the UC \ion{H}{2} region coincident with G45.07+0.13:DRT04 2 has water and OH masers distributed in a linear fashion parallel to this outflow angle, and running through the UC \ion{H}{2} region peak. Hunter, Phillips, \& Menten (1997) claim that this strongly supports the theoretical ideas that a protostellar jet from a massive star can power H$_2$O masers. Hofner \& Churchwell (1996) also claim that these water masers are located along the flow axis and show good correspondence with the outflow velocities. They too claim this is strong evidence that the water masers are taking part in the bipolar outflow. The OH masers are also linearly distributed here, with a position angle near perpendicular to the water masers and outflow axis. One may postulate that these OH masers delineate a circumstellar disk, however the masers are offset from the UC \ion{H}{2} and mid-infrared peak by 0.5$\arcsec$ (though could be considered coincident given our astrometric accuracy), and there is no hint of elongation in the thermal dust distribution. In any case, this source provides compelling evidence that at least some water masers are associated with jets from young stellar sources. In this case, the water masers would exist close to the stellar source and exhibit linear patterns radial to the stellar source.

Other possible sources in this survey where water masers may be tracing outflow in this manner are G00.55-0.85:DRT04 1, G10.62-0.38:DRT04 2, G19.61-0.23:DRT03 3, and G43.80-0.13:DRT04 1. For G00.55-0.85, there is a string of water masers pointing radially away from source 1 towards source 3. Though CO has been detected toward this site (Plume, Jaffe, \& Evans 1992), there is no other information available concerning CO outflow. One notices when looking at source 3 that the OH masers are situated at a position angle perpendicular to the water masers, but near the edge of the UC\ion{H}{2} region. These OH masers may be shock-excited by the outflow (as traced by the water masers) impinging upon source 3. As for G10.62-0.38, the situation is a little more complicated. The string of masers in this case appear to lie in a line that goes through the peak of mid-infrared source 2. The radio emission is coincident with the water and OH masers here, and not the mid-infrared sources. Since outflows can contain partially ionized gas ($\sim$10\%, Masson \& Chernin 1993), they can be observed at radio wavelengths longer than 1 cm (Rodriguez 1994). In this case we may be seeing the water masers and the ionized outflow seen in radio continuum emission emanating from source 2. For G19.61-0.23:DRT03 3, it appears that the line of water masers here follow well the elongated peak in the ammonia emission clump found there. Therefore, the water masers are more likely associated with the the molecular clump that an outflow. In the case of G43.80-0.13:DRT04 1, the water maser distribution is elongated almost perpendicular to the OH maser distribution. However the centers of the two are offset by $\sim$2$\arcsec$ and without any other wavelength information, it is difficult to postulate on whether the water masers here are tracing an outflow.

The second way an outflow can stimulate water maser emission is, as was explained above, by the shock of the outflowing gas on surrounding material. Elitzur (1992) describes a model where water masers are locally created and pumped by the interaction of the outflow from a young star with clumps or inhomogeneities in the surrounding cloud. Unfortunately, this is a difficult hypothesis to try to prove with the mid-infrared survey alone. One would have to look for water masers offset from a mid-infrared source, but could not be sure if they are associated with an outflow or simply associated with, say, an embedded source too cool to detect in the mid-infrared. One would need auxiliary information, such as CO outflow maps or shock-excited H$_2$ images, in order to be confident that the water masers are in fact associated with shocks from outflows.

\subsubsection{Possible Associations of Hydroxyl Masers}

Though the emphasis of this work is on the relationship of water masers to the mid-infrared sources in the survey, we have accurate OH maser positions for 22 of the 26 sites in the survey. We can therefore comment briefly on the apparent relationships between OH masers and mid-infrared sources.

Sometimes OH masers are separated from water masers and sometimes they are intermixed, and many times they are intermixed  both spatially and in velocity. There does not appear to be anything in particular about the nature of the mid-infrared sources that differentiates this, though we caution that the sample size of the survey is small. For instance, when OH masers are associated with a different source than the water masers, there are apparently no outstanding characteristics of the mid-infrared sources that differentiate them from one another.

As mentioned in the introduction, it has been suggested that the shock fronts in the expanding \ion{H}{2} regions around young massive stars may provide a habitable zone for OH masers. This idea was apparently supported by several observations of OH masers projected on the circumference of \ion{H}{2} regions, like the observations of W3(OH) by Reid et al. (1980), and by subsequent observations of OH masers in \ion{H}{2} regions by Ho et al. (1983) and Baart \& Cohen (1985), among others. In addition there are the hypotheses that OH masers are excited by, and trace the locations of, embedded protostellar objects, or that OH masers exist in the warm and dense material of circumstellar disks around young stars. Because gas and dust are well-mixed in compact \ion{H}{2} regions, and because embedded protostellar objects and circumstellar disks also are known to be dusty objects, we expect for all three of the above scenarios that OH masers should be spatially coincident with mid-infrared sources. However, given the sources in this survey we certainly cannot say that OH masers are always projected on UC \ion{H}{2} regions and/or mid-infrared sources. On the other hand, the median separation of OH masers to mid-infrared sources (7800 AU, or 1$\farcs$8) implies a close physical relationship to the mid-infrared sources. Since the mid-infrared sources in this survey have typical radii of 1$\arcsec$ to 2$\arcsec$, this would imply that a significant portion of the OH masers are directly associated with the mid-infrared sources.

However, there are several cases where the OH masers (as well as water masers) do not appear to be associated with any mid-infrared or radio source. Perhaps these are pin-pointing the locations of extremely cool and severely embedded protostars, so that no mid-infrared emission can be observed. Another possibility is that these apparently solitary masers are produced by passing shocks inside the molecular cloud, created for instance by distant or more evolved outflowing sources.

\subsection{The Relationship between Mid-Infrared Sources and Radio Continuum and IRAS Data}

Table 4 lists the derived mid-infrared spectral types, and the radio spectral types derived from the radio flux.  In all cases the mid-infrared spectral type is later than the radio spectral type. Occasionally the two derivations closely approximated each other, with the closest being G35.20-0.74:DRT04 1 with a radio derived spectral type of B2.4 and a mid-infrared derived spectral type of B3.1. The largest discrepancy is for G34.26+0.15:DRT03 5, where the radio derived spectral type is B0.7 and the mid-infrared derived spectral type is B7.0. However, because the mid-infrared derived spectral types are in all cases lower than the radio derived spectral types, it can be concluded that the mid-infrared spectral types derived for the sites where no radio emission are found, are good estimates of the \textit{latest} spectral type the stellar source could be. From Table 4 it can be seen that, with the exception of G09.619+0.193:DPT00 1, all of the sources in these fields are O and B types, confirming that the water and OH masers, are indeed present where massive stars form. On the other hand, the high rate of detection of mid-infrared sources at the location of the masers where there is no radio continuum does not exclude the possibility that many of these maser sites may be associated with young intermediate mass stars, as the mid-infrared derived spectral types suggest.

The closest IRAS sources to the water maser sites are tabulated in Table 5, and were compiled from the IRAS Point Source Catalog (PSC). The actual separations between the IRAS source centers and the mid-infrared sources in this survey are between a few arcseconds and two arcminutes. Only one field, G12.68-0.18, had an IRAS source more than 3\arcmin\ away and so this source is considered to not have an IRAS counterpart.

Table 5 has a column that lists the extrapolated value for the IRAS flux at 11.7 microns, and an interpolated value of the IRAS flux at 20.8 microns. Therefore one can compare the flux measurements from IRAS, which had approximately 75$\arcsec$$\times$ 1$\farcm$5 spatial resolution, and the flux in the whole $61\arcsec$$\times$ 61$\arcsec$ field of view of MIRLIN. Inspection of Table 5 shows that, in most cases, the IRAS fluxes are much larger than the fluxes seen in the \textit{entire} MIRLIN field of view. The \textit{single} source observed at the location of water masers that comes closest in its 11.7 $\mu$m flux to that see in the IRAS observations is 35.20-0.74:DRT04 1. This source alone can account for 97\% of the extrapolated IRAS flux at 11.7 $\micron$ and 68\% of the IRAS flux interpolated at 20.8 $\micron$. Hence it can be concluded that the IRAS measurements do not accurately approximate the true mid-infrared fluxes for the sources directly coincident with the maser emission. The IRAS measurements are probably measuring flux from sources that lie outside our field of view, as well as extended emission around these sources. Therefore the single source associated with the H$_{2}$O maser cannot be modeled correctly with the fluxes obtained by IRAS for the whole region surrounding the source. Therefore spectral energy distributions, like those compiled by Testi et al. (1994) using their observed near-infrared fluxes and the IRAS 12, 25, 60 and 100 $\mu$m fluxes, inaccurately describe the maser source alone. For instance, derived luminosities and spectral types for these young stellar objects using IRAS (e.g., Wood \& Churchwell 1989; Testi et al. 1994; Walsh et al. 1997) are in most cases extreme upper limits.

\subsection{Relationship between Mid-Infrared Sources and Near-Infrared Data}

We will be describe in detail our own near-infrared observations of these maser sites in a forthcoming paper. For now, however, we will make a simple comparison of the sources we found at 10 $\micron$ to those found in near-infrared surveys of water maser sites already published by Testi et al. (1994, 1998). There are eight maser sites coincident in our surveys, G00.55-0.85, G09.62-0.19, G10.62-0.38, G12.68-0.18, G16.59-0.05, G19.61-0.23, G28.86+0.07, and G35.20-1.74. These fields contain reddened near-infrared sources as determined by Testi et al. (1994, 1998) and are thought to be associated with the masers. The K-band positions of these sources are plotted as squares at their proper positions in the figures in this paper. It can be seen that only three of the eight K-band sources actually correspond with a mid-infrared source to within 2$\arcsec$ (they claim their astrometry is good to 1$\arcsec$). One of these, G35.20-1.74, was the only site visited by both Testi et al. (1994, 1998) surveys, and was corrected by 4$\arcsec$ in the 1998 publication. Even so, the other two sites common to our mid-infrared data from the 1998 survey do not seem to be anymore positionally coincident than the four sites that are common in the 1994 survey. This discrepancy could be because: 1) we might be looking at different parts of the same extended object (which seems to be the case for G19.61-0.23), 2) either the mid-infrared or near-infrared astrometry is inaccurate, or 3) we are not looking at the same object (which is obviously the case for G12.68-0.18). In regard to this final point, it should be stated that Testi et al. (1994, 1998) caution that their survey may have some contamination from field sources, which is not a problem in the mid-infrared.

It was found in our survey that mid-infrared sources were detected within 4$\arcsec$ of 37 of the 44 centers of water maser emission (84\%). Testi et al. (1998) states identification of near-infrared \emph{K} sources to within 5$\arcsec$ in 73\% of the cases of all of their survey. Hence it may be concluded from this information that near-infrared sources, like cm radio continuum sources (\S5.1), are not as closely linked as mid-infrared sources are to water masers.

In seven of the eight cases the near-infrared source found by Testi et al. (1994, 1998) corresponds to a mid-infrared source to within 5$\arcsec$. We have assumed for the purposes of the next section that both near and mid-infrared sources are indeed the same object and that our mid-infrared astrometry is correct. Since these objects are thought to be young stellar objects, we can employ the spectral indices of these objects to categorize their evolutionary state. The spectral index is defined by Lada (1987) as $\alpha_{IR}$=dlog($\lambda$F$_{\lambda}$)/dlog($\lambda$),where the difference is taken between the near-infrared and mid-infrared.

Here we use the \emph{K} near-infrared magnitudes of Testi et al. (1994, 1998) and the \emph{N} flux density values we observed. Positive values of 0 $<$ $\alpha_{IR}$ $<$ +3 are considered to be Class I sources which are thought to be the earliest phases of protostellar evolution, where a star and disk are embedded in an infalling dust envelope. Class II objects are defined as having -2 $<$ $\alpha_{IR}$ $<$ 0, and are thought to be a newly born pre-main sequence star surrounded by a nebular, optically thick disk. Class III objects are pre-main-sequence or near-main-sequence stars, possibly with planetary or stellar companions and an optically thin disk and have values of -3 $<$ $\alpha_{IR}$ $<$ -2. These classes represent an evolutionary progression in the birth of a star. Recently, Andr\'{e}, Ward-Thompson, \& Barsony (1993) suggested a new class of sources to be added to those given by Lada (1987) called Class 0. This category represents sources so deeply embedded that little emission would be detected at near- and mid-infrared wavelengths. These objects have unusually strong sub-millimeter emission and are now believed to be the youngest observable stages of stellar birth and are little more than dense molecular cores (Andr\'{e} \& Montmerle 1994).

Because we are detecting these objects in the near- and mid-infrared, we would not expect to be observing Class 0 objects in these cases. However, the spectral indices for all but G16.59-0.05:DRT04 1 suggest that these objects are extremely young. The sources at G00.55-0.85:DRT04 2, G10.62-0.38:DRT04 4, G19.61-0.23:DRT03 3, G28.86+0.07:DRT04 1 and G35.20-1.74:DRT04 1 have a values equal to 2.64, 2.13, 2.96, 2.35, 3.29, respectively. These can be classified as extreme Class I objects, whereas G16.59-0.05:DRT04 1, though apparently slightly more evolved with a smaller a value of 1.08, it would still be considered a Class I object in this scheme. G35.20-1.74:DRT04 1 has a positive value of $\alpha_{IR}$ that is greater than 3, implying that this source is an extreme Class I object, perhaps even a Class 0 source. In the final case of G09.62-0.19, the near-infrared source seen near the maser location has been shown not to be the same as any of the sources seen in the mid-infrared (De Buizer et al. 2003) and so no spectral index was derived for this source.

\subsection{Comparison to the Methanol Maser Selected Mid-Infrared Survey}

Twenty-one sites of massive star formation with methanol maser emission were imaged at 10 and 18 $\micron$ by De Buizer, Pi{\~ n}a, \& Telesco (2000). The purpose of that survey, like the one presented here, was to explore the relationships between masers and mid-infrared sources in star forming regions. The detection rate of mid-infrared sources in fields containing methanol maser emission was found to be 67\%. This is comparable to the detection rate of 77\% towards water maser seen in the present survey. However this similarity was to be expected, since 22 of the 26 water maser sources in our survey are contained in the methanol maser survey of Caswell et al. (1995) and are known to contain 6.7 GHz methanol masers on the field (within the 10$\arcsec$ astrometric accuracy of the observations). Similarly many of the sources in the methanol maser-selected sample are now known to have water masers as well. Though it would appear that many maser species (water, methanol, and OH) appear in the same star forming regions, they are very often tracing different sources or different parts of sources (or apparently no sources at all), and it is this closer spatial relationship that we have tried to explore in both surveys.

The main goal of the methanol maser selected survey was to investigate the idea that methanol masers are tracing circumstellar accretion disks around young massive stars. Ten of the twenty-one sites in that survey contained methanol masers distributed in a linear fashion. Norris et al. (1993, 1998) pointed out that methanol masers in linear distributions constituted a statistically large portion of the methanol maser population and thus an explanation of these linear arrangements needed to be made. Norris et al. (1993, 1998) argued that such linear arrangements were possible if the masers were tracing the dense material of an edge-on circumstellar disk. Further evidence that this may be the case came from a few linear methanol maser distributions in their survey that had gradients in the radial velocities of the maser spots along the maser distribution, indicating that the masers were tracing some sort of rotating medium. Of the 10 sites of linearly distributed methanol masers in the mid-infrared survey of De Buizer, Pi{\~ n}a, \& Telesco (2000), there were 8 detections, 3 of which have sources that are resolved and elongated in their thermal dust emission at the same position angle as the maser distribution. De Buizer, Pi{\~ n}a, \& Telesco (2000) argued that these elongated mid-infrared objects may in fact be circumstellar disks. These circumstellar disk candidates seemed to lend credibility to the hypothesis that linearly distributed methanol masers exist in, and delineate circumstellar accretion disks around young massive stars. However, follow-up observations in the mid-infrared at high spatial resolution by De Buizer et al. (2002) of one of these elongated sources revealed it to be three individual mid-infrared sources arranged in a linear fashion, and not a disk. Further observations by De Buizer (2003) searching for outflow perpendicular to the angle of the accretion disks as given by the methanol masers, showed a higher proclivity of outflow tracers parallel to the methanol maser distribution, in direct conflict with the maser/accretion disk hypothesis. De Buizer (2003) suggests that linearly distributed methanol masers may instead be tracing outflows from young massive protostars instead of accretion disks. Furthermore, Walsh et al. (1998) performed an extensive survey of methanol maser sites and found that, while there were indeed a large number of sites of methanol masers in linear arrangements, very few had velocity gradients along their distributions and therefore these velocity gradients were not a characteristic feature of methanol masers. Walsh et al. (1998) instead described how such linear arrangements of methanol masers may be created by shock fronts.

In any case, linear maser distributions are thought to be a distinguishing feature of methanol masers. However, are methanol masers the only maser species that have maser spots arranged in linear fashions? Inspection of the water and OH distributions for the sites in this survey have shown that water masers and OH masers both have a significant fraction of their distributions arranged linearly. Elongated and/or linear water maser distributions are apparent in G00.38+0.04, G00.55-0.85, G10.62-0.38, G19.61-0.23, G33.13-0.09, G35.03+0.35, G43.80-0.13, G48.61+0.02, G49.49-0.39, and Cepheus A HW2 (Figure 22). Elongated and/or linear OH maser distributions are apparent in G00.55-0.85, G16.59-0.05, G28.86+0.07, G32.74-0.07, G33.13-0.09, G34.26+0.15, G35.03+0.35, G35.20-1.74, G40.62-0.14, G43.80-0.13, and G45.07+0.13 (Figure 23). Therefore, in this survey the percentage of linear maser distributions for water masers is 38\% (10/26) and 50\% (11/22) for OH masers. This percentage of linearly distributed water masers is identical to the 37\% of the population of linear distributions in the methanol maser survey by Walsh et al. (1998), and the 38\% in the methanol maser survey of Phillips et al. (1998). In fact, with 50\% of the population of OH masers groups in this survey being linearly distributed, it would appear linearity is most common in OH masers, rather than methanol masers. Furthermore, recent VLBI water maser observations toward several star-forming regions are not only showing linear structures at scales of several AU, but they also are exhibiting well defined line-of-sight velocity gradients (e.g., Cepheus A: Torrelles et al. 2001; W3 IRS5: Imai, Deguchi, \& Sasao (2002); NGC2071: Seth, Greenhill, \& Holder (2002); W75N(B): Torrelles et al. 2003). Therefore, we conclude that linear maser distributions (and even velocity gradients) are not unique to methanol masers and it would appear that linear distributions of masers are common to all the molecular maser species we have surveyed. Furthermore, many of the maser distributions in this water and OH maser survey certainly do not appear to be associated with circumstellar disks. We have water masers distributed in lines radially to some mid-infrared sources (e.g. G00.55-0.85 and G10.62-0.38), and we have OH masers in lines delineating the edges of UC \ion{H}{2} regions (e.g. G00.55-0.85 and G34.26+0.15). The former may be more likely associated with outflow and the latter with the shock front of an expanding \ion{H}{2} region, rather than circumstellar disks. Therefore, while there still may exist certain remote and individual cases were masers (either methanol, water, or OH) may indeed be tracing circumstellar disks, it is unlikely that linearly distributed methanol masers are, as a population, delineating circumstellar disks around young massive stars.

\section{Conclusions}

We observed 26 regions of massive star formation in the mid-infrared that contain water and OH maser emission in order to study the relationship between the stellar sources, dust, and maser emission. From our observations we draw the following main conclusions:

\begin{enumerate}

\item Water masers are found to be associated in some cases with young embedded stars. This was seen in De Buizer et al. (2003) and there are a handful of other candidate embedded sources found in this larger survey. These embedded sources may be HMPOs or perhaps even young embedded intermediate mass stars, given their mid-infrared luminosities.

\item Some distributions of water masers with respect to the mid-infrared and/or radio sources on the field would seem to indicate that some water masers are indeed associated with outflows from these objects.

\item IRAS fluxes are not a good indicator of the spectral energy distribution of the sources of maser emission. Since nearly two-thirds of the fields where mid-infrared sources were detected exhibit multiple mid-infrared sources (14/20), and the remaining third, several have nearby sources seen at other wavelengths (e.g., radio or sub-mm sources), it can be concluded that massive stars are in general gregarious by nature and form in a clustered way. This means that even the angular resolutions of the Midcourse Space Experiment (MSX) satellite ($\sim$5$\arcsec$ at 8 $\mu$m, $\sim$13$\arcsec$ at 21 $\mu$m), though many times better than IRAS, will suffer confusion when observing these regions. Even the Spitzer Space Telescope ($\sim$2$\arcsec$ at 8 $\mu$m, $\sim$6$\arcsec$ at 24 $\mu$m) will have confusion problems at longer wavelengths.

\item For the handful of sources where near-infrared observations were made by Testi et al. (1994, 1998) we conclude based upon the near- to mid-infrared slope of their spectral energy distribution that they are very young embedded objects (Class I or 0). We will explore the nature of the near-infrared emission from these regions using our own observations as well as those from 2MASS in a following paper.

\item Mid-infrared emission seems to be more closely associated with water and OH maser emission than cm radio continuum emission from UC \ion{H}{2} regions, and to a lesser extent, near-infrared emission.

\item Methanol masers are not the only maser population with linearly distributed maser groups. We find that the percentage of linearly distributed water masers is the same (38\%) as the percentage of linearly distributed methanol masers, and that OH masers in our survey are even more prone to linear distributions (50\%). Given the mid-infrared morphologies of the sources associated with the masers, and the outflow observations of De Buizer (2003), we conclude that the vast majority of linearly distributed masers are not tracing circumstellar disks.

\item This mid-infrared survey and that of De Buizer, Pi{\~ n}a, \& Telesco(2000) have confirmed that maser emission in general can trace a variety of phenomena associated with massive stars including shocks, outflows, infall and circumstellar disks. No one maser species is linked exclusively to one particular process or phenomenon. This implies that the conditions for maser excitation are not very stringent, given that each of the phenomena just mentioned would be associated with different densities, temperatures, and molecular abundances.

\item The three different masers types (water, OH, and methanol) do not seem to be associated with different early evolutionary stages of massive stars. Instead it appears that they all trace a variety of stellar phenomena throughout many early stages of massive stellar evolution. This conclusion is based on the fact that the maser sources exhibit the presence or absence of emission at near-infrared, mid-infrared, submillimeter, and radio wavelengths in a variety of combinations and degrees. We also argue that the masers with no associated radio continuum emission could be accounted for by excitation from a small subset of the population of young intermediate mass stars.

\end{enumerate}

\acknowledgments
This research is based on observations made at NASA's InfraRed Telescope Facility by Gemini staff, supported by the Gemini Observatory, which is operated by the Association of Universities for Research in Astronomy, Inc., on behalf of the international Gemini partnership.

Facilities: IRTF(OSCIR), IRTF(MIRLIN).

\clearpage

\begin{deluxetable}{lccccc}
\tablewidth{0pt}
\tabletypesize{\small}
\tablecaption{Coordinates of Water Maser Reference Features \label{tbl-1}}
\tablehead{
\colhead{Target Name} & \colhead{R.A. (B1950)\tablenotemark{a}}   & \colhead{Decl. (B1950)\tablenotemark{a}}  & \colhead{R.A. (J2000)\tablenotemark{b}}   & \colhead{Decl. (J2000)\tablenotemark{b}} & \colhead{Mid-IR?\tablenotemark{c}}
}
\startdata
G00.38$+$0.04                      &17 43 11.23 &$-$28 34 34.0  &17 46 21.34  &$-$28 35 40.6 &N\\
G00.55$-$0.85                      &17 47 03.83 &$-$28 53 39.5  &17 50 14.46  &$-$28 54 29.1 &Y\\
G09.62$+$0.19\tablenotemark{d}     &18 03 16.08 &$-$21 31 59.2  &18 06 16.22  &$-$21 31 38.3 &Y\\
G10.62$-$0.38                      &18 07 30.56 &$-$19 56 28.8  &18 10 28.56  &$-$19 55 49.5 &Y\\
G11.94$-$0.62\tablenotemark{d}     &18 11 03.70 &$-$18 54 18.4  &18 14 00.32  &$-$18 53 23.6 &Y\\
G12.21$-$0.10\tablenotemark{d}     &18 09 43.76 &$-$18 25 06.7  &18 12 39.76  &$-$18 24 17.7 &N\\
G12.68$-$0.18                      &18 10 59.17 &$-$18 02 42.5  &18 13 54.68  &$-$18 01 48.1 &Y\\
G16.59$-$0.05                      &18 18 18.05 &$-$14 33 17.7  &18 21 09.10  &$-$14 31 51.5 &Y\\
G19.61$-$0.23                      &18 24 50.25 &$-$11 58 31.7  &18 27 38.09  &$-$11 56 37.2 &Y\\
G28.86$+$0.07                      &18 41 08.28 &$-$03 38 34.5  &18 43 46.22  &$-$03 35 29.9 &Y\\
G32.74$-$0.07                      &18 48 47.81 &$-$00 15 43.5  &18 51 21.86  &$-$00 12 06.2 &N\\
G33.13$-$0.09                      &18 49 34.25 &$+$00 04 32.2  &18 52 07.91  &$+$00 08 12.8 &N\\
G34.26$+$0.15                      &18 50 46.36 &$+$01 11 13.9  &18 53 18.76  &$+$01 14 59.6 &Y\\
G35.03$+$0.35                      &18 51 29.12 &$+$01 57 30.7  &18 54 00.64  &$+$02 01 19.4 &N\\
G35.20$-$0.74                      &18 55 41.05 &$+$01 36 31.1  &18 58 12.98  &$+$01 40 37.6 &Y\\
G35.20$-$1.74                      &18 59 13.24 &$+$01 09 13.5  &19 01 45.68  &$+$01 13 35.0 &Y\\
G35.58$-$0.03                      &18 53 51.38 &$+$02 16 29.4  &18 56 22.55  &$+$02 20 28.1 &Y\\
G40.62$-$0.14                      &19 03 35.43 &$+$06 41 57.2  &19 06 01.63  &$+$06 46 36.9 &Y\\
G43.80$-$0.13                      &19 09 30.98 &$+$09 30 46.8  &19 11 54.05  &$+$09 35 51.2 &Y\\
G45.07$+$0.13                      &19 11 00.40 &$+$10 45 43.1  &19 13 22.06  &$+$10 50 53.6 &Y\\
G45.47$+$0.05                      &19 12 04.42 &$+$11 04 11.0  &19 14 25.74  &$+$11 09 25.9 &Y\\
G45.47$+$0.13                      &19 11 45.97 &$+$11 07 02.9  &19 14 07.23  &$+$11 12 16.6 &N\\
G48.61$+$0.02                      &19 18 12.93 &$+$13 49 44.7  &19 20 31.19  &$+$13 55 24.9 &Y\\
G49.49$-$0.39                      &19 21 26.32 &$+$14 24 41.8  &19 23 43.98  &$+$14 30 35.2 &Y\\
G75.78$+$0.34\tablenotemark{d}     &20 19 51.88 &$+$37 17 00.3  &20 21 43.98  &$+$37 26 38.0 &N\\
Cepheus A HW2\tablenotemark{e}     &22 54 19.04 &$+$61 45 47.4  &22 56 17.97  &$+$62 01 49.9 &Y\\
\enddata

\tablenotetext{a}{Maser reference positions are from Forster and Caswell (1989) unless otherwise noted.}
\tablenotetext{b}{Converted from the B1950 coordinates. The transformation from Besselian dates in the FK4 system to Julian dates in the FK5 system are accurate to $\sim$0$\farcs$1.}
\tablenotetext{c}{Detection of a mid-infrared source within 5$\arcsec$ of any water maser at any of the observed wavelengths.}
\tablenotetext{d}{B1950 coordinates are from Hofner \& Churchwell (1996).}
\tablenotetext{e}{B1950 coordinates are from Torrelles et al. (1996).}
\end{deluxetable}

\clearpage

\begin{deluxetable}{lccccc}
\tabletypesize{\scriptsize}
\tablewidth{0pt}
\tablecaption{Flux Densities Derived from the Mid-Infrared Observations \label{tbl-2}}
\tablehead{
\colhead{Source Name} &\colhead{Maser}  &\colhead{F$_{N}$} &\colhead{F$_{11.7\micron}$} &\colhead{F$_{18.1\micron}$} &\colhead{F$_{20.8\micron}$} \\
 &\colhead{Assoc.\tablenotemark{a}}  &\colhead{(Jy)} &\colhead{(Jy)} &\colhead{(Jy)} &\colhead{(Jy)}
}
\startdata
G00.38$+$0.04:DRT04 1   &     &\nodata          &1.33$\pm$0.13     &\nodata          &1.56$\pm$0.36    \\
G00.55$-$0.85:DRT04 1   &w    &0.27$\pm$0.02    &0.30$\pm$0.05     &$<$0.37          &1.30$\pm$0.33    \\
G00.55$-$0.85:DRT04 2   &     &1.90$\pm$0.05    &3.11$\pm$0.29     &13.2$\pm$1.49    &31.4$\pm$4.26    \\
G00.55$-$0.85:DRT04 3   &h    &2.60$\pm$0.06    &4.28$\pm$0.39     &19.4$\pm$2.18    &47.2$\pm$6.31    \\
G00.55$-$0.85:DRT04 4   &     &1.06$\pm$0.04    &1.01$\pm$0.13     &6.44$\pm$0.74    &12.08$\pm$1.98   \\
G09.62$+$0.19:DRT03 1   &     &\nodata          &0.20$\pm$0.06     &\nodata          &\nodata          \\
G09.62$+$0.19:DRT03 2   &     &\nodata          &1.56$\pm$0.26     &\nodata          &\nodata          \\
G09.62$+$0.19:DRT03 3   &     &\nodata          &18.6$\pm$1.69     &\nodata          &\nodata          \\
G09.62$+$0.19:DRT03 4   &     &\nodata          &0.34$\pm$0.06     &\nodata          &\nodata          \\
G09.62$+$0.19:DRT03 5   &     &\nodata          &0.14$\pm$0.04     &\nodata          &\nodata          \\
G09.62$+$0.19:DRT03 6   &     &\nodata          &0.17$\pm$0.05     &\nodata          &\nodata          \\
G09.62$+$0.19:DRT03 7   &     &\nodata          &0.24$\pm$0.06     &\nodata          &\nodata          \\
G09.619$+$0.193:DPT00 1 &     &0.33$\pm$0.02    &0.43$\pm$0.05     &3.17$\pm$0.39    &\nodata          \\
G09.62$+$0.19:DRT03 9   &     &\nodata          &0.07$\pm$0.02     &\nodata          &\nodata          \\
G10.62$-$0.38:DRT04 1   &     &0.05$\pm$0.01    &0.10$\pm$0.03     &0.38$\pm$0.07    &\nodata          \\
G10.62$-$0.38:DRT04 2   &w    &0.26$\pm$0.01    &0.23$\pm$0.03     &1.97$\pm$0.24    &\nodata          \\
G10.62$-$0.38:DRT04 3   &     &0.07$\pm$0.01    &0.19$\pm$0.04     &$<$0.37          &\nodata          \\
G10.62$-$0.38:DRT04 4   &h    &0.14$\pm$0.01    &0.20$\pm$0.04     &0.50$\pm$0.08    &\nodata          \\
G10.62$-$0.38:DRT04 5   &     &\nodata          &0.91$\pm$0.12     &\nodata          &11.1$\pm$1.55    \\
G11.94$-$0.62:DRT03 1   &w    &\nodata          &0.18$\pm$0.04     &\nodata          &0.83$\pm$0.28    \\
G11.94$-$0.62:DRT03 2   &     &\nodata          &1.11$\pm$0.14     &\nodata          &5.43$\pm$0.80    \\
G11.94$-$0.62:DRT03 3   &     &\nodata          &1.28$\pm$0.16     &\nodata          &7.82$\pm$1.20    \\
G11.94$-$0.62:DRT03 4   &     &\nodata          &1.44$\pm$0.17     &\nodata          &11.3$\pm$1.64    \\
G11.94$-$0.62:DRT03 5   &     &\nodata          &1.45$\pm$0.18     &\nodata          &11.1$\pm$1.53    \\
G11.94$-$0.62:DRT03 6   &     &\nodata          &1.21$\pm$0.14     &\nodata          &8.70$\pm$1.43    \\
G11.94$-$0.62:DRT03 7   &     &\nodata          &$<$0.12           &\nodata          &5.01$\pm$1.04    \\
G12.21$-$0.10           &     &\nodata          &$<$0.12           &\nodata          &$<$0.75          \\
G12.68$-$0.18:DRT04 1   &w    &0.18$\pm$0.01    &0.19$\pm$0.04     &0.78$\pm$0.12    &3.17$\pm$0.50    \\
G12.68$-$0.18:DRT04 2   &     &0.27$\pm$0.02    &0.50$\pm$0.10     &$<$0.37          &3.28$\pm$0.57    \\
G16.59$-$0.05:DRT04 1   &w,h  &0.03$\pm$0.01    &$<$0.12           &0.23$\pm$0.06    &0.77$\pm$0.24    \\
G16.59$-$0.05:DRT04 2   &     &0.17$\pm$0.02    &0.49$\pm$0.06     &2.98$\pm$0.35    &6.62$\pm$0.98    \\
G19.61$-$0.23:DRT03 1   &     &\nodata          &0.58$\pm$0.07     &1.39$\pm$0.15    &3.66$\pm$0.72    \\
G19.61$-$0.23:DRT03 2   &     &2.64$\pm$0.08    &3.04$\pm$0.30     &9.20$\pm$1.07    &17.4$\pm$2.53    \\
G19.61$-$0.23:DRT03 3   &w,h  &9.47$\pm$0.21    &11.6$\pm$1.06     &52.8$\pm$5.95    &82.9$\pm$11.3    \\
G19.61$-$0.23:DRT03 4   &     &4.06$\pm$0.12    &3.36$\pm$0.38     &18.3$\pm$2.09    &33.5$\pm$4.61    \\
G19.61$-$0.23:DRT03 5   &     &4.02$\pm$0.14    &4.75$\pm$0.45     &21.2$\pm$2.43    &20.9$\pm$3.21    \\
G19.61$-$0.23:DRT03 6   &     &0.08$\pm$0.01    &$<$0.12           &1.08$\pm$0.18    &2.70$\pm$0.85    \\
G28.86$+$0.07:DRT04 1   &w,h  &1.95$\pm$0.06    &1.93$\pm$0.18     &6.38$\pm$0.73    &13.8$\pm$2.01    \\
G28.86$+$0.07:DRT04 2   &     &0.09$\pm$0.01    &0.13$\pm$0.06     &0.71$\pm$0.12    &3.73$\pm$0.82    \\
G32.74$-$0.07           &     &$<$0.03          &$<$0.12           &\nodata          &$<$0.75          \\
G33.13$-$0.09           &     &$<$0.03          &$<$0.12           &\nodata          &$<$0.75          \\
G34.26$+$0.15:DRT04 1   &w,h  &4.52$\pm$0.12    &3.28$\pm$0.31     &\nodata          &19.8$\pm$2.85    \\
G34.26$+$0.15:DRT03 5   &     &0.74$\pm$0.04    &0.64$\pm$0.08     &\nodata          &5.15$\pm$1.21    \\
G35.03$+$0.35           &     &$<$0.03          &\nodata           &\nodata          &\nodata          \\
G35.20$-$0.74:DRT04 1   &w,h  &2.81$\pm$0.08    &2.91$\pm$0.27     &\nodata          &76.8$\pm$10.2    \\
G35.20$-$1.74:PFL97 MIR1   &w,h  &2.33$\pm$0.06    &2.57$\pm$0.23     &\nodata          &22.5$\pm$2.99    \\
G35.20$-$1.74:PFL97 MIR3   &     &\nodata          &31.6$\pm$2.83     &\nodata          &293$\pm$38.5     \\
G35.20$-$1.74:PFL97 MIR5   &     &\nodata          &4.02$\pm$0.44     &\nodata          &28.5$\pm$4.05    \\
G35.58$-$0.03:DRT04 1   &w,h  &1.55$\pm$0.06    &1.79$\pm$0.18     &\nodata          &22.1$\pm$2.96    \\
G40.62$-$0.14:DRT04 1   &w,h  &0.93$\pm$0.03    &1.16$\pm$0.12     &\nodata          &21.5$\pm$2.96    \\
G43.80$-$0.13:DRT04 1   &w,h  &1.00$\pm$0.04    &0.57$\pm$0.06     &\nodata          &14.0$\pm$1.94    \\
G43.80$-$0.13:DRT04 2   &     &0.48$\pm$0.03    &0.83$\pm$0.10     &\nodata          &9.12$\pm$1.38    \\
G45.07$+$0.13:DRT03 1   &     &0.89$\pm$0.05    &1.40$\pm$0.13     &\nodata          &15.1$\pm$2.13    \\
G45.07$+$0.13:DRT03 2   &w,h  &20.4$\pm$0.43    &21.7$\pm$1.94     &\nodata          &102$\pm$13.5     \\
G45.07$+$0.13:DRT03 3   &w,h  &4.26$\pm$0.10    &5.50$\pm$0.49     &\nodata          &19.7$\pm$2.78    \\
G45.47$+$0.05:DRT04 1   &     &0.29$\pm$0.04    &0.15$\pm$0.04     &\nodata          &8.71$\pm$1.22    \\
G45.47$+$0.13           &     &$<$0.03          &$<$0.12           &\nodata          &$<$0.75          \\
G48.61$+$0.02:DRT04 1   &     &\nodata          &1.97$\pm$0.20     &\nodata          &15.1$\pm$2.12    \\
G48.61$+$0.02:DRT04 2   &     &0.12$\pm$0.02    &0.54$\pm$0.08     &\nodata          &2.62$\pm$0.52    \\
G48.61$+$0.02:DRT04 3   &     &0.07$\pm$0.02    &$<$0.12           &\nodata          &$<$0.75          \\
G48.61$+$0.02:DRT04 4   &     &0.06$\pm$0.02    &$<$0.12           &\nodata          &$<$0.75          \\
G49.49$-$0.39:DRT04 1   &     &0.29$\pm$0.03    &1.06$\pm$0.12     &\nodata          &6.47$\pm$0.93    \\
G75.78$+$0.34           &     &\nodata          &$<$0.12           &\nodata          &$<$0.75          \\
Cepheus A HW2:DRT04 1   &w    &\nodata          &2.22$\pm$0.30     &\nodata          &\nodata          \\
\enddata
\tablecomments{
All values quoted with a ``$<$'' are non-detections quoted with the typical 3-$\sigma$ upper limit on a point source detection for that filter and instrument.}
\tablenotetext{a} {The letter ``w'' denotes the sources most likely associated with the water masers, and ``h'' denotes the sources most likely associated with the hydroxyl masers. Sources are labeled 1, 2, 3, etc. for each field. These are the IAU recommended names which are in the form Glll.ll$\pm$b.bb:DRT04 \#, where \# is the source number. Some of these sources already have names, as discussed in the text and given in the table.}
\end{deluxetable}

\clearpage

\begin{deluxetable}{lcccccc}
\tabletypesize{\scriptsize}
\tablewidth{0pt}
\tablecaption{Physical Parameters of the Mid-Infrared Sources \label{tbl-3}}
\tablehead{
\colhead{Source Name} &\colhead{Maser} &\colhead{D\tablenotemark{b}}    & \colhead{T$_{Dust}$} & \colhead{$\tau _{11.7 \micron}$\tablenotemark{c}} & \colhead{A$_{V}$\tablenotemark{c}} & \colhead{L$_{MIR}$}\\
  &\colhead{Assoc.\tablenotemark{a}} &\colhead{(kpc)}    & \colhead{(K)} &  &  & \colhead{(L$_{\sun}$)}
}
\startdata
G00.38$+$0.04:DRT04 1\tablenotemark{o} &    &10.1\tablenotemark{d} &300/236   &$\infty$/0.00    &$\infty$/0.0   &1490/1140   \\ 
G00.55$-$0.85:DRT04 1\tablenotemark{o} &w   &9.1\tablenotemark{d}  &170/148  &$\infty$/0.01  &$\infty$/0.2  &662/409 \\ 
G00.55$-$0.85:DRT04 2 &    &9.1\tablenotemark{d}  &120  &0.12  &4.3  &9303  \\ 
G00.55$-$0.85:DRT04 3 &h   &9.1\tablenotemark{d}  &116  &0.10  &3.6  &15470 \\ 
G00.55$-$0.85:DRT04 4 &    &9.1\tablenotemark{d}  &112  &0.05  &1.6  &4343  \\ 
G09.619$+$0.193:DPT00 1\tablenotemark{o,p} &    &0.7\tablenotemark{e}  &132/121  &$\infty$/0.07  &$\infty$/2.3  &16/1.7 \\ 
G10.62$-$0.38:DRT04 1\tablenotemark{n,p} &    &6.5\tablenotemark{d}  &178/158  &\nodata  &\nodata  &117/71  \\ 
G10.62$-$0.38:DRT04 2\tablenotemark{o,p} &w   &6.5\tablenotemark{d}  &145/131  &$\infty$/0.02  &$\infty$/0.08  &745/377  \\ 
G10.62$-$0.38:DRT04 4\tablenotemark{n,p} &h   &6.5\tablenotemark{d}  &197/173\tablenotemark{n}  &\nodata  &\nodata   &149/98 \\ 
G10.62$-$0.38:DRT04 5 &    &6.5\tablenotemark{d}  &125  &0.01  &0.3  &1657 \\ 
G11.94$-$0.62:DRT03 1\tablenotemark{o} &w   &4.2\tablenotemark{f}  &167/146  &$\infty$/0.01  &$\infty$/0.2  &90/55 \\ 
G11.94$-$0.62:DRT03 2 &    &4.2\tablenotemark{f}  &144  &0.00  &0.1  &356   \\ 
G11.94$-$0.62:DRT03 3 &    &4.2\tablenotemark{f}  &136  &0.00  &0.1  &498   \\ 
G11.94$-$0.62:DRT03 4 &    &4.2\tablenotemark{f}  &128  &0.02  &0.7  &709  \\ 
G11.94$-$0.62:DRT03 5 &    &4.2\tablenotemark{f}  &128  &0.01  &0.3  &697   \\ 
G11.94$-$0.62:DRT03 6 &    &4.2\tablenotemark{f}  &130  &0.00  &0.1  &547   \\ 
G12.68$-$0.18:DRT04 1\tablenotemark{o} &w   &4.9\tablenotemark{g}  &123/112  &$\infty$/0.07  &$\infty$/2.5  &526/237 \\ 
G12.68$-$0.18:DRT04 2 &    &4.9\tablenotemark{g}  &144  &0.01  &0.2  &216   \\ 
G16.59$-$0.05:DRT04 1\tablenotemark{n,q} &w,h &4.7\tablenotemark{h}  &116/107\tablenotemark{n}  &\nodata  &\nodata   &153/63 \\ 
G16.59$-$0.05:DRT04 2 &    &4.7\tablenotemark{h}  &113  &0.01  &0.2  &528   \\ 
G19.61$-$0.23:DRT03 1 &    &4.0\tablenotemark{i}  &134  &0.00  &0.1  &211   \\ 
G19.61$-$0.23:DRT03 2 &    &4.0\tablenotemark{i}  &138  &0.02  &0.7  &1014 \\ 
G19.61$-$0.23:DRT03 3 &w,h &4.0\tablenotemark{i}  &131  &0.07  &2.5  &4749 \\ 
G19.61$-$0.23:DRT03 4 &    &4.0\tablenotemark{i}  &121  &0.02  &0.7  &1905 \\ 
G19.61$-$0.23:DRT03 5 &    &4.0\tablenotemark{i}  &148  &0.02  &0.7  &1268 \\ 
G19.61$-$0.23:DRT03 6\tablenotemark{p} &    &4.0\tablenotemark{i}  &104  &0.04  &1.4  &166  \\ 
G28.86$+$0.07:DRT04 1 &w,h &8.5\tablenotemark{j}  &131  &0.14  &4.9  &3584 \\ 
G28.86$+$0.07:DRT04 2 &    &8.5\tablenotemark{j}  &98   &0.04  &1.4  &1101 \\ 
G34.26$+$0.15:DRT04 1 &w,h &4.2\tablenotemark{j}  &136  &0.04  &1.4  &1265 \\ 
G34.26$+$0.15:DRT03 5 &    &4.2\tablenotemark{j}  &127  &0.06  &2.1  &324  \\ 
G35.20$-$0.74:DRT04 1 &w,h &2.3\tablenotemark{j}  &100  &0.30  &10.5 &1665 \\ 
G35.20$-$1.74:PFL97 MIR1\tablenotemark{o} &w,h &3.1\tablenotemark{f}  &139/126  &$\infty$/0.29  &$\infty$/10.1  &2740/780 \\ 
G35.20$-$1.74:PFL97 MIR3 &    &3.1\tablenotemark{f}  &124  &0.24  &8.4  &10130\\ 
G35.20$-$1.74:PFL97 MIR5 &    &3.1\tablenotemark{f}  &131  &0.02  &0.7  &978  \\ 
G35.58$-$0.03:DRT04 1 &w,h &10.2\tablenotemark{m}  &116  &0.18  &6.3  &8345  \\ 
G40.62$-$0.14:DRT04 1 &w,h &2.3\tablenotemark{j}  &107  &0.37  &12.9 &439  \\ 
G43.80$-$0.13:DRT04 1 &w,h &9.0\tablenotemark{m}  &101  &0.21  &7.3  &4544  \\ 
G43.80$-$0.13:DRT04 2 &    &9.0\tablenotemark{m}  &118  &0.01  &0.3   &2633  \\ 
G45.07$+$0.13:DRT03 1 &    &9.7\tablenotemark{k}  &119  &0.05  &1.8  &5071 \\ 
G45.07$+$0.13:DRT03 2\tablenotemark{o} &w,h &9.7\tablenotemark{k}  &166/149  &$\infty$/0.60  &$\infty$/21.0  &59400/36920 \\ 
G45.07$+$0.13:DRT03 3\tablenotemark{o} &w,h &9.7\tablenotemark{k}  &181/158  &$\infty$/0.08  &$\infty$/2.8  &11400/7401 \\ 
G45.47$+$0.05:DRT04 1 &    &8.3\tablenotemark{l}  &87   &0.10  &3.5  &3007 \\ 
G48.61$+$0.02:DRT04 1 &    &11.5\tablenotemark{j} &143  &0.01  &0.3  &7403  \\ 
G48.61$+$0.02:DRT04 2\tablenotemark{n} &    &11.5\tablenotemark{j} &164/144\tablenotemark{n}  &\nodata  &\nodata   &2150/1290 \\ 
G49.49$-$0.39:DRT04 1\tablenotemark{o} &    &7.3\tablenotemark{l}  &153/136  &$\infty$/0.05  &$\infty$/1.8  &2200/1248 \\ 
\enddata
\tablecomments{All physical parameters were derived using the 11.7 $\mu$m and 20.8 $\mu$m flux densities unless otherwise noted.}
\tablenotetext{a}  {A ``w'' denotes the sources closest to the water masers, and a ``h'' denotes the sources closest to the hydroxyl masers.}
\tablenotetext{b} {Distances are from: d) Walsh et al. (1997); e) Wink, Altenhoff, \& Mezger (1982); f) Hofner \& Churchwell (1996); g) this work; h) Codella, Testi, \& Cesaroni (1997); i) Genzel \& Downes (1977); j) Forster \& Caswell (1989); k) Wood \& Churchwell (1989); l) Kuchar \& Bania (1994); and m) Watson et al. (2003)}
\tablenotetext{c}  {These values are emission optical depth at 11.7 $\micron$ and corresponding extinction in the visible.}
\tablenotetext{n}  {Because of the low S/N for this source, we cannot determine if it is resolved or not. All derived values for this source are given in the form BB/OT, where BB is the lower limit (blackbody) size of the source, and OT is the optically thin upper limit on size.}
\tablenotetext{o}  {This source is unresolved. Derived values are given in the form BB/UL, where BB is the lower limit (blackbody) size of the source, and UL is the upper limit on size given by our resolution.}
\tablenotetext{p}  {Physical parameters were calculated using the 11.7 $\mu$m and 18.1 $\mu$m flux densities.}
\tablenotetext{q}  {Physical parameters were calculated using the N band and 20.8 $\mu$m flux densities.}
\end{deluxetable}

\clearpage

\begin{deluxetable}{lcccccc}
\tablewidth{0pt}
\tabletypesize{\scriptsize}
\tablecaption{Radio Continuum Flux and Derived Spectral Types \label{tbl-4}}
\tablehead{
\colhead{Source}    &\colhead{$\nu_{radio}$($\lambda_{radio}$)}  &\colhead{Radio}   &\colhead{Radio F$_{\nu}$}   &\colhead{Radio} &\colhead{Mid-IR} \\
\colhead{Name}    &\colhead{GHz(cm)}  &\colhead{Reference\tablenotemark{a}}   &\colhead{(mJy)}   &\colhead{Sp. Type} &\colhead{Sp. Type}
}
\startdata
G00.38$+$0.04:DRT04 1   &8.5(3.5) &FC2000  &$<$0.7     &$<$B1.2   &B3.2/B3.8     \\
G00.55$-$0.85:DRT04 1   &8.5(3.5) &FC2000  &?\tablenotemark{b}  &\nodata   &B5.6/B6.3     \\
G00.55$-$0.85:DRT04 2   &8.5(3.5) &FC2000  &36         &O9.9      &B1.2          \\
G00.55$-$0.85:DRT04 3   &8.5(3.5) &FC2000  &167        &O8.7      &B1.0          \\
G00.55$-$0.85:DRT04 4   &8.5(3.5) &FC2000  &$<$1.9     &$<$B0.9   &B2.5          \\
G09.619$+$0.193:DPT00 1 &8.5(3.5) &FC2000  &$<$0.7     &$<$B3.2   &A4.3/A8.8     \\
G10.62$-$0.38:DRT04 1   &5.0(6.0) &WC1989  &$<$4.2     &$<$B0.9   &B8.6/B9.5     \\
G10.62$-$0.38:DRT04 2   &5.0(6.0) &WC1989  &$<$4.2     &$<$B0.9   &B5.4/B6.5     \\
G10.62$-$0.38:DRT04 4   &5.0(6.0) &WC1989  &$<$4.2     &$<$B0.9   &B8.2/B8.9     \\
G10.62$-$0.38:DRT04 5   &5.0(6.0) &WC1989  &$<$4.2     &$<$B0.9   &B3.1          \\
G11.94$-$0.62:DRT03 1   &5.0(6.0) &WC1989  &$<$1.2     &$<$B0.7   &B9.1/A0.0     \\
G11.94$-$0.62:DRT03 2   &5.0(6.0) &WC1989  &Merged\tablenotemark{c}   &\nodata    &B6.7   \\
G11.94$-$0.62:DRT03 3   &5.0(6.0) &WC1989  &Merged\tablenotemark{c}   &\nodata    &B6.0    \\
G11.94$-$0.62:DRT03 4   &5.0(6.0) &WC1989  &Merged\tablenotemark{c}   &\nodata    &B5.5    \\
G11.94$-$0.62:DRT03 5   &5.0(6.0) &WC1989  &Merged\tablenotemark{c}   &\nodata    &B5.5    \\
G11.94$-$0.62:DRT03 6   &5.0(6.0) &WC1989  &Merged\tablenotemark{c}   &\nodata    &B5.9    \\
G12.68$-$0.18:DRT04 1   &8.5(3.5) &FC2000  &$<$0.5      &$<$B2.0  &B5.9/B7.7     \\
G12.68$-$0.18:DRT04 2   &8.5(3.5) &FC2000  &$<$0.5      &$<$B2.0  &B7.8          \\
G16.59$-$0.05:DRT04 1   &8.5(3.5) &FC2000  &0.3         &B2.2     &B8.2/B9.7     \\
G16.59$-$0.05:DRT04 2   &8.5(3.5) &FC2000  &$<$0.3      &$<$B2.2  &B5.9          \\
G19.61$-$0.23:DRT03 1   &4.9(6.1) &G1998   &$<$2.7      &$<$B1.4  &B7.9          \\
G19.61$-$0.23:DRT03 2   &4.9(6.1) &G1998   &530         &O9.1     &B4.2          \\
G19.61$-$0.23:DRT03 3   &4.9(6.1) &G1998   &1370        &O8.4     &B2.3          \\
G19.61$-$0.23:DRT03 4   &4.9(6.1) &G1998   &2650        &O7.7     &B3.0          \\
G19.61$-$0.23:DRT03 5   &4.9(6.1) &G1998   &890         &O8.7     &B3.5          \\
G19.61$-$0.23:DRT03 6   &4.9(6.1) &G1998   &?\tablenotemark{d}    &\nodata  &B8.1          \\
G28.86$+$0.07:DRT04 1   &23.0(1.3)&C1997   &$<$0.34     &$<$B1.6   &B2.8        \\
G28.86$+$0.07:DRT04 2   &23.0(1.3)&C1997   &$<$0.34     &$<$B1.6   &B3.9        \\
G34.26$+$0.15:DRT04 1   &5.0(6.0) &WC1989  &1527        &O8.3     &B3.5          \\
G34.26$+$0.15:DRT03 5   &5.0(6.0) &WC1989  &34          &B0.7     &B7.0          \\
G35.20$-$0.74:DRT04 1   &15.0(2.0)&HL1988  &0.8         &B2.4     &B3.1          \\
G35.20$-$1.74:PFL97 MIR1   &5.0(6.0) &WC1989  &$<$2.3      &$<$B1.7  &B3.0/B5.2     \\
G35.20$-$1.74:PFL97 MIR3   &5.0(6.0) &WC1989  &1529        &O8.7     &B1.2          \\
G35.20$-$1.74:PFL97 MIR5   &5.0(6.0) &WC1989  &$<$2.3      &$<$B1.7  &B4.3          \\
G35.58$-$0.03:DRT04 1   &8.3(3.6) &KCW1994 &197         &O8.5     &B1.4          \\
G40.62$-$0.14:DRT04 1   &5.0(6.0) &HM1993  &3.8         &B1.7     &B6.2          \\
G43.80$-$0.13:DRT04 1   &8.3(3.6) &KCW1994 &62.6        &O9.5     &B2.4          \\
G43.80$-$0.13:DRT04 2   &8.3(3.6) &KCW1994 &$<$0.7      &$<$B1.2  &B3.0          \\
G45.07$+$0.13:DRT03 1   &5.0(6.0) &WC1989  &$<$1.0      &$<$B1.1  &B2.2          \\
G45.07$+$0.13:DRT03 2   &5.0(6.0) &WC1989  &$<$1.0      &$<$B1.1  &O9.8/B0.5     \\
G45.07$+$0.13:DRT03 3   &5.0(6.0) &WC1989  &142         &O8.8     &B1.1/B1.5     \\
G45.47$+$0.05:DRT04 1   &5.0(6.0) &WC1989  &$<$1.8      &$<$B1.0  &B2.9          \\
G48.61$+$0.02:DRT04 1   &8.3(3.6) &KCW1994 &76.9        &O8.9     &B1.5          \\
G48.61$+$0.02:DRT04 2   &8.3(3.6) &KCW1994 &4.2         &B0.7     &B3.0/B3.4     \\
G49.49$-$0.39:DRT04 1   &5.0(6.0) &ZH1997  &$<$12.0     &$<$B0.7   &B3.0/B3.5    \\
\enddata
\tablecomments{All cm radio continuum flux values quoted with a ``$<$'' are 3$\sigma$ upper limits. Corresponding radio spectral types with a ``$<$'' are also upper limits, i.e. the real spectral type of the source is later than the one listed.}
\tablenotetext{a}{References are: FC2000-Forster \& Caswell (2000); WC1989-Wood \& Churchwell (1989); G1998-Garay et al. (1998); C1997-Codella, Testi, \& Cesaroni (1997); KCW1994-Kurtz, Churchwell, \& Wood (1994); HL1988-Heaton \& Little (1988); HM1993-Hughes \& MacLeod (1993); and ZH1997-Zhang \& Ho (1997).}
\tablenotetext{b}{The source was observed to have radio continuum emission but no flux density was given by the authors.}
\tablenotetext{c}{The UC \ion{H}{2} region as seen in cm continuum emission breaks up into individual sources in the mid-infrared. Integrated flux density of of the UC \ion{H}{2} region is 879.4 mJy at 6 cm.}
\tablenotetext{d}{Extended cm radio continuum is present at the location of the masers, but there was no well-defined source.}
\end{deluxetable}

\clearpage

\begin{deluxetable}{lccccc}
\tablewidth{0pt}
\tabletypesize{\scriptsize}
\tablecaption{Total Integrated Flux Density Observed in the MIRLIN Field of View and Corresponding IRAS Flux Density Measurements \label{tbl-5}}
\tablehead{
\colhead{Field} & \colhead{MIRLIN 11.7/20.8 $\mu$m}   & \colhead{IRAS PSC}  & \colhead{Sep.\tablenotemark{a}} & \colhead{IRAS 12/25 $\mu$m}  & \colhead{IRAS 11.7/20.8 $\mu$m} \\
\colhead{Name}       & \colhead{(Jy)}            & \colhead{Name}   &(arcmin)   & \colhead{(Jy)}               & \colhead{(Jy)}
}
\startdata
G00.38$+$0.04  &1.3/1.6         &17432$-$2835  &1.40    &25.02/246.37   &21.65/184.13 \\
G00.55$-$0.85  &15.2/123        &17470$-$2853  &0.48    &42.38/5475.57  &32.28/2547.0 \\
G09.62$+$0.19  &21.3/--    &18032$-$2032  &0.65    &38.63/292.40   &33.87/229.22 \\
G10.62$-$0.38  &2.2/38.3        &18075$-$1956  &0.27    &23.48/148.44   &20.77/120.20 \\
G11.94$-$0.62  &10.2/73.8       &18110$-$1854  &0.01    &13.57/222.10   &11.45/151.30 \\
G12.21$-$0.10  &$<$0.1/$<$0.8   &18097$-$1825A &0.86    &2.47/9.62      &2.24/8.50    \\
G12.68$-$0.18  &2.1/11.5        &NO IRAS       &$>$3    &\nodata        &\nodata      \\
G16.59$-$0.05  &0.7/11.5        &18182$-$1433  &0.30    &2.49/35.35     &2.11/24.71   \\
G19.61$-$0.23  &38.6/211        &18248$-$1158  &0.67    &47.84/406.90   &41.70/312.31 \\
G28.86$+$0.07  &2.2/22.7        &18411$-$0338  &0.03    &8.59/105.18    &7.35/75.55   \\
G32.74$-$0.07  &$<$0.01/$<$0.8  &18487$-$0015  &0.46    &3.47/7.87      &3.23/7.65    \\
G33.13$-$0.09  &$<$0.01/$<$0.8  &18496$+$0004  &0.66    &3.25/24.11     &2.85/18.97   \\
G34.26$+$0.15  &72.4/286        &18507$+$0110  &0.45    &140.18/1106.12 &122.65/860.59\\
G35.03$+$0.35  &$<$0.03\tablenotemark{b}/--      &18515$+$0157  &0.93    &5.37/42.2      &4.70/32.86   \\
G35.20$-$0.74  &3.3/82.4        &18556$+$0136  &0.06    &4.26/217.26    &3.40/120.04  \\
G35.20$-$1.74  &48.8/344        &18592$+$0108  &0.56    &114.48/1022.63 &99.55/777.95 \\
G35.58$-$0.03  &1.7/24.4        &18538$+$0216  &0.28    &6.01/77.07     &5.13/54.90   \\
G40.62$-$0.14  &1.2/22.3        &19035$+$0641  &0.14    &2.27/66.71     &1.86/40.81   \\
G43.80$-$0.13  &1.5/27.0        &19095$+$0930  &0.21    &5.30/129.12    &4.38/81.76   \\
G45.07$+$0.13  &28.6/175        &19110$+$1045  &0.03    &57.62/494.34   &50.21/378.82 \\
G45.47$+$0.05  &0.2/12.6        &19120$+$1103  &1.00    &78.78/640.53   &68.83/495.66 \\
G45.47$+$0.13  &9.7/115         &19117$+$1107  &0.37    &37.30/304.00   &32.58/235.14 \\
G48.61$+$0.02  &8.1/36.2        &19181$+$1349  &0.76    &24.93/175.30   &21.94/139.26 \\
G49.49$-$0.39  &227/1550        &19213$+$1424  &1.01    &424.24/4344.36 &366.39/3224.0\\
G75.78$+$0.34  &$<$0.1/$<$0.8   &20198$-$3716  &0.92    &73.80/480.00   &65.20/386.73 \\
Cepheus A HW2  &2.2/--     &22543$+$6145  &0.18    &11.24/820.30   &8.81/424.15  \\
\enddata
\tablecomments{For column 5, the color-corrected IRAS flux densities at 12 and 25 $\mu$m were fit with a blackbody curve from which an extrapolated value was found at 11.7 $\mu$m and an interpolated value was found at 20.8 $\mu$m.}
\tablenotetext{a}  {Separation between maser reference feature and IRAS PSC source coordinates.}
\tablenotetext{b}  {This value is the N band flux density of the field. No 11.7 $\mu$m data was taken.}
\end{deluxetable}

\clearpage

\begin{figure}
\epsscale{0.85}
\plotone{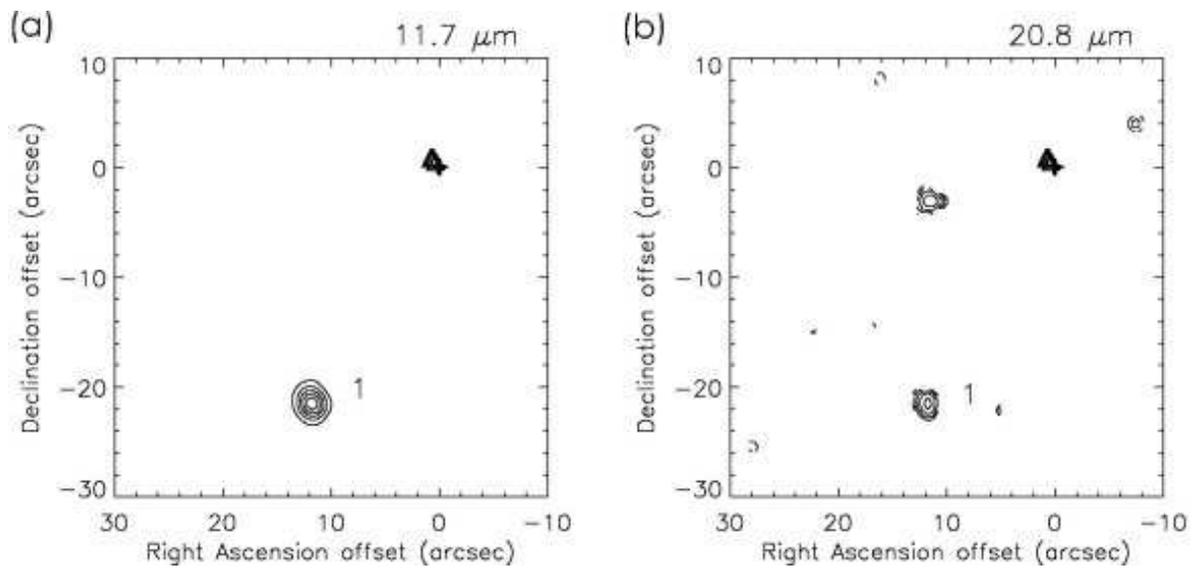}
\caption{Contour plots of the region of G00.38+0.04 observed at (a) 11.7 smoothed to a resolution of 1$\farcs$9 (b) 20.8 $\mu$m smoothed to a resolution of 2$\farcs$0. Unless otherwise noted, in Figures 1-20 crosses represent the water maser positions and triangles represent the OH maser positions from Forster \& Caswell (1989); squares represent the location of near-infrared sources (if any) from Testi et al. (1994, 1998); and the origin of each figure is the position of the water maser reference feature given in Table 1.  \label{fig1}}
\end{figure}

\clearpage

\begin{figure}
\epsscale{0.85}
\plotone{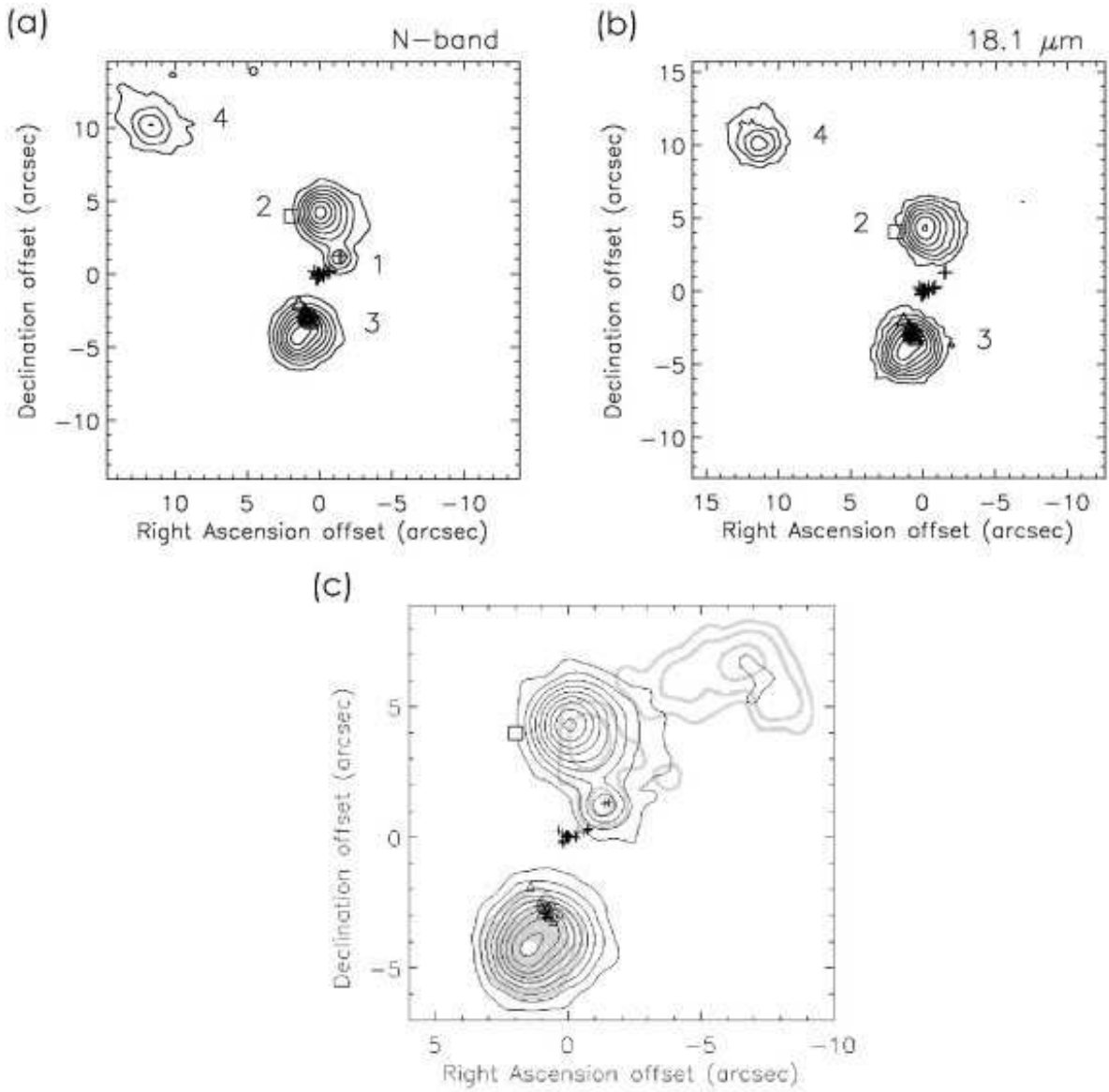}
\caption{Contour plots of the region of G00.55-0.85 observed at (a) N-band smoothed to a resolution of 1$\farcs$4 and (b) 18.1 $\mu$m smoothed to a resolution of 1$\farcs$7. (c) This plot shows the N-band contours once again, overlaid with 3.5 cm contours (thick gray) from Walsh et al. (1998). See Figure 1 for explanation of symbols. \label{fig2}}
\end{figure}

\clearpage

\begin{figure}
\epsscale{0.85}
\plotone{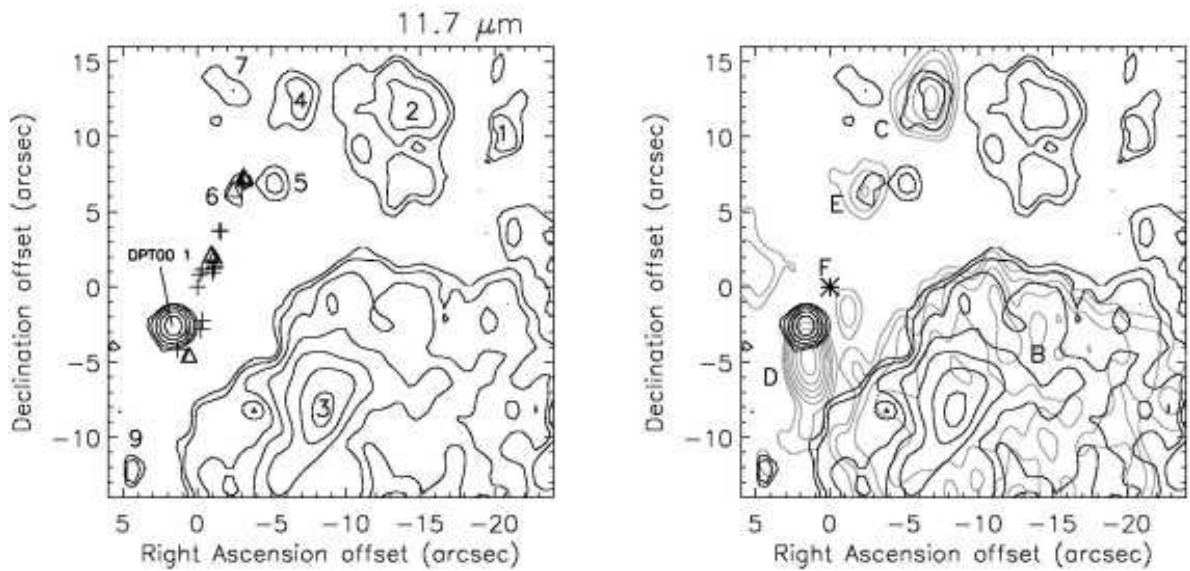}
\caption{Contour plots of the G9.62+0.19 region observed at (a) 11.7$\micron$ smoothed to a resolution of 1$\farcs$7 with the mid-infrared source G9.619+0.193:DPT00  1 labeled. (b) Comparison between the 3.5 cm radio continuum image (gray contours) of Phillips et al. (1998) with the 11.7$\micron$ IRTF image (black contours). Radio components as defined by Garay et al. (1993) are labeled. The position of the ammonia hot core is shown as a asterisk and labeled F. See Figure 1 for explanation of symbols. \label{fig3}}
\end{figure}

\clearpage

\begin{figure}
\epsscale{0.85}
\plotone{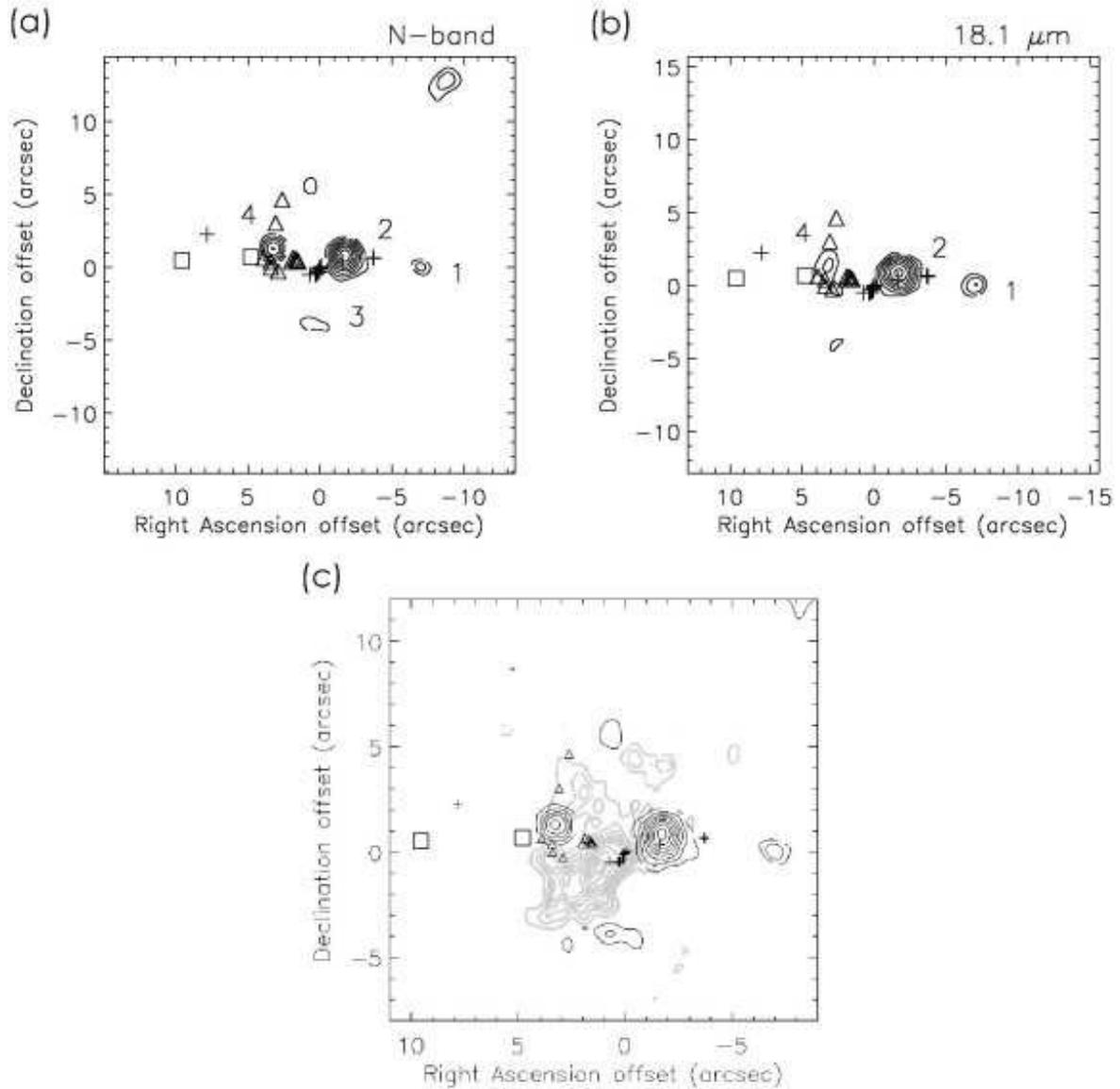}
\caption{Contour plots of the region of G10.62-0.38 observed at (a) N-band smoothed to a resolution of 1$\farcs$5 and (b) 18.1 $\mu$m smoothed to a resolution of 1$\farcs$8. (c) This plot shows the N-band contours with the 6 cm contours (thick gray) from Wood \& Churchwell (1989). See Figure 1 for explanation of symbols. Source 5 is not visible on the field. \label{fig4}}
\end{figure}

\clearpage

\begin{figure}
\epsscale{0.85}
\plotone{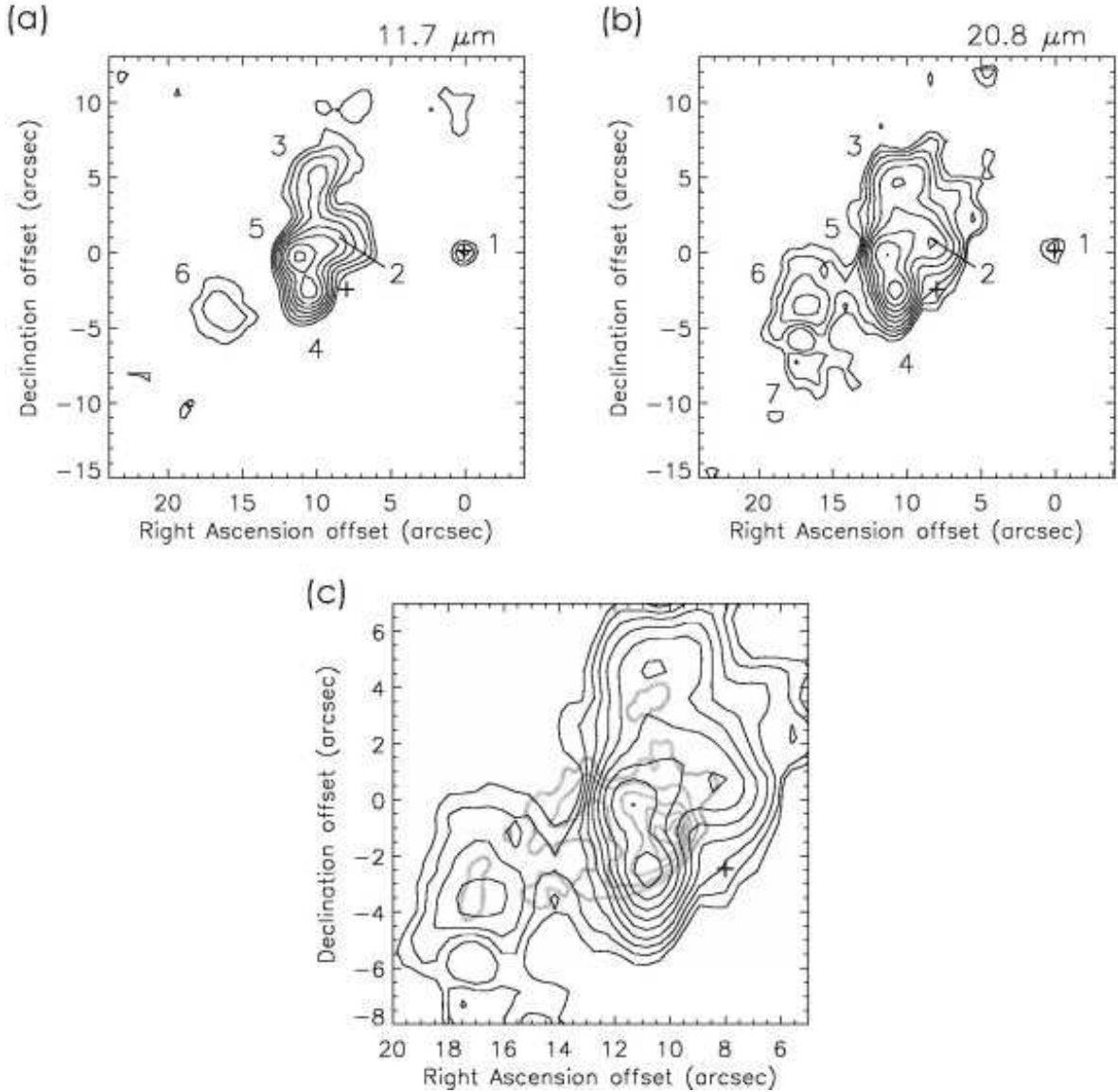}
\caption{Contour plots of the region of G11.94-0.62 observed at (a) 11.7 $\mu$m smoothed to a resolution of 1$\farcs$6 and (b) 20.8 $\mu$m smoothed to a resolution of 2$\farcs$0. (c) Contour plot of the  20.8 $\mu$m image of G11.94-0.62 (black contours) overlaid with the 2 cm radio continuum image (gray contours) of Hofner \& Churchwell (1996). See Figure 1 for explanation of symbols.\label{fig5}}
\end{figure}

\clearpage

\begin{figure}
\epsscale{0.85}
\plotone{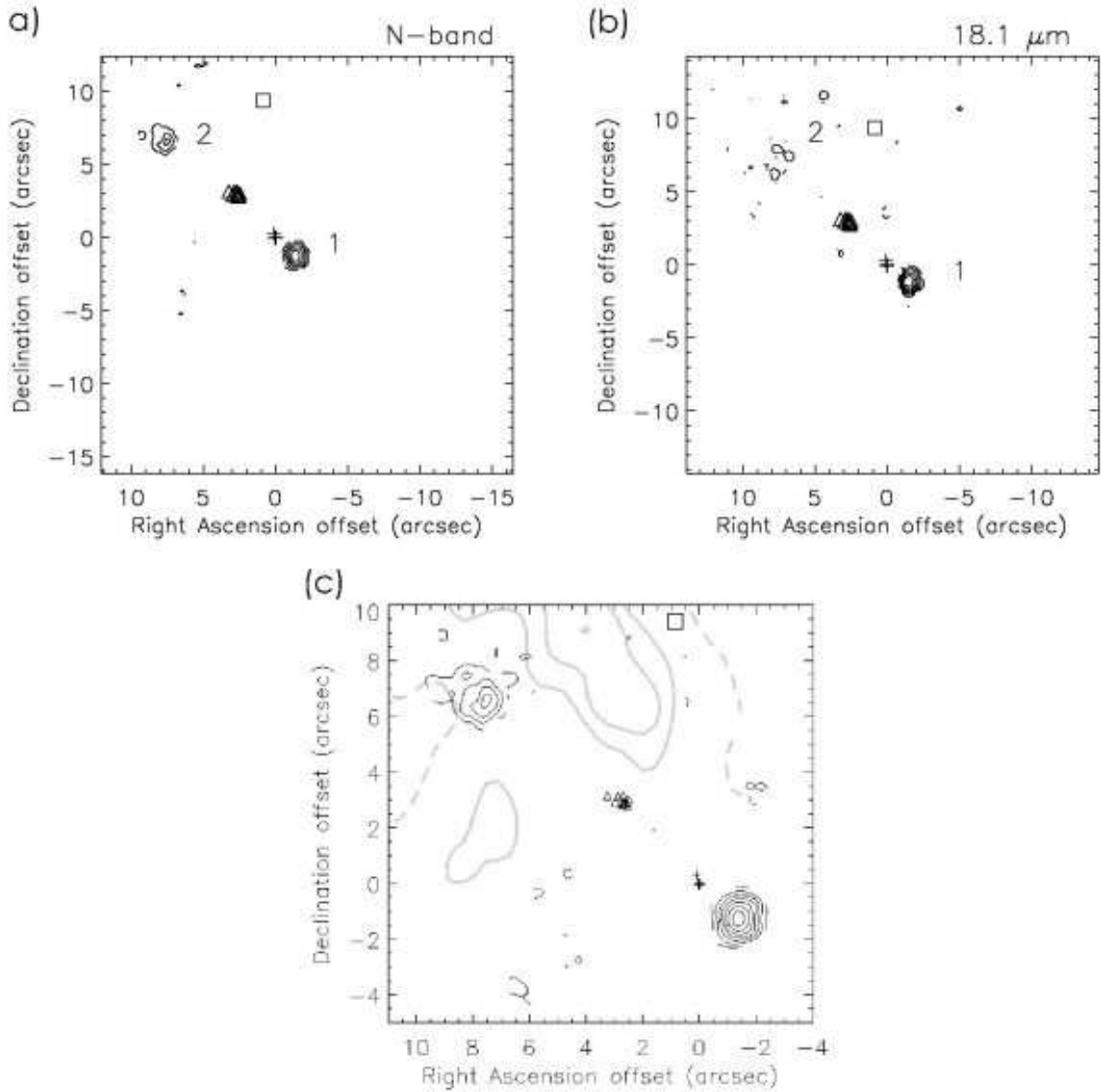}
\caption{Contour plots of the region of G12.68-0.18 observed at (a) N-band smoothed to a resolution of 1$\farcs$3 and (b) 18.1 $\mu$m smoothed to a resolution of 1$\farcs$7. (c) This plot shows the N-band contours overlaid with the 1.3 cm radio map (thick gray) of Codella, Testi, \& Cesaroni (1997). See Figure 1 for explanation of symbols. \label{fig6}}
\end{figure}

\clearpage

\begin{figure}
\epsscale{0.85}
\plotone{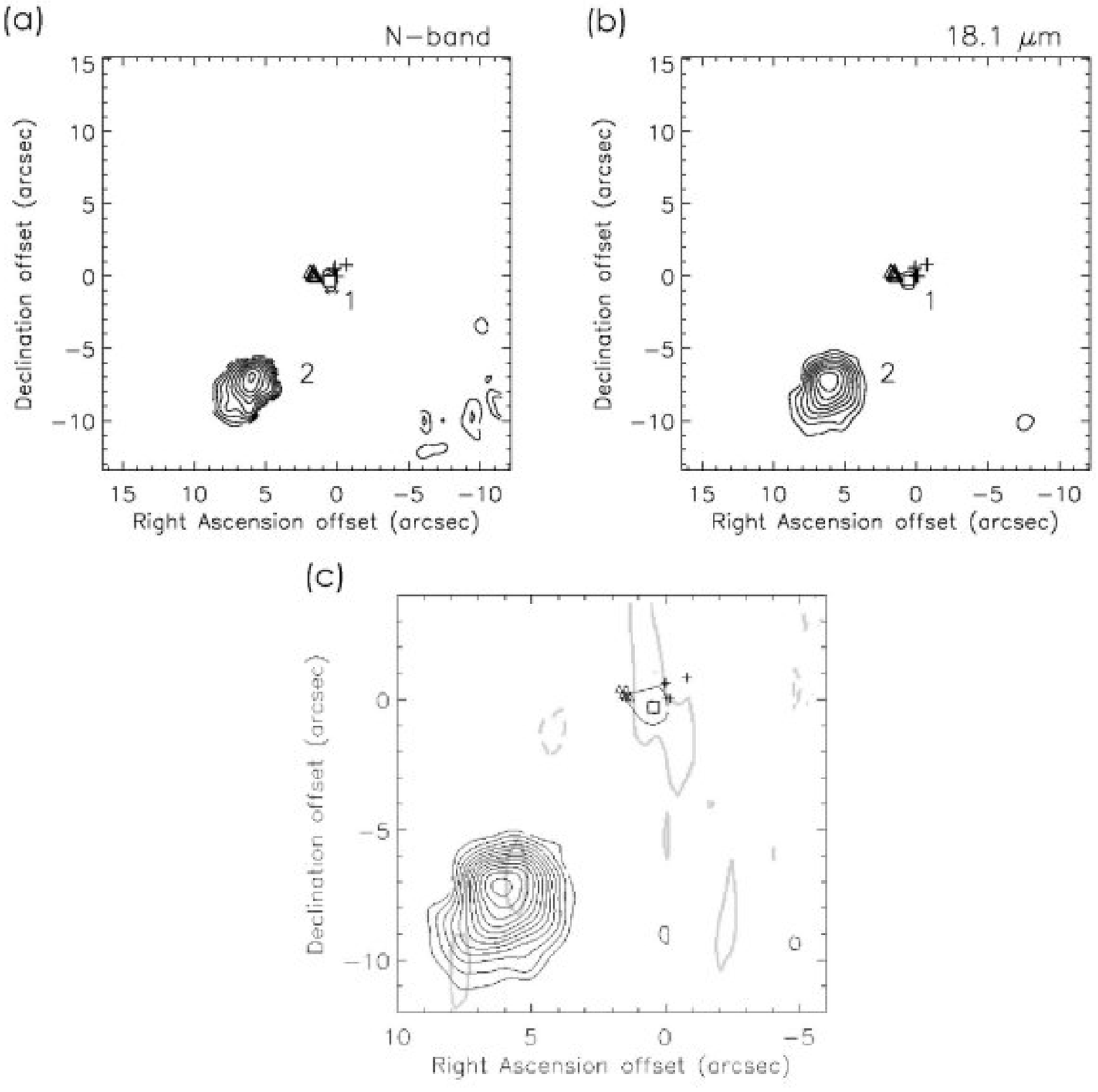}
\caption{Contour plots of the region of G16.59-0.05 observed at (a) N-band smoothed to a resolution of 1$\farcs$6 and (b) 18.1 $\mu$m smoothed to a resolution of 1$\farcs$9. (c) This plot shows the 18.1 $\micron$ contours overlaid with the 3 cm contours (thick gray) from Forster \& Caswell (2000). See Figure 1 for explanation of symbols. \label{fig7}}
\end{figure}

\clearpage

\begin{figure}
\epsscale{0.85}
\plotone{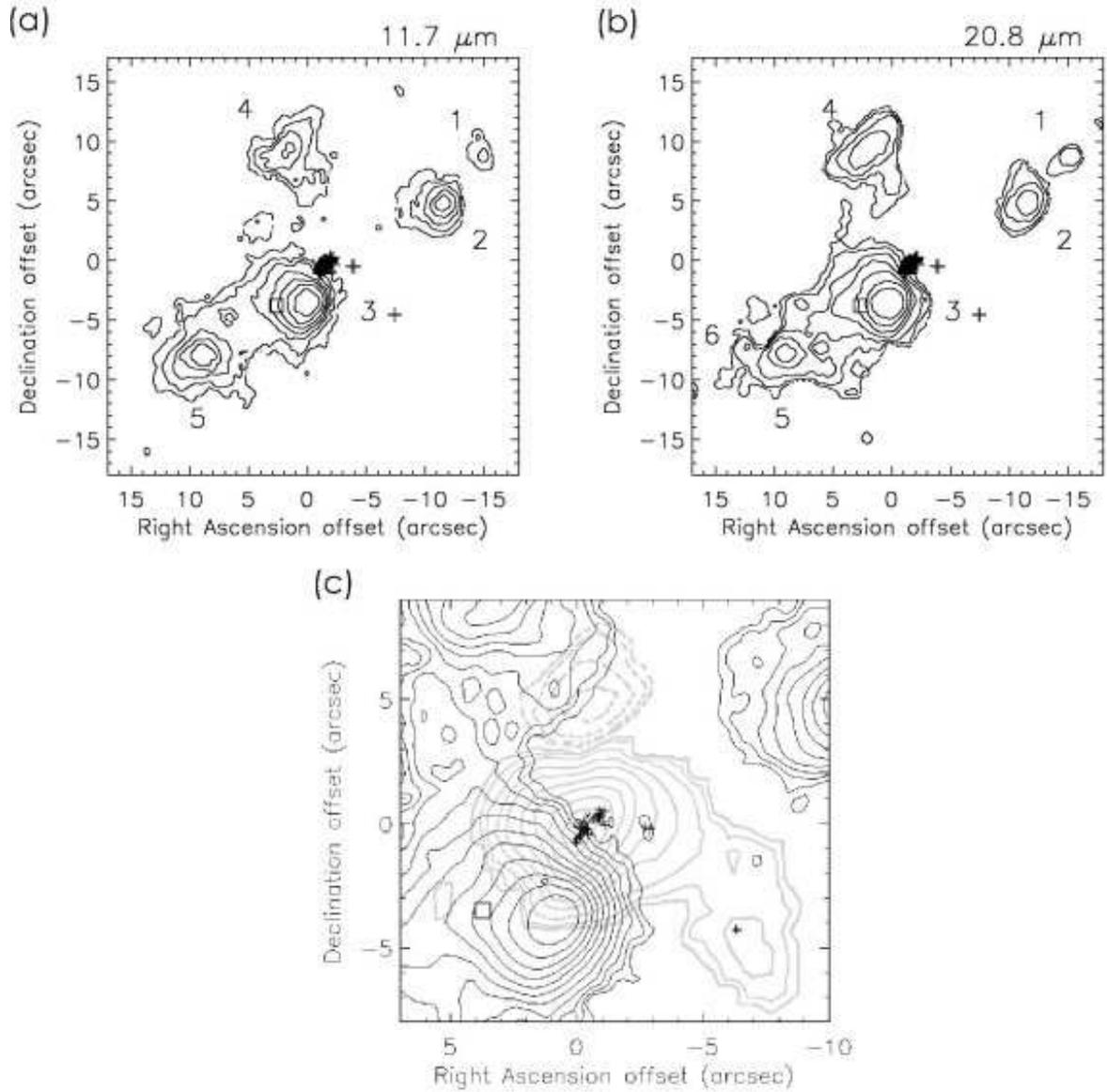}
\caption{Contour plots of the region of G19.61-0.23 observed at (a) 11.7 $\mu$m smoothed to a resolution of 1$\farcs$5 and (b) 20.8 $\mu$m smoothed to a resolution of 1$\farcs$9. (c) This plot shows the shows the 18.1 $\mu$m contours with the ammonia contours (thick gray) from Garay et al. (1998). See Figure 1 for explanation of symbols. \label{fig8}}
\end{figure}

\clearpage

\begin{figure}
\epsscale{0.85}
\plotone{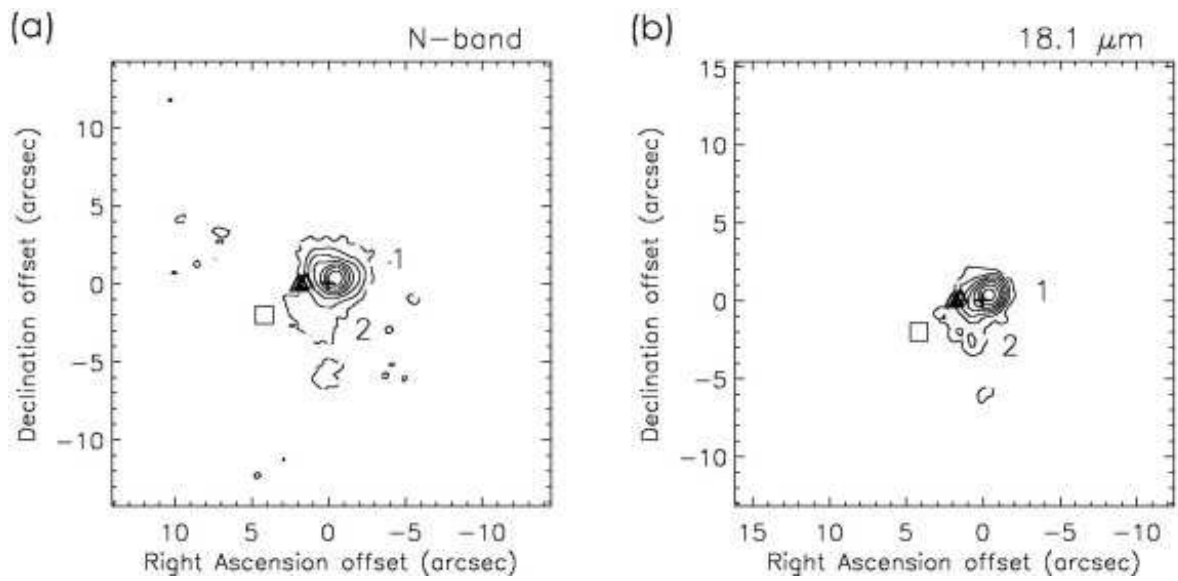}
\caption{Contour plots of the region of G28.86+0.07 observed at (a) N-band smoothed to a resolution of 1$\farcs$3 and (b) 18.1 $\mu$m smoothed to a resolution of 1$\farcs$7. See Figure 1 for explanation of symbols. \label{fig9}}
\end{figure}

\clearpage

\begin{figure}
\epsscale{0.85}
\plotone{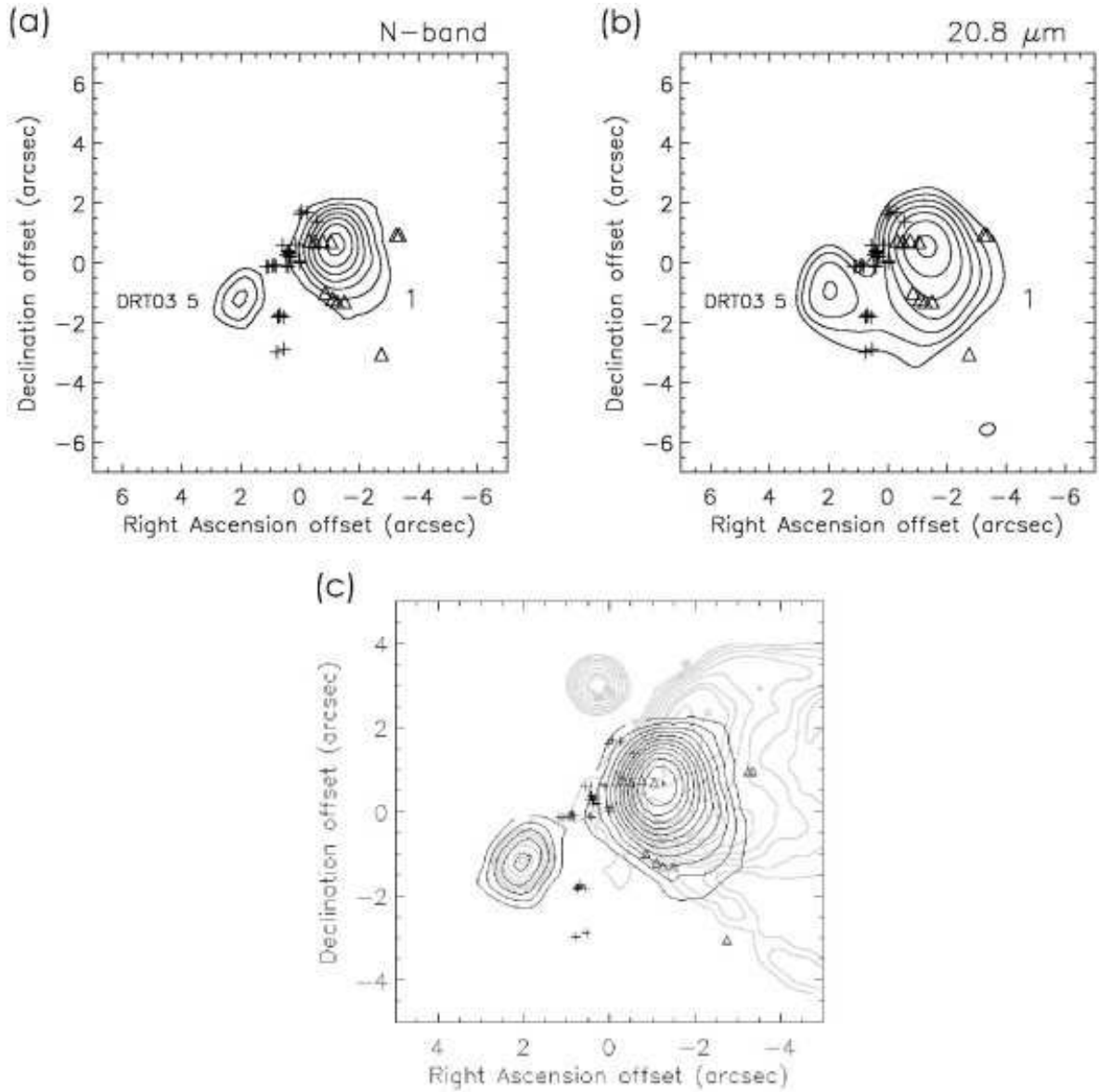}
\caption{Contour plots of the region of G34.26+0.15 observed at (a) N-band smoothed to a resolution of 1$\farcs$3 and (b) 20.8 $\mu$m smoothed to a resolution of 2$\farcs$1. (c) This plot shows the N-band contours overlaid with the 2 cm radio continuum contours (thick gray) from Gaume, Fey, \& Claussen (1994). See Figure 1 for explanation of symbols.\label{fig10}}
\end{figure}

\clearpage

\begin{figure}
\epsscale{0.85}
\plotone{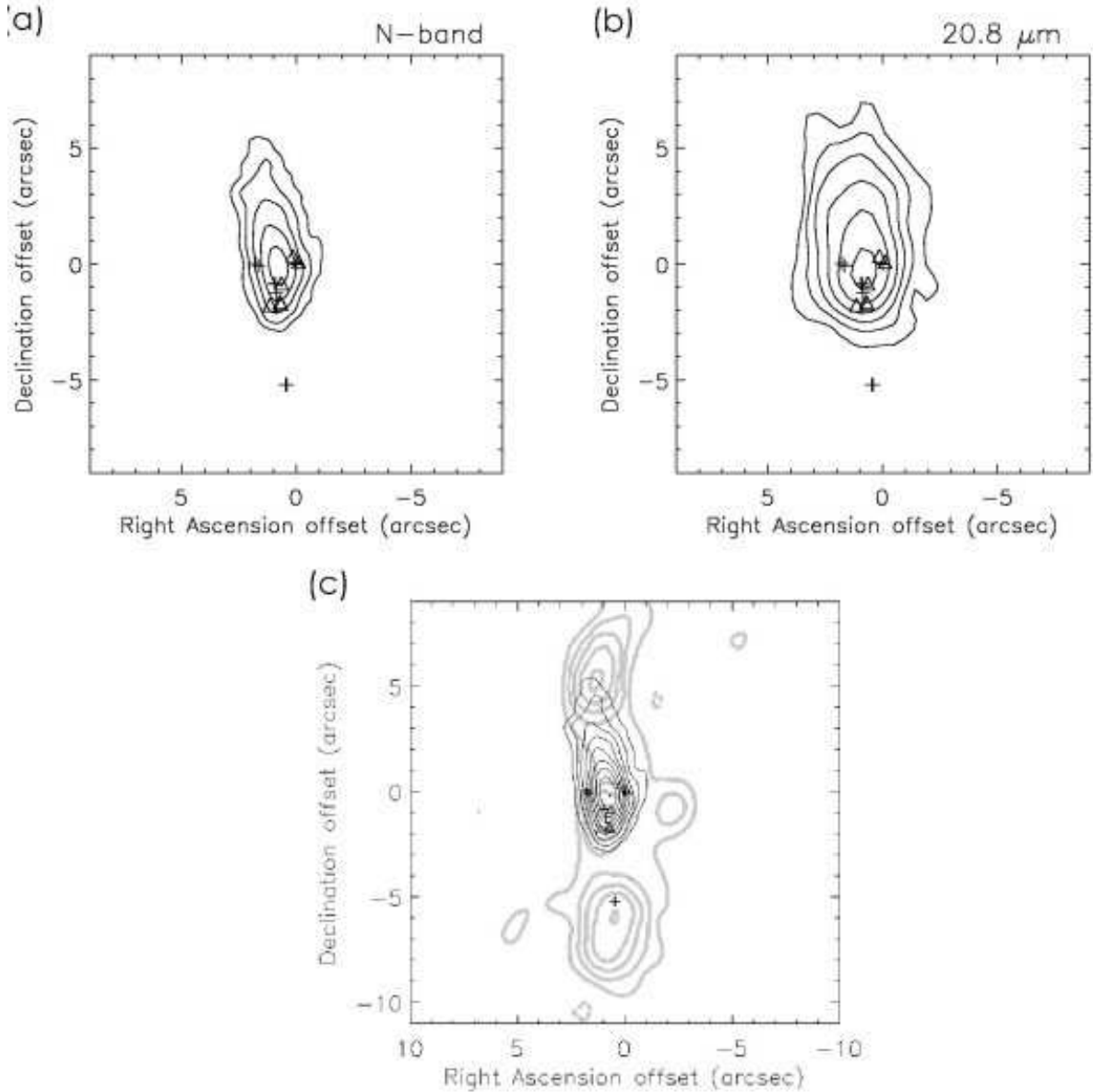}
\caption{Contour plots of the region of G35.20-0.74 observed at (a) N-band smoothed to a resolution of 1$\farcs$4 and (b) 20.8 $\mu$m smoothed to a resolution of 2$\farcs$0. (c) This plot shows the N-band contours overlaid with the 2 cm radio contours from Heaton \& Little (1988). See Figure 1 for explanation of symbols.\label{fig11}}
\end{figure}

\clearpage

\begin{figure}
\epsscale{0.85}
\plotone{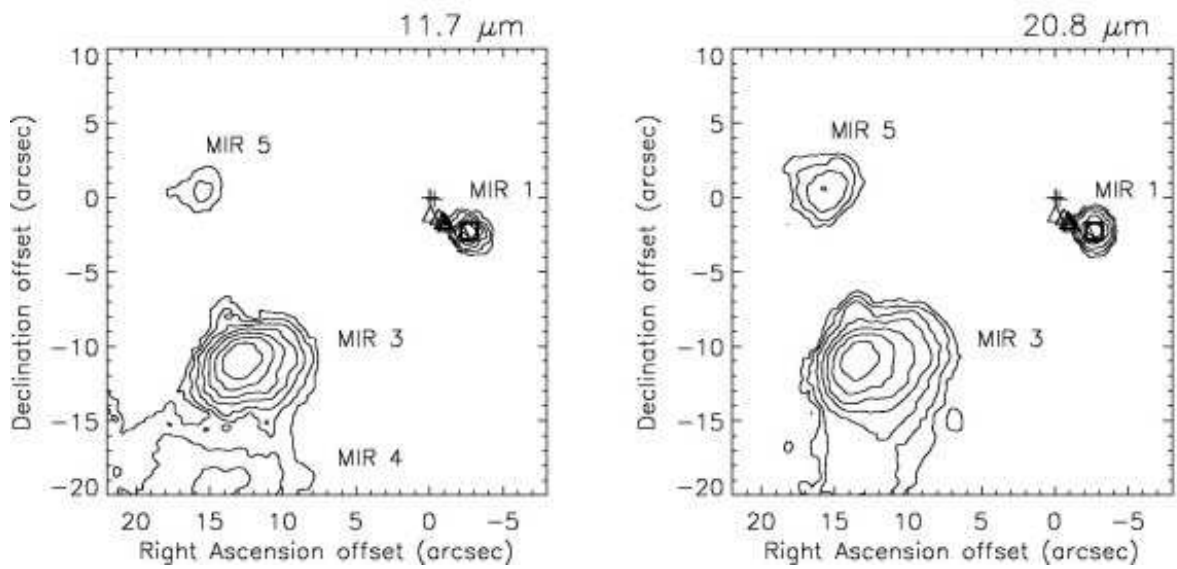}
\caption{Contour plots of the region of G35.20-1.74 observed at (a) 11.7 $\mu$m smoothed to a resolution of 1$\farcs$5 and (b) 20.8 $\mu$m smoothed to a resolution of 1$\farcs$8. Mid-infrared source names are from Persi et al. (1997). See Figure 1 for explanation of symbols.\label{fig12}}
\end{figure}

\clearpage

\begin{figure}
\epsscale{0.85}
\plotone{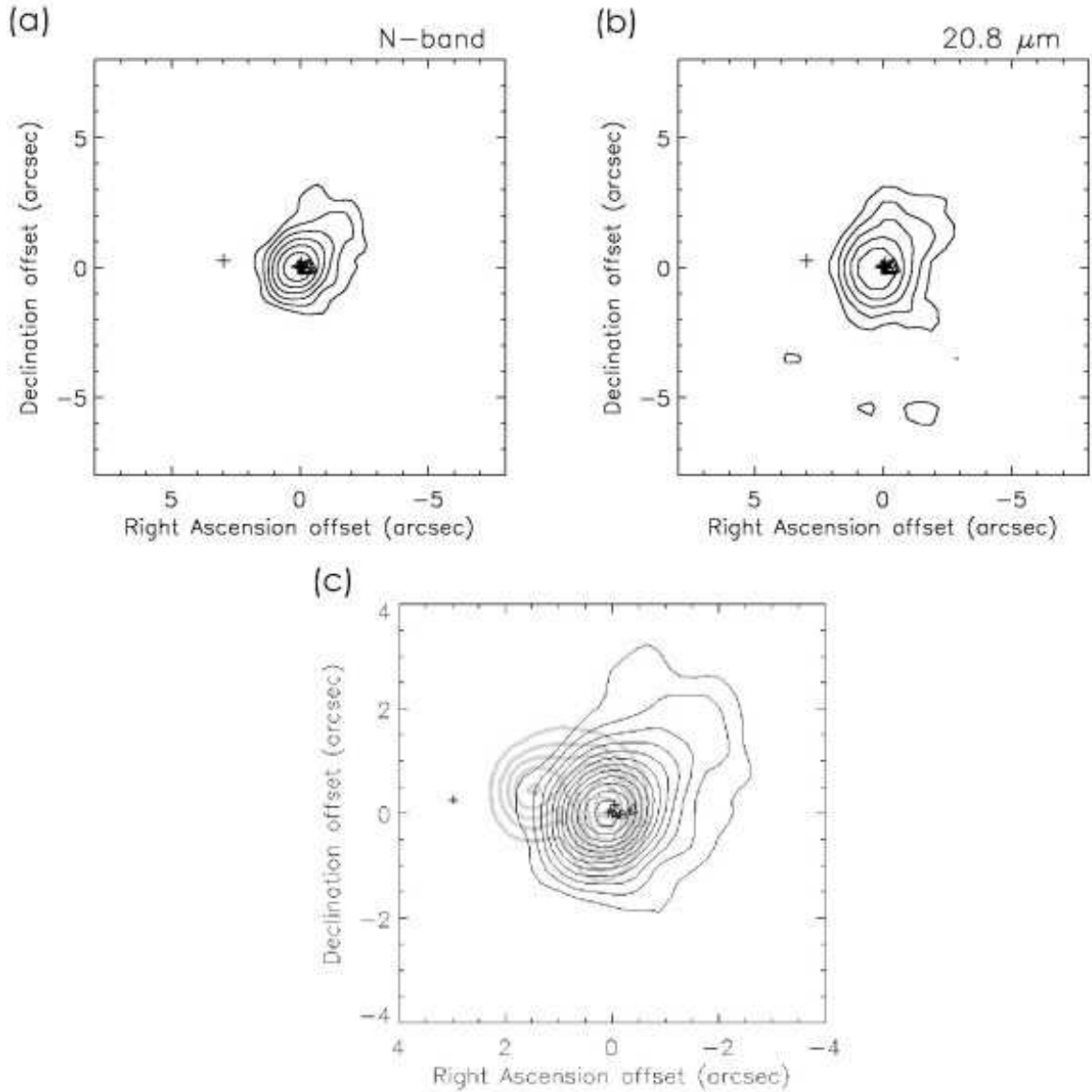}
\caption{Contour plots of the region of G35.58-0.03 observed at (a) N-band smoothed to a resolution of 1$\farcs$4 and (b) 20.8 $\mu$m smoothed to a resolution of 1$\farcs$8. (c) This plot shows the N-band contours overlaid with the 3.6 cm contours (thick gray) from Kurtz, Churchwell, \& Wood (1994). See Figure 1 for explanation of symbols. \label{fig13}}
\end{figure}

\clearpage

\begin{figure}
\epsscale{0.85}
\plotone{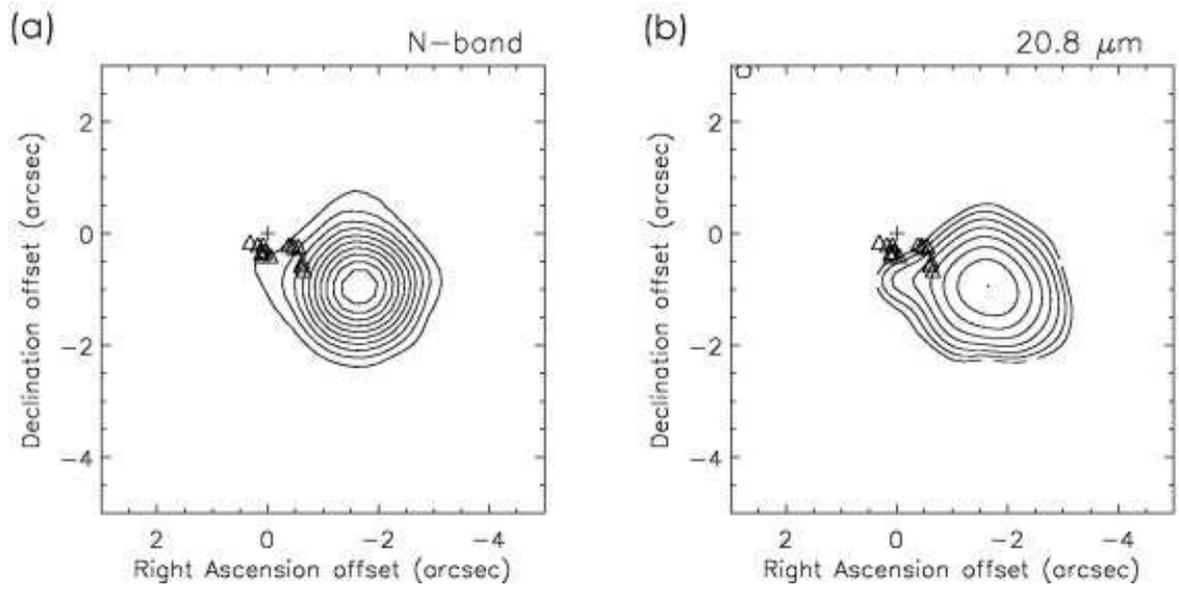}
\caption{Contour plots of the region of G40.62-0.14 observed at (a) N-band smoothed to a resolution of 1$\farcs$4 and (b) 20.8 $\mu$m smoothed to a resolution of 1$\farcs$8. See Figure 1 for explanation of symbols. \label{fig14}}
\end{figure}

\clearpage

\begin{figure}
\epsscale{0.85}
\plotone{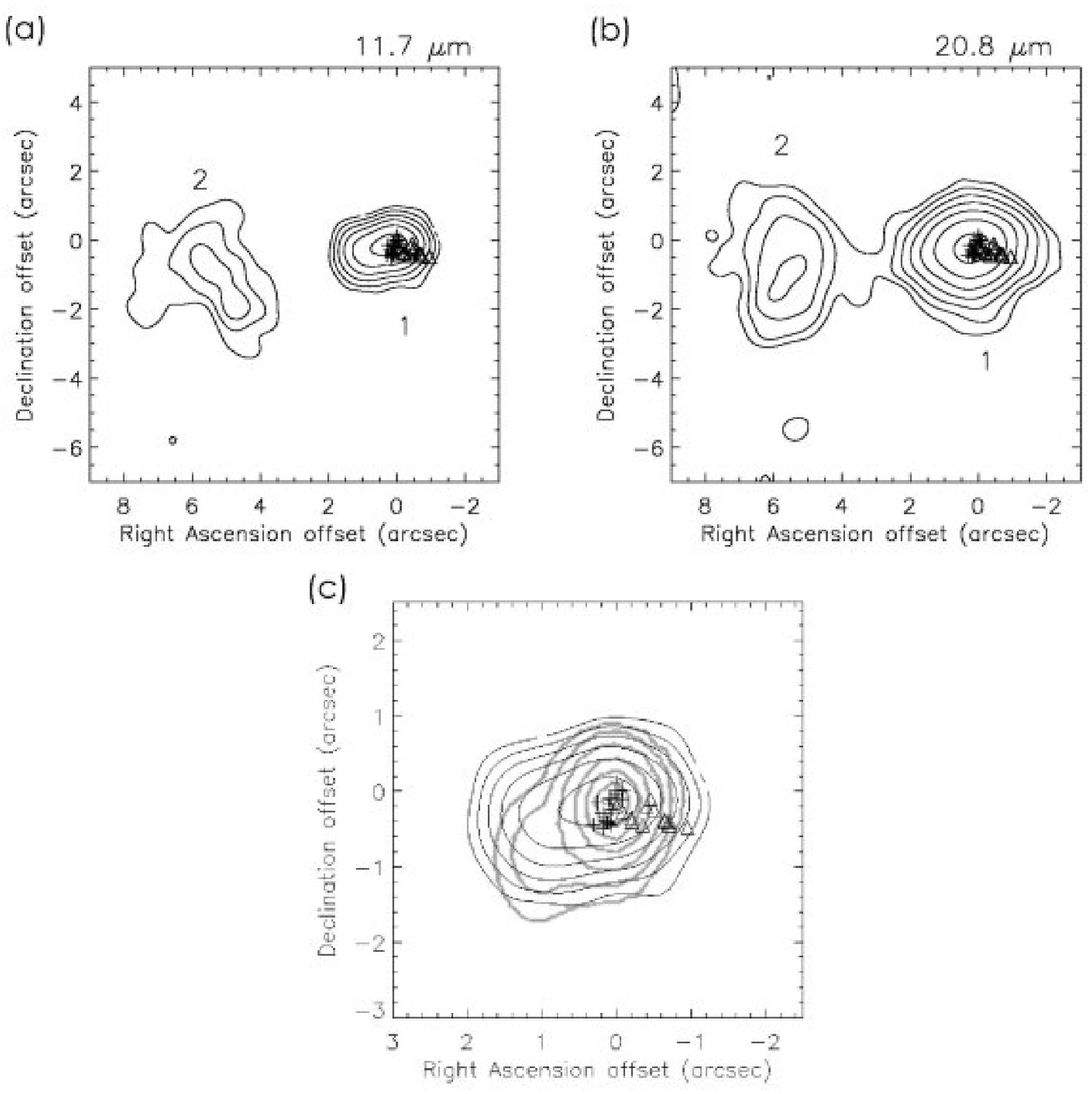}
\caption{Contour plots of the region of G43.80-0.13 observed at (a) 11.7 $\mu$m smoothed to a resolution of 1$\farcs$5 and (b) 20.8 $\mu$m smoothed to a resolution of 1$\farcs$8. (c) This plot shows the N-band contours overlaid with the 3.6 cm radio contours (thick gray) from Kurtz, Churchwell, \& Wood (1994). See Figure 1 for explanation of symbols.\label{fig15}}
\end{figure}

\clearpage

\begin{figure}
\epsscale{0.85}
\plotone{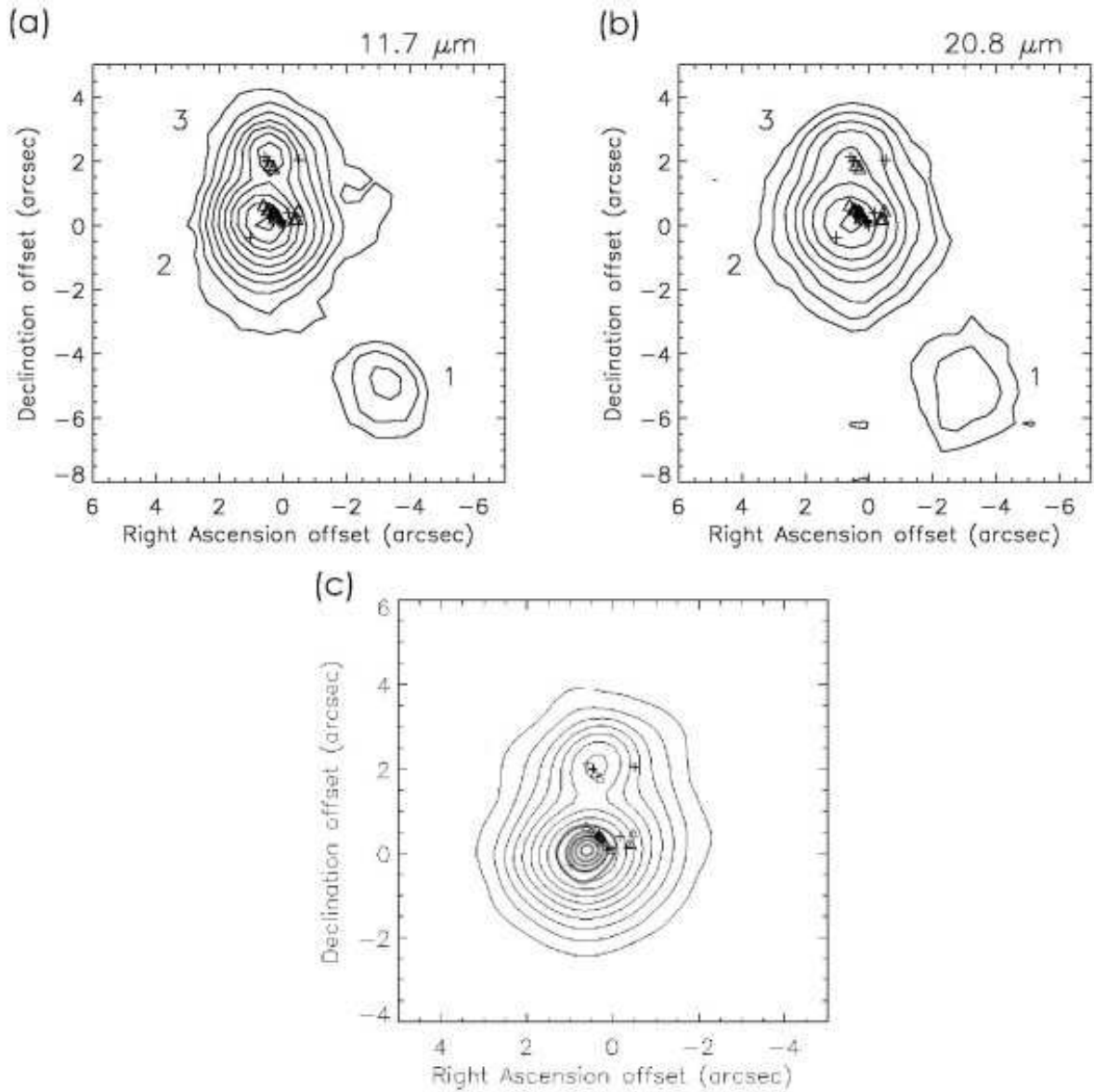}
\caption{Contour plots of the region of G45.07+0.13 observed at (a) 11.7 $\mu$m with no smoothing (resolution of 1$\farcs$3) and (b) 20.8 $\mu$m with no smoothing (resolution of 1$\farcs$7). (c) This plot shows the N-band contours overlaid with the 2 cm radio continuum emission (gray) from Hofner \& Churchwell (1996). See Figure 1 for explanation of symbols.\label{fig16}}
\end{figure}

\clearpage

\begin{figure}
\epsscale{0.85}
\plotone{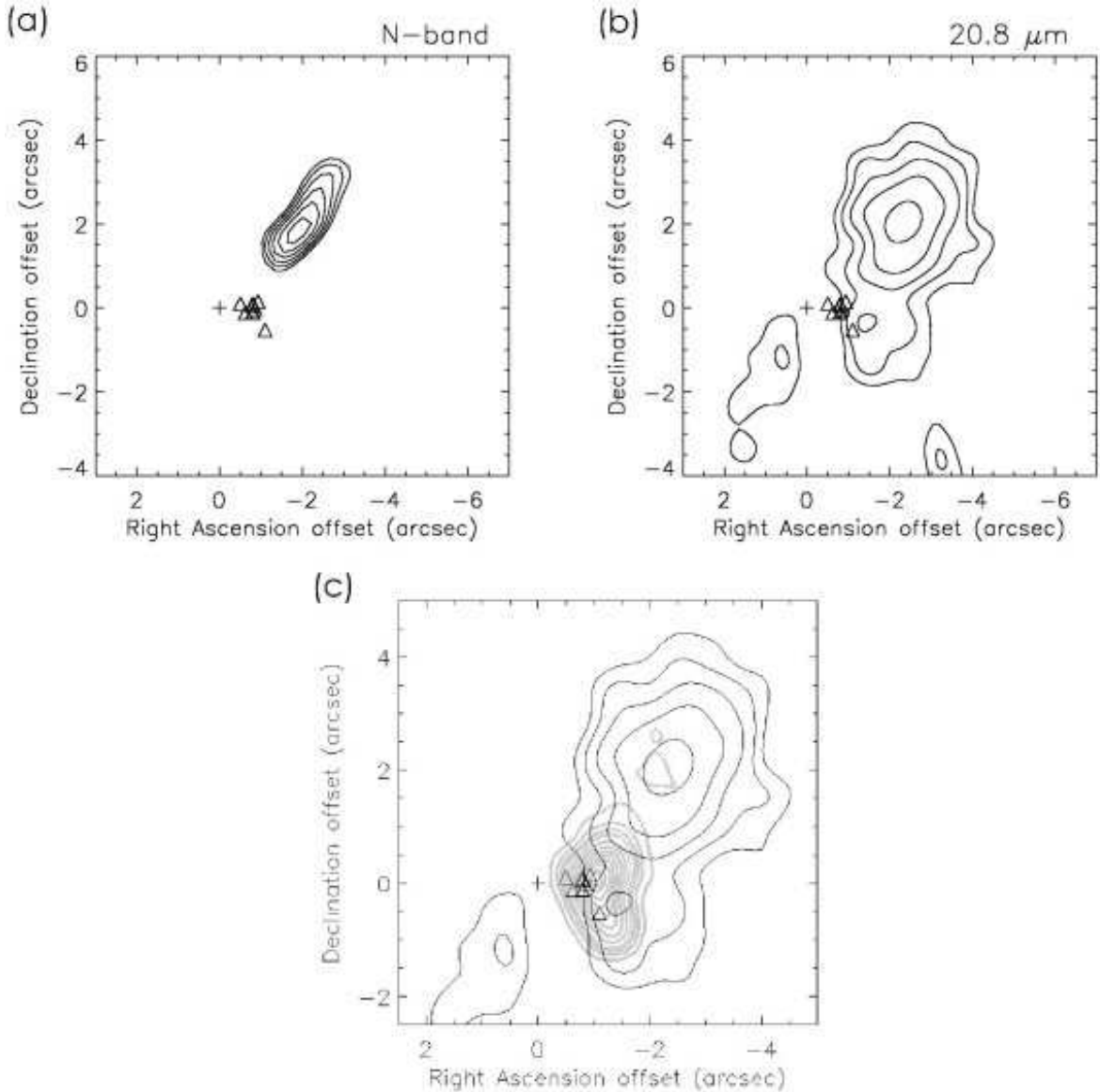}
\caption{Contour plots of the region of G45.47+0.05 observed at (a) N-band smoothed to a resolution of 1$\farcs$6 and (b) 20.8 $\mu$m smoothed to a resolution of 1$\farcs$8. (c) This plot shows the 20.8 $\mu$m contours overlaid with of the 6 cm radio continuum contours (thick gray) from Wood \& Churchwell (1989). See Figure 1 for explanation of symbols.\label{fig17}}
\end{figure}

\clearpage

\begin{figure}
\epsscale{0.85}
\plotone{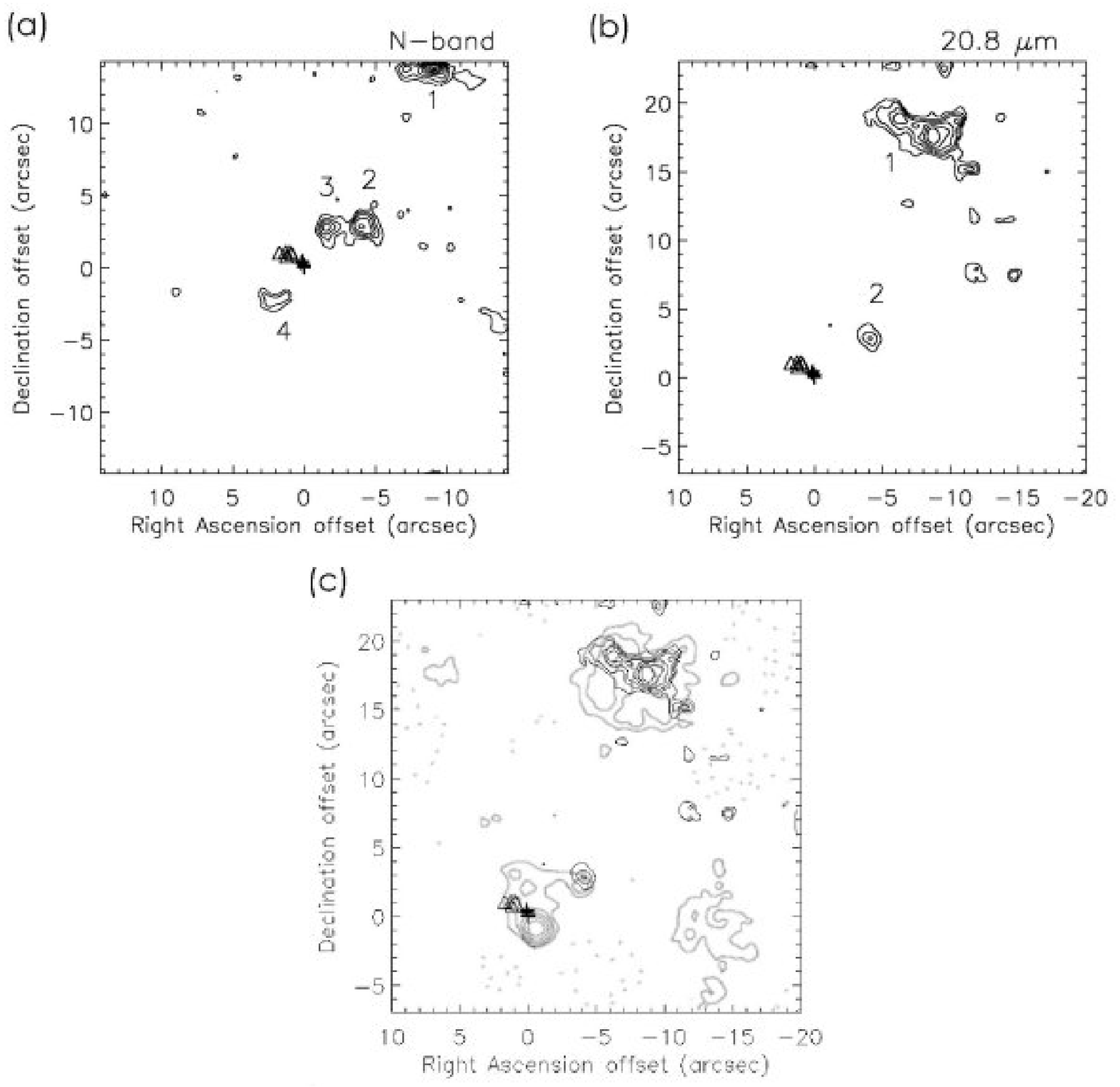}
\caption{Contour plots of the region of G48.61+0.02 observed at (a) N-band smoothed to a resolution of 1$\farcs$4 and (b) 20.8 $\mu$m smoothed to a resolution of 2$\farcs$5. (c) This plot shows the 20.8 $\mu$m contours overlaid with of the 3.6 cm radio continuum contours (thick gray) of Kurtz et al. (1999).  See Figure 1 for explanation of symbols.\label{fig18}}
\end{figure}

\clearpage

\begin{figure}
\epsscale{0.85}
\plotone{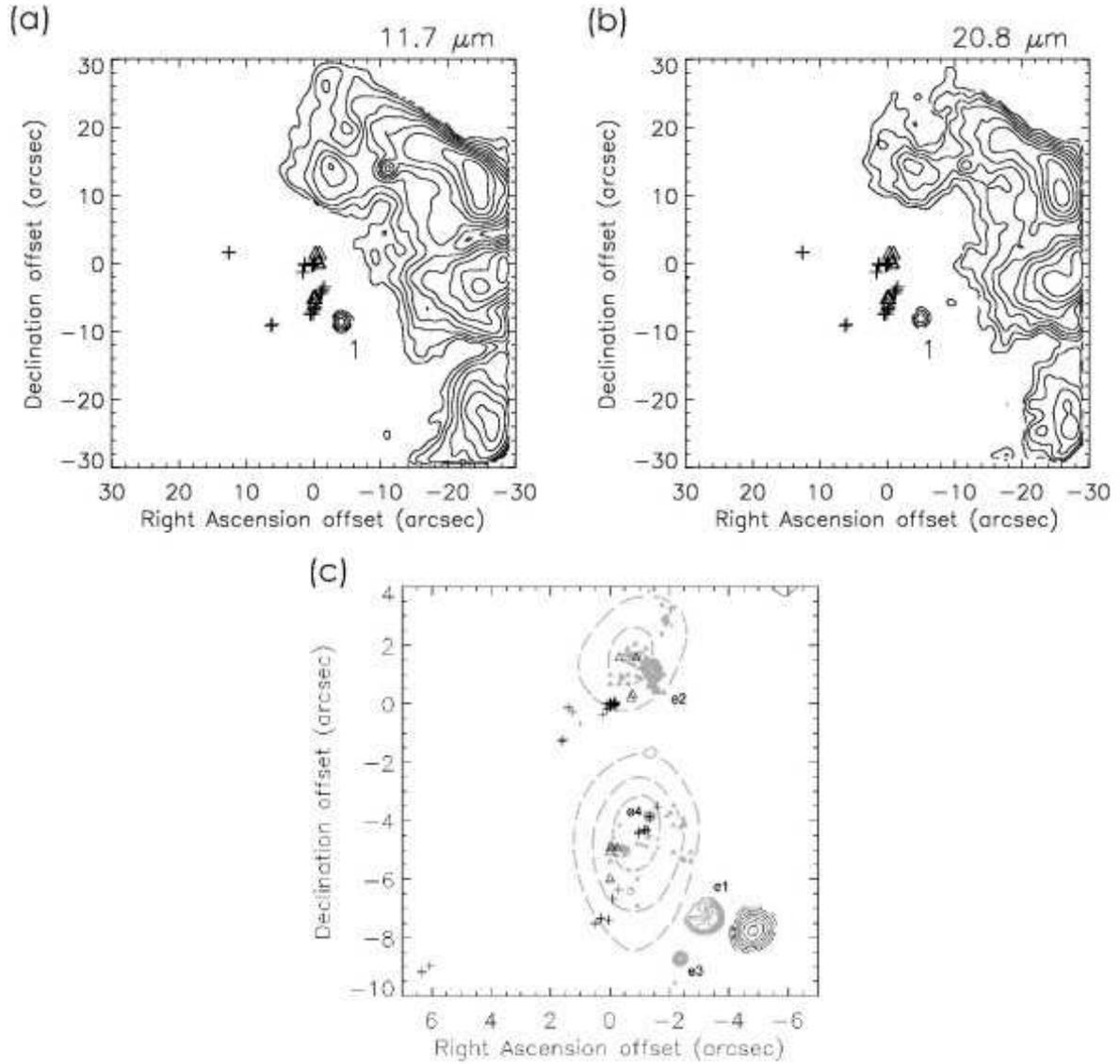}
\caption{Contour plots of the region of G49.49-0.39 observed at (a) 11.7 $\mu$m smoothed to a resolution of 1$\farcs$5 and (b) 20.8 $\mu$m smoothed to a resolution of 1$\farcs$8. (c) This plot shows the N-band contours overlaid with of the 3.6 cm radio continuum (continuous gray) contours of Gaume, Johnston, \& Wilson (1993), and the ammonia contours (broken gray) from Ho et al. (1983). See Figure 1 for explanation of symbols.  Other gray symbols are other water and OH masers as observed by of Gaume, Johnston, \& Wilson (1993).\label{fig19}}
\end{figure}

\clearpage

\begin{figure}
\epsscale{0.85}
\plotone{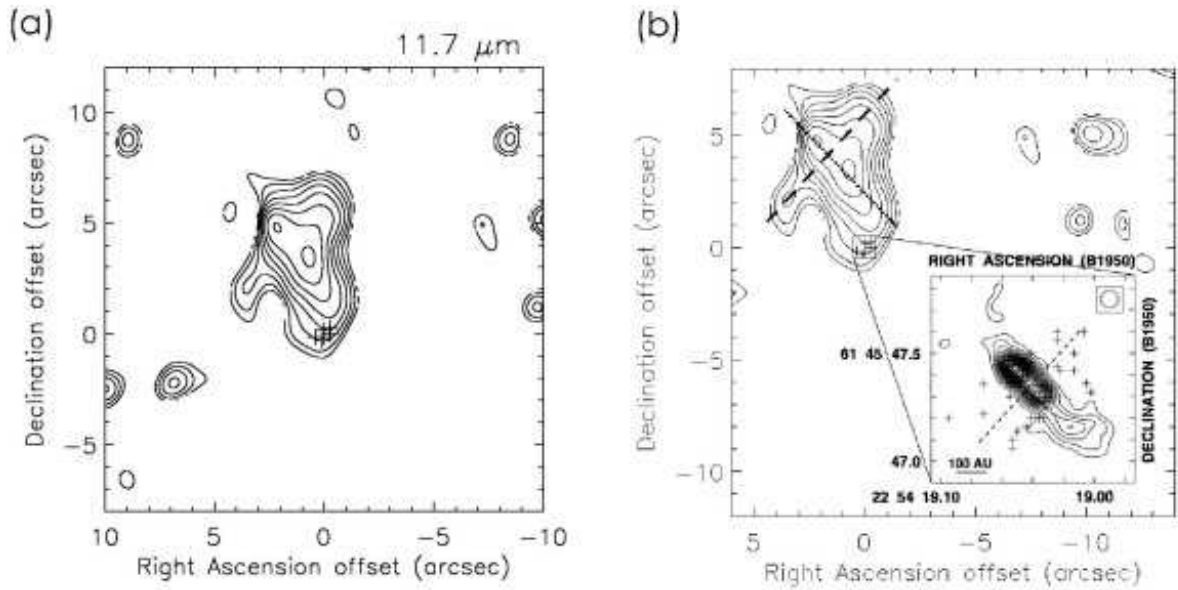}
\caption{Contour plot of the region of Cepheus A HW2 observed at (a) 11.7 $\mu$m smoothed to a resolution of 1$\farcs$5. (b) This plot again shows the 11.7 $\mu$m image of the region with a inset showing the region around the water masers magnified. This inset is from Torrelles et al. (1996) and shows the 1.3 cm continuum emission contours as well as a more detailed view of the distribution of water masers. The dashed and dotted lines delineate how the mid-infrared source appears to be elongated both parallel to the water maser distribution and parallel to the radio continuum emission, though these features appear offset from the mid-infrared source center.  \label{fig20}}
\end{figure}

\clearpage

\begin{figure}
\epsscale{0.85}
\plotone{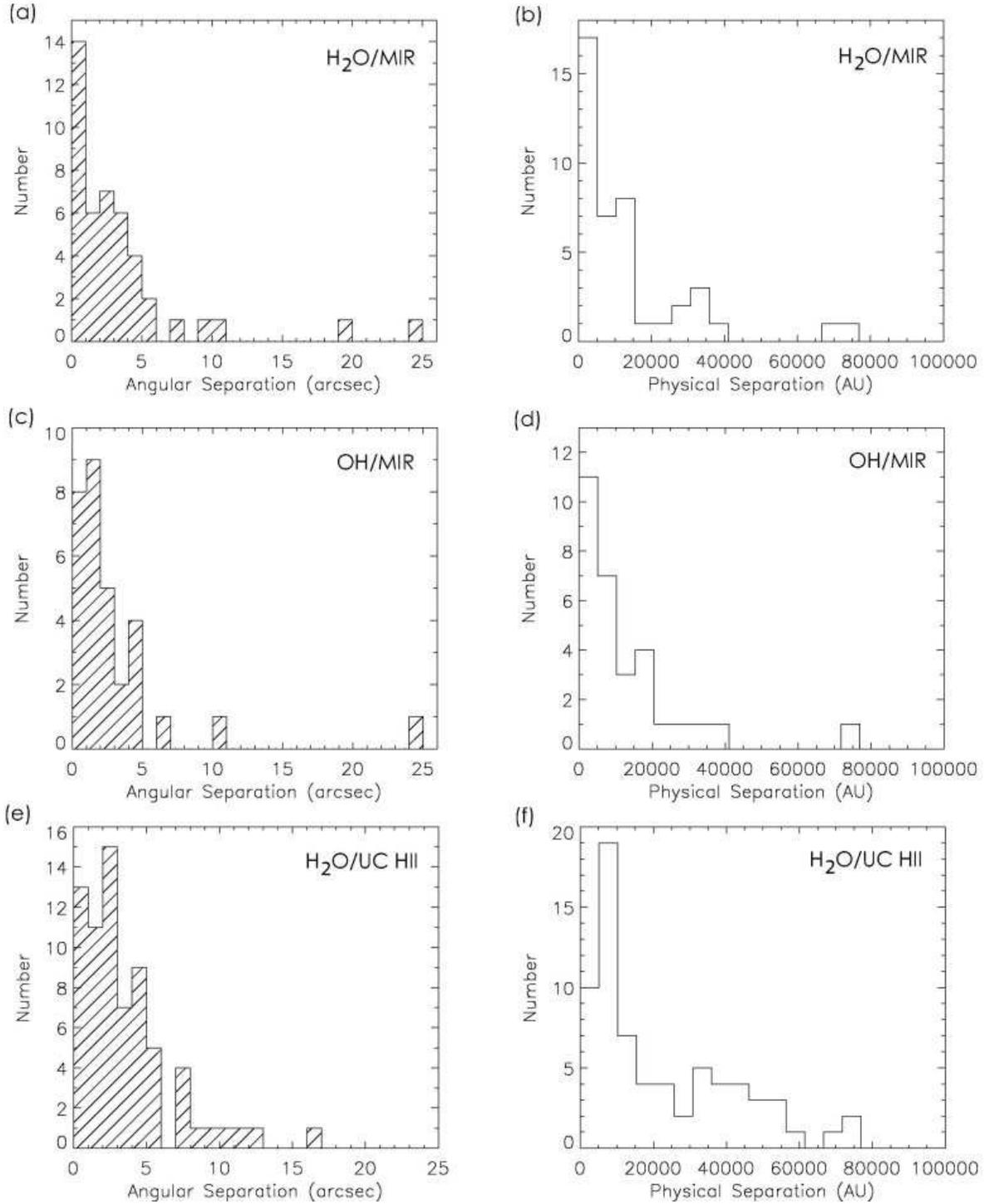}
\caption{Separations between masers and mid-infrared and UC \ion{H}{2} region source peaks. The top two histograms show the water maser number as it is related to distance from the nearest mid-infrared source in terms of (a) angular distance and (b) physical distance. The middle two histograms show the same for the OH masers, again in terms of (c) angular distance and (d) physical distance. The bottom two histograms show water maser numbers with respect to distance from UC \ion{H}{2} regions (from Hofner \& Churchwell 1996) in terms of (e) angular distance and (f) physical distance. \label{fig21}}
\end{figure}

\clearpage

\begin{figure}
\epsscale{0.85}
\plotone{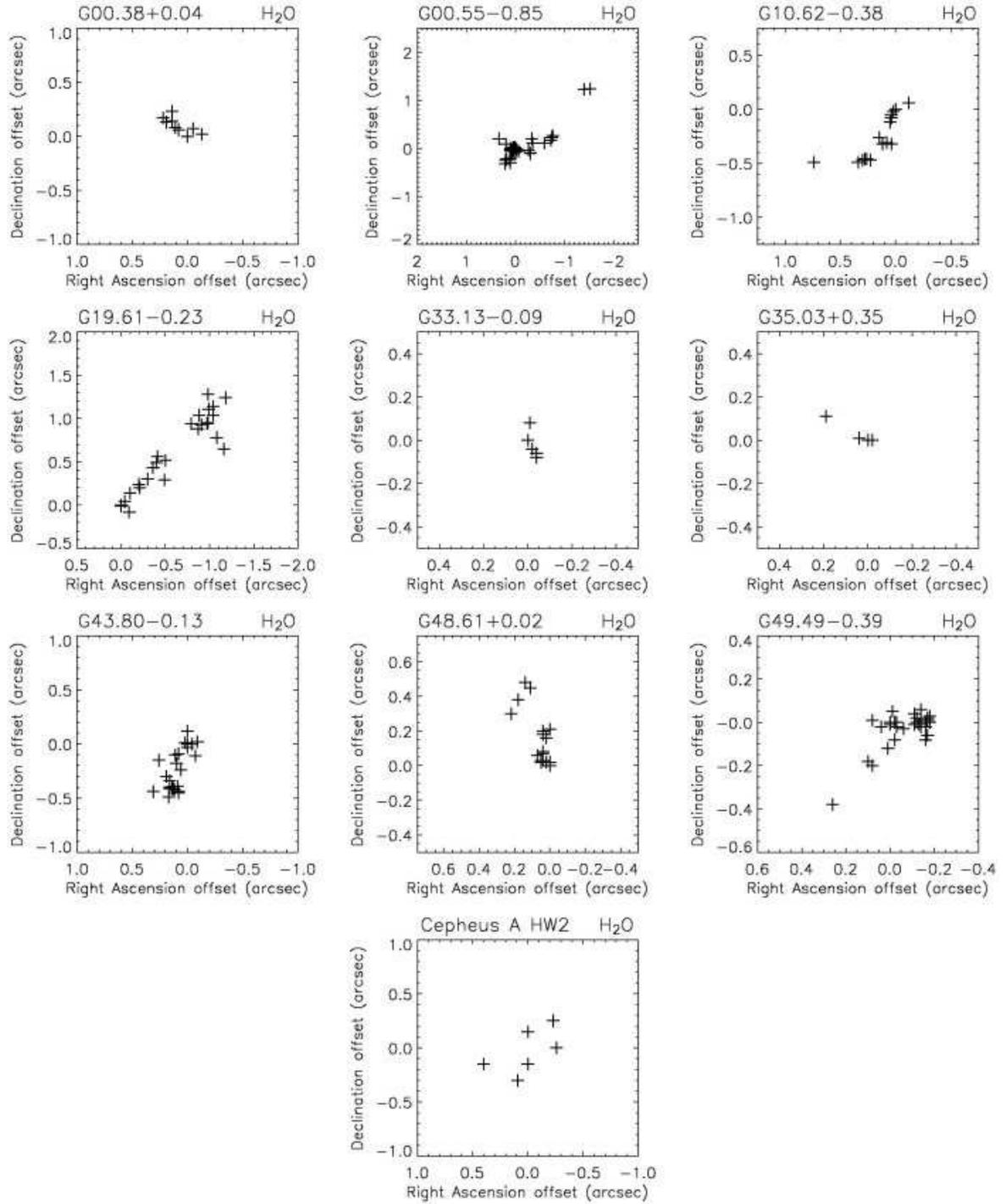}
\caption{All 10 of the water maser sites in this survey that appear to have their masers arranged in a linear or elongated fashion similar to that seen in methanol maser groups of Norris et al. (1993). The elongated distribution of masers in G48.61+0.02 may be two separate linear distributions. The water masers of Cepheus A HW2, previously thought to delineate a circumstellar disk are actually the most dispersed of the linear distributions. \label{fig22}}
\end{figure}

\clearpage

\begin{figure}
\epsscale{0.85}
\plotone{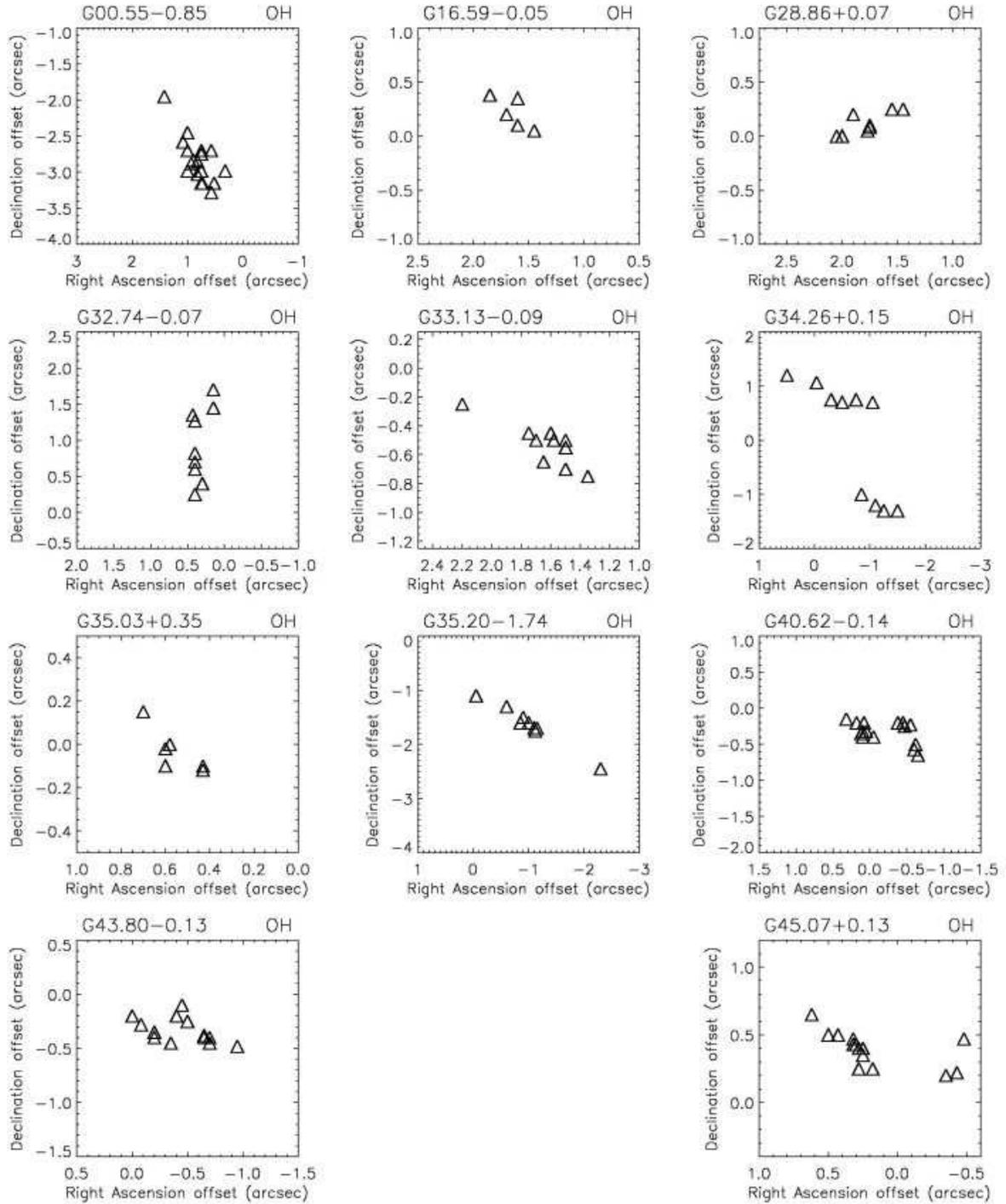}
\caption{All 11 of the hydroxyl maser sites in this survey that appear to have their masers arranged in a linear or elongated fashion similar to that seen in methanol maser groups of Norris et al. (1993). G34.26+0.15 has two hydroxyl maser groups that are linearly distributed separated by $\sim$2$\arcsec$. The elongated distribution of masers in G40.62-0.14 may be two separate linear distributions as well. There are two nearby groups of hydroxyl masers in G45.07+0.13, the western group being linearly distributed. \label{fig23}}
\end{figure}

\clearpage

\end{document}